\documentclass[reqno,9pt]{amsart}

\usepackage{amsthm, amsmath, amssymb, amsfonts, graphicx}
\usepackage[OT4]{fontenc}
\usepackage[utf8]{inputenc}
\usepackage[english]{babel}
\usepackage{bbm}
\usepackage[inline]{enumitem}
\usepackage{float}
\restylefloat{table}
\usepackage[colorlinks=true, pdfstartview=FitV, linkcolor=blue,
            citecolor=blue, urlcolor=blue]{hyperref}
\usepackage[usenames]{color} 
\usepackage[round]{natbib}

\usepackage{graphicx}
\usepackage[font=sl,labelfont=bf,width=\textwidth]{caption}
\usepackage{color}

\DeclareMathOperator{\sgn}{sgn}

\newcommand{\stab}{{$\alpha$-stable }}

\DeclareMathOperator*{\argmax}{argmax}
\DeclareMathOperator*{\argmin}{argmin}

\usepackage{multirow}
\usepackage{amsthm}
\newtheorem{theorem}{Theorem}

\newtheorem{proposition}[theorem]{Proposition}

\newtheorem{remark}[theorem]{Remark}

\def\bR{\mathbb{R}}
\def\bN{\mathbb{N}}
\usepackage{bbm}

\usepackage{amssymb}

\usepackage{placeins} 
\usepackage{graphicx}

\usepackage{setspace}
\usepackage[foot]{amsaddr}

\begin{document}

\title[Estimation of stability index for symmetric $\alpha$-stable distribution]{Estimation of stability index for symmetric $\alpha$-stable distribution using quantile conditional variance ratios}

\author{Kewin P\k{a}czek$^{\ast}$}
\author{Damian Jelito$^{\ast}$}
\author{Marcin Pitera$^{\ast}$}
\author{Agnieszka Wy\l{}oma\'{n}ska$^{\dagger}$}
\address{$^{\ast}$Institute of Mathematics, Jagiellonian University, S. {\L}ojasiewicza 6, 30-348 Krak{\'o}w, Poland}
\address{$^{\dagger}$Faculty of Pure and Applied Mathematics, Hugo Steinhaus Center, Wroc{\l}aw University of Science and Technology, Wyspia{\'n}skiego 27, 50-370 Wroc{\l}aw, Poland}
\email{kewin.paczek@im.uj.edu.pl, damian.jelito@uj.edu.pl, marcin.pitera@uj.edu.pl, agnieszka.wylomanska@pwr.edu.pl}

\maketitle

\vspace{-1cm}
\begin{abstract}
The class of \stab distributions is widely used in various applications, especially for modelling heavy-tailed data. Although the \stab distributions have been used in practice for many years, new methods for identification, testing, and estimation are still being refined and new approaches are being proposed. The constant development of new statistical methods is related to the low efficiency of existing algorithms, especially when the underlying sample is small or the distribution is close to Gaussian. In this paper we propose a new estimation algorithm for stability index, for samples from the symmetric \stab distribution. The proposed approach is based on a quantile conditional variance ratio. We study the statistical properties of the proposed estimation procedure and show empirically that our methodology often outperforms other commonly used estimation algorithms. Moreover, we show that our statistic extracts unique sample characteristics that can be combined with other methods to refine existing methodologies via ensemble methods. Although our focus is set on the symmetric \stab case, we demonstrate that the considered statistic is  insensitive to the skewness parameter change, so that our method could be also used in a more generic framework. For completeness, we also show how to apply our method on real data linked to plasma physics.
\vspace{0.2cm}

\end{abstract}

%%%%%%%%%%%%%%%%%%%%%%%%%%%%%%%%%%%%%%%%%%%%%%%%%%%%%%%%%%%%%%%%%%%%%%%%%%%%%%%%%%%%%%%%%%%%%%%%%%%%%%

\section{Introduction}
The \stab distributions are an important modelling tool as they constitute a domain of attraction for sums of independent random variables. That is, if the sum of independent and identically distributed (i.i.d.) random variables converges in distribution, then the limiting distribution belongs to the $\alpha$-stable family by the generalized Central Limit Theorem, see \cite{levy1924theorie}, \cite{khinchine1936lois}, \cite{shao}, and \cite{stable} for classical references. The family of \stab probability distribution is often used to describe the generalized white noise, without any assumptions imposed on the underlying moments; note that the sums of i.i.d. \stab random variables retain the shape of the original distribution, which is known as the stability property, see \cite{feller1966} and \cite{limit1}.

The first applications of the \stab distributions were related to generic financial data description and telephone line noise modelling, see \cite{mand} and \cite{stuck}. Since 1990s, the \stab distributions have found numerous other applications linked e.g. to financial markets, telecommunications,  condition monitoring, physics, biology, medicine, and climate dynamics, see \cite{fin_new1,fin_new2,mon_new1,mon_new2,mon_new3,phys_new1,phys_new2,phys21,phys22,phys23}, and references therein. See also \cite{shao22,alek_book,non_gauss,nolan_book,biol_new1,biol_new2,phys35,phys33} for classical positions on \stab distributed processes.

In general, the \stab distributions are characterized by four parameters corresponding to location, scale, tail structure, and symmetry. In this paper we consider symmetric \stab distributions and focus on the modelling of the {\it tail index} $\alpha \in(0,2]$, also called the {\it stability index}, responsible for the tail power-law behavior of the \stab distribution, see~\cite{nolan_book}. For $\alpha<2$, the \stab distribution belongs to the wide class of the heavy-tailed distributions and in this case the corresponding random variable has infinite variance; the smaller the value of $\alpha$, the heavier the tail. On the other hand, for $\alpha=2$, the \stab distribution reduces to the Gaussian distribution. Also, it should be noted that in general the probability density function (PDF) of \stab distribution is not given in an explicit form, and the distribution structure is typically expressed via its characteristic function.

In this paper we consider the problem of statistical estimation of the stability index for the symmetric \stab distribution. The estimation problem has been studied by many authors for more than 50 years and one can find various approaches in the literature. The most recognized methods are linked to quantile methods, see e.g. \cite{fama,est3_new,dominicy,Huixia,leitch}, techniques based on the characteristic functions, see e.g.  \cite{press,weronr,est4_new,kogon,HASSANNEJAD,arad,Paulson}, maximum likelihood, see e.g. \cite{dumouchel,Nolan2001,MITTNIK,Matsui,Brorsen},  Hill's estimators, see e.g. \cite{Picket1998,Pickands1975,DeHaan1980,Dekkers1990,Resnick1997}, and other techniques tailored for the \stab distributions, see e.g. \cite{lombardi,Escobar2016,teimouri,garcia2011estimation,nikias1,Sathe,DEHAAN199939, Matusi2018}. Also, we refer to \cite{Akgiray} for a comparative study of the estimation algorithms for the $\alpha$-stable distribution. 

While most of the considered estimators have nice theoretical properties and their asymptotic behavior is well studied, they are not very effective when applied to practical data. This is mainly due to the fact that for \stab distributed samples the estimation algorithms to be effective require a large sample sizes  as well as substantial computational power. This is especially visible when the stability index is close to $2$, i.e. when the \stab distribution tends to the Gaussian distribution. The related problem of discriminating the \stab distribution from the Gaussian one was discussed in the literature e.g. in \cite{krzysiek0,krzysiek1,PitCheWyl2021,iskander2020}.

To overcome the problem of small sample size, lack of the explicit form of PDF, and infinite variance of the underlying sample, the approach proposed in this paper is based on the quantile conditional variance (QCV) class of statistics. The QCV based methods have been recently proposed mainly in the context of goodness-of-fit testing or to explain empirical phenomena such as the 20/60/20 rule, see \cite{JelPit2018} and \cite{JawPit2015}. Also, this methodology has been recently applied for the local damage detection based on the signals with heavy-tailed background noise, or studies the asymptotic behavior of empirical processes  see \cite{HebZimWyl2020,HebZimPitWyl2019,Ghoudi2018}. 

Although the theoretical variance for the \stab distribution with $\alpha<2$ is infinite, the QCV always exists and can be used to characterize the \stab distribution up to the location, see \cite{JawPit2020}. Here, we extend the \stab distribution goodness-of-fit methodology proposed in \cite{PitCheWyl2021} and show that it can be efficiently used for \stab tail index estimation in the symmetric case. While our approach can be generally classified as the quantile method, we show that it is more effective than the commonly used algorithms proposed in \cite{est3_new}. This is mainly due to the fact that QCV gathers information about multiple quantiles at the same time and better reflects the tail structure. Moreover, the empirical QCV based ratio statistics proposed in this paper are easy to implement as they are based on sample QCV. This makes our approach effective for small samples and computationally efficient. In addition, our data study shows that the QCV approach could be successfully applied to estimate the tail index in the near-Gaussian environment, i.e. when the true stability index is close to $2$. As already mentioned, in this paper we focus on the tail index estimation for the symmetric \stab distributions. However, we also demonstrate that the approach proposed is in fact insensitive to the skewness parameter changes and could be easily extended to a general setting. Finally, it is worth mentioning that the QCV approach extracts sample characteristics different to the ones extracted by other estimation procedures such as the regression methods. Consequently, the QCV approach could also be used to refine the existing frameworks, e.g. via ensemble learning, see~\cite{Pol2006} and references therein.

The rest of the paper is organized as follows. In Section \ref{S:sym_alfa_stab} we recall the definition of the symmetric \stab distribution as well as the main facts related to the corresponding tail behavior. Moreover, we discuss the characterization of the \stab distribution via the conditional variance.   Next, in Section \ref{S:cond.variance.fit} we analyze the sample QCV for the \stab distribution and introduce the estimation technique based on the QCV approach.  In Section \ref{simul} we present the simulation study where for Monte Carlo simulated samples we demonstrate the efficiency of the methodology  proposed. In this section we compare the results with the benchmark methods. By simulation study we also demonstrate the robustness of the methodology in question for the skewness parameters change.  Finally, in Section \ref{real_data}, we analyze real data  obtained in experiments on the controlled thermonuclear fusion and show that the methodology  proposed efficiently  discriminates the \stab distributions with $\alpha$ close to $2$.
In the last section we conclude the paper.

%%%%%%%%%%%%%%%%%%%%%%%%%%%%%%%%%%%%%%%%%%%%%%%%%%%%%%%%%
\section{The symmetric $\alpha$-stable distribution}\label{S:sym_alfa_stab}

There are many ways to define the family of $\alpha$-stable distributions. In this paper we use the {\it 0-parametrization} and follow the characteristic function approach, see Section 1.3 in \cite{nolan_book} in which, other parametrizations are stated. Namely, we say that the random variable $X$ follows the $\alpha$-stable distribution with parameters $\alpha \in (0,2]$, $\beta \in [-1,1]$, $c > 0$, $\mu \in \mathbb{R}$ and write $X \sim S(\alpha, \beta,c,\mu)$ if its characteristic function is given by 
\begin{equation}\label{eq:fchar}
    \phi(u) = \mathbb{E}( \exp(iuX)) =
  \begin{cases} 
   \exp\left(-c^\alpha|u|^\alpha \left(1 + i\beta \sgn(u) \tan \frac{\pi \alpha}{2}\left(|cu|^{1-\alpha} - 1 \right)\right) + i \mu u \right) & \text{if } \alpha \neq 1,\\
   \exp\left(-c|u| \left(1 + i\beta \frac{2}{\pi}\sgn(u) \log\left(c|u|\right) \right) + i\mu u\right) & \text{if } \alpha = 1.
  \end{cases}
\end{equation}
The parameter $\alpha\in (0,2]$ is linked to the stability index, i.e. it determines the rate at which the tails of the distribution diminish; see Theorem~\ref{th:approx_Nolan} or~\cite{nolan_book} for details. Note that for $\alpha = 2$, the $\alpha$-stable distribution simplifies to the Gaussian distribution. The parameter $\beta\in [-1,1]$ is used to quantify the skewness of the heavy-tailed distribution; however, note that the classical Pearson's skewness coefficient is undefined for $\alpha<2$. In particular, for $\beta>0$ ($\beta<0$, respectively) the $\alpha$-stable distribution is skewed to the right (left, respectively) and if $\beta = 0$, then the $\alpha$-stable distribution is symmetric around the location parameter $\mu\in \bR$. The coefficient $c>0$ is the scale parameter, see \cite{stable,alek_book}.

In the following, we focus on the normalized and symmetric $\alpha$-stable distributions, i.e. we assume that $X\sim S(\alpha,0,1,0)$ with $\alpha\in (0,2]$. For brevity, we write $X\sim S\alpha S$ and say that $X$ follows the symmetric $\alpha$-stable distribution. For any $\alpha\in (0,2]$, we use $F_{\alpha}(\cdot)$, $f_{\alpha}(\cdot)$, and $Q_{\alpha}(\cdot)$, to denote the cumulative distribution function (CDF), the probability density function (PDF), and the quantile function (inverse CDF) of the symmetric $\alpha$-stable distribution, respectively. Note that the explicit formula for PDF of $S\alpha S$ is known only for the Gaussian distribution ($\alpha=2$) and the Cauchy distribution ($\alpha=1$). However, it is possible to provide an integral representation of $f_{\alpha}(\cdot)$. Namely, from Theorem 3.2 in~\cite{nolan_book}, for $\alpha\in (0,2)$, we have
\begin{equation}
    f_\alpha(x) = \frac{1}{\pi} \int_0^\infty \cos(xt) \exp(-t^\alpha)dt, \quad x\in \mathbb{R}.
\end{equation}

\subsection{Tails of the symmetric $\alpha$-stable distribution}

\begin{figure}[htp!]
\begin{center}
\includegraphics[width=0.39\textwidth]{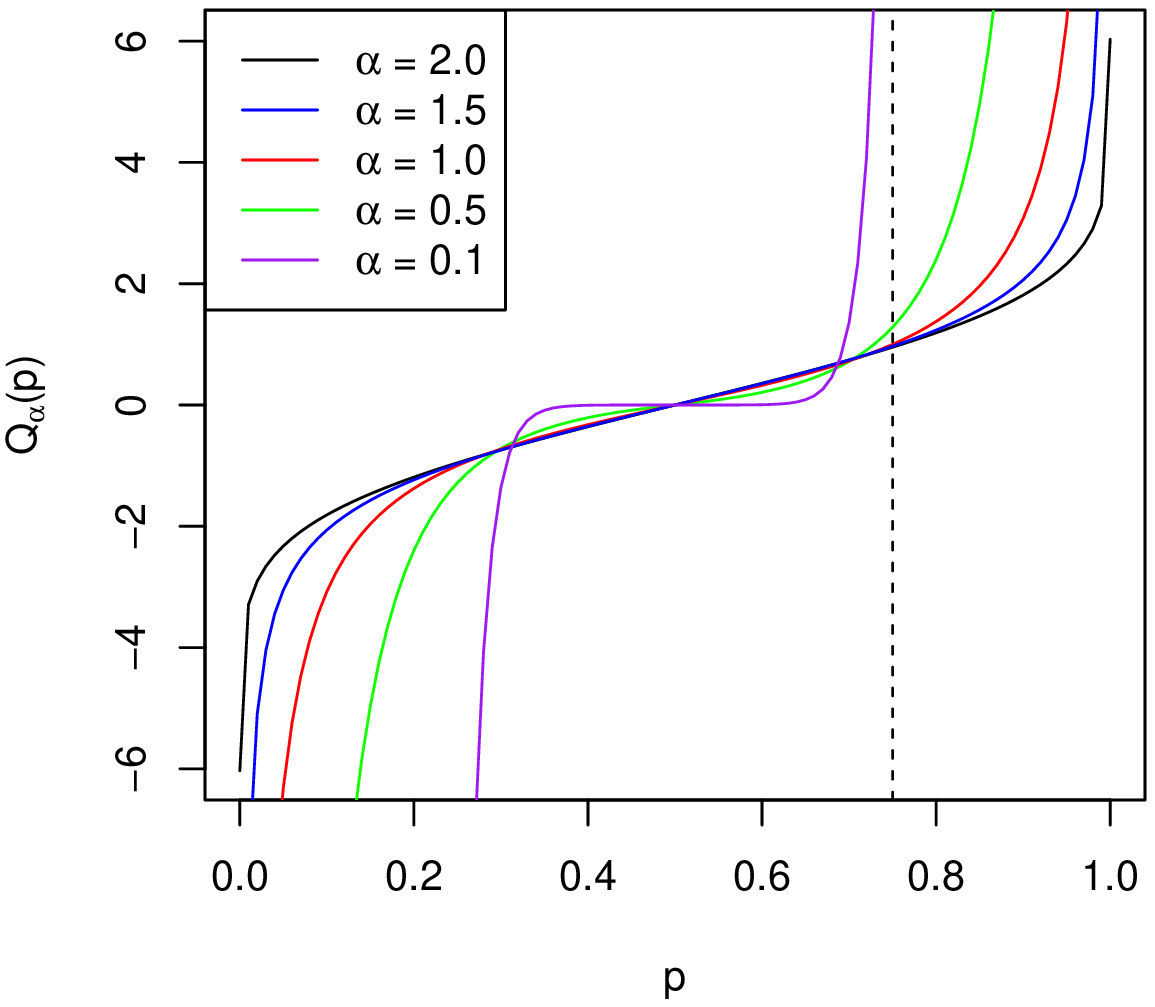}
\includegraphics[width=0.39\textwidth]{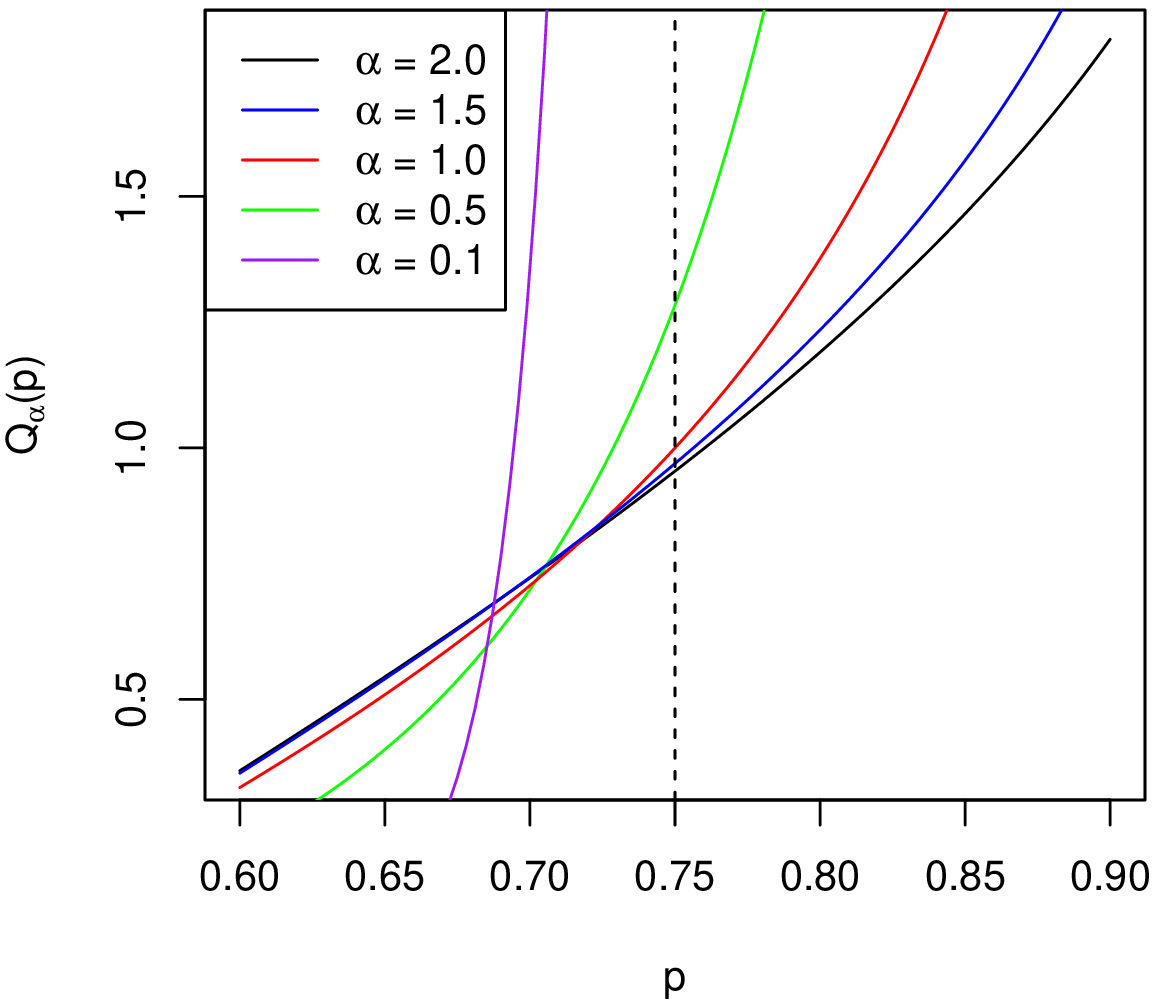}
\end{center}
\caption{Plots of the quantile functions $Q_{\alpha}(\cdot)$ for $\alpha\in \{0.1,0.5,1,1.5,2\}$. The left panel shows the full domain $p\in (0,1)$, while the right panel shows the sub-domain $p\in (0.6,0.9)$.
The dashed vertical line corresponds to the quantile level $p_0=0.75$. Note that, for $p\geq 0.75$, we can see that the quantile  lines $(p,Q_{\alpha}(p))$ do not cross each other which indicates that quantile values are monotone with respect to the tail index $\alpha$, as stated in Proposition~\ref{pr:1}.}
\label{fig:quantiles}
\end{figure}

Let us now focus on the asymptotic behavior of the symmetric $\alpha$-stable distribution with $\alpha\in (0,2)$. The tails of $S\alpha S$ distribution could be characterized using power law dynamics with an additional exponent factor linked to the parameter $\alpha$. This is formalized in Theorem~\ref{th:approx_Nolan}, where we use simplified notation $f(x)\overset{x\to x_0}{\sim} g(x)$ whenever $\lim_{x\to x_0}\frac{f(x)}{g(x)}=1$ and $\Gamma(\cdot)$ to denote the Gamma function; see Theorem 1.2 in~\cite{nolan_book} for details and the proof.
\begin{theorem}\label{th:approx_Nolan}
For any $\alpha\in (0,2)$, $X\sim S\alpha S$, and $c_\alpha := \sin(\frac{\pi \alpha}{2}) \Gamma(\alpha)/\pi$, we get
\begin{equation}
\mathbb{P}[X > x] \overset{x\to \infty}{\sim} c_\alpha x^{-\alpha}\quad \text{and} \quad
f_{\alpha}(x) \overset{x\to \infty}{\sim} \alpha x^{-(\alpha +1)}.
\end{equation}
\end{theorem}
Next, we show the asymptotic behavior of the quantile function of the symmetric $\alpha$-stable distribution. While this could be seen as a complementary result to Theorem~\ref{th:approx_Nolan}, we decided to present the proof for its completeness.
\begin{proposition}\label{pr:quantile_tail}
For any $\alpha\in (0,2)$ and $\bar{c}_{\alpha}:=\left(\frac{\Gamma(\alpha)\sin(\pi \alpha/2)}{\pi} \right)^{1/\alpha}$, we get
\[
Q_{\alpha}(p) \overset{p\to 1^-}{\sim} \bar{c}_{\alpha}(1-p)^{-1/\alpha}.
\]
\end{proposition}
\begin{proof}
Let us fix $\alpha\in (0,2)$. Let $X\sim S\alpha S$, let $\bar{F}_{\alpha}(\cdot):=1-F_{\alpha}(\cdot)$ be the survival function of $X$, and let $\bar{Q}_{\alpha}(\cdot)$ be the inverse of $\bar{F}_{\alpha}(\cdot)$. From Theorem~\ref{th:approx_Nolan}, we get $\lim_{x\to\infty} \frac{\bar{F}_{\alpha}(x)}{c_{\alpha}x^{-\alpha}}=1$. Thus, setting $x=\bar{Q}_{\alpha}(p)$ for $p\in (0,1)$, we get
$
\lim_{p\to 0^+} \frac{\bar{F}_{\alpha}(\bar{Q}_{\alpha}(p))}{c_{\alpha}\left(\bar{Q}_{\alpha}(p)\right)^{-\alpha}}=1.
$
Now, recalling that $\bar{F}_{\alpha} \left(\bar{Q}_{\alpha}(p)\right)=p$, we know this is equivalent to $\lim_{p\to 0^+} \frac{p}{c_{\alpha}\left(\bar{Q}_{\alpha}(p)\right)^{-\alpha}}=1$.
By considering the $\alpha$-root of the inverted limit, we know that 
$
\lim_{p\to 0^+} \frac{\bar{Q}_{\alpha}(p)}{\bar{c}_{\alpha}p^{-1/\alpha}}=1.
$
Finally, noting that $\bar{Q}_{\alpha}(p)={Q}_{\alpha}(1-p)$ for $p\in (0,1)$, we get
\[
\lim_{p\to 1^-} \frac{Q_{\alpha}(p)}{\bar{c}_{\alpha}(1-p)^{-1/\alpha}}=1,
\]
which concludes the proof.
\end{proof}
Using Proposition~\ref{pr:quantile_tail} we can show that, for quantile levels $p$ high enough, the quantiles lines $(p,Q_{\alpha}(p))$ are monotone with respect to the parameter $\alpha$; see Proposition~\ref{pr:1} for details and Figure~\ref{fig:quantiles} for a numerical illustration. This shows that the observation stated in Theorem~\ref{th:approx_Nolan}, which links the parameter $\alpha$ to the tail thickness, could be expressed in terms of the quantile functions.  Note that this result could be associated with the concept of the first order tail stochastic dominance, see~\cite{Zie2001,OrtLanPet2016,BedLocMar2021} for some other results on the stochastic ordering in reference to $\alpha$-stable distributions.
\begin{proposition}\label{pr:1}
For any $\alpha_1,\alpha_2\in (0,2]$ satisfying $\alpha_1<\alpha_2$  there exists $p_0\in (1/2,1)$ such that for any $p\in [p_0,1)$ we get
\begin{equation}\label{eq:pr:1:0}
Q_{\alpha_1}(p)>Q_{\alpha_2}(p).
\end{equation}
\end{proposition}
\begin{proof}
Let $\alpha_1,\alpha_2\in (0,2]$ be such that $\alpha_1<\alpha_2$. For brevity, we assume that $\alpha_2<2$; the case $\alpha_2=2$ could be treated using a similar logic. In order to prove \eqref{eq:pr:1:0}, it is enough to show that
\begin{equation}\label{eq:pr:1:1}
\lim_{p\to 1^-} \frac{Q_{\alpha_1}(p)}{Q_{\alpha_2}(p)}=\infty.
\end{equation}
Indeed, from~\eqref{eq:pr:1:1} it follows that there exists some $p_0\in(1/2,1)$ such that for any $p\geq p_0$ we have  $\frac{Q_{\alpha_1}(p)}{Q_{\alpha_2}(p)}>1$, which shows~\eqref{eq:pr:1:0}. Let us now prove \eqref{eq:pr:1:1}. Using Proposition~\ref{pr:quantile_tail}, we get
\[
\lim_{p\to 1^-} \frac{Q_{\alpha_1}(p)}{Q_{\alpha_2}(p)} = \lim_{p\to 1^-} \left[\frac{Q_{\alpha_1}(p)}{\bar{c}_{\alpha_1}(1-p)^{-1/\alpha_1}}\cdot \frac{\bar{c}_{\alpha_2}(1-p)^{-1/\alpha_2}}{Q_{\alpha_2}(p)}\cdot\frac{\bar{c}_{\alpha_1}(1-p)^{-1/\alpha_1}}{\bar{c}_{\alpha_2}(1-p)^{-1/\alpha_2}}\right]=\frac{\bar{c}_{\alpha_1}}{\bar{c}_{\alpha_2}}\lim_{p\to 1^-} (1-p)^{1/\alpha_2-1/\alpha_1}.
\]
Recalling that $\alpha_1<\alpha_2$ we get $1/\alpha_2-1/\alpha_1<0$ and consequently $
\lim_{p\to 1^-} (1-p)^{1/\alpha_2-1/\alpha_1}=\infty$.
Noting that $\frac{\bar{c}_{\alpha_1}}{\bar{c}_{\alpha_2}}>0$, we conclude the proof.
\end{proof}

\begin{remark}\label{rem:po.ind}
From simulations, we get that the constant $p_0\in (1/2,1)$ from Proposition~\ref{pr:1} can be chosen independently of $\alpha_1$ and $\alpha_2$, and it satisfies the inequality $p_0\leq 0.75$; see Figure~\ref{fig:quantiles}. The difficulty of analytical calculation of $p_0$ could be linked to the fact that the proof of Proposition~\ref{pr:1} is based on the tail approximation of the quantile function; cf. Theorem~\ref{th:approx_Nolan} and Proposition~\ref{pr:quantile_tail}.  Since we were not able to find $p_0$ explicitly, we deduced it from the numerical studies. We refer to Theorem~1.2 in~\cite{nolan_book} as well as to the following remarks for more information on technical calculation difficulties for $\alpha$-stable distribution tails; see also the discussion after Equation 3.2 in~\cite{fama}.
\end{remark}

\begin{remark}
From Proposition~\ref{pr:1} and  Remark~\ref{rem:po.ind} we see that the quantile functions values are monotone with respect to the tail index for sufficiently large quantile levels. This result should be compared with~\cite{Zie2001}, where the full-domain dispersive ordering result for the stable distributions is shown. 
However, this result is stated in the 2-parametrization while in this paper we follow the 0-parametrization. In fact, as seen in Figure~\ref{fig:quantiles}, the full-domain ordering does not hold for the 0-parametrization, so that the result from~\cite{Zie2001} cannot be extended to our setting. We refer to Proposition 3.7 in~\cite{nolan_book} as well as to the following discussion for more details about various parametrizations of the $\alpha$-stable distribution family.
\end{remark}

\subsection{Quantile conditional variance for $\alpha$-stable distributed random variables.}

In this section, we discuss the conditional variance characterization of the symmetric $\alpha$-stable distributions. Let us consider a generic $X\sim S\alpha S$ with some $\alpha \in (0,2]$. For any $0<a<b<1$, we define the quantile conditioning set $M_{\alpha}(a,b):=\{X\in( Q_{\alpha}(a),Q_{\alpha}(b))\}$. The quantile conditional variance (QCV) of $X$ on $M_{\alpha}(a,b)$ is given by
\begin{equation}\label{eq:cond_var}
\sigma^2_{\alpha}(a,b):=\mathbb{E}[X^2| M_{\alpha}(a,b)]-(\mathbb{E}[X| M_{\alpha}(a,b)])^2.
\end{equation}
Note that $\sigma^2_{\alpha}(a,b)$ is simply the quantile trimmed variance of $X$, i.e. the conditional variance on the quantile interval $( Q_{\alpha}(a),Q_{\alpha}(b)]$. Also, note that for any $\alpha \in (0,2]$ and $0<a<b<1$, the quantile conditional variance $\sigma^2_{\alpha}(a,b)$ is well-defined and finite. Also, the family $(\sigma^2_{\alpha}(a,b))_{a,b}$ characterizes the $\alpha$-stable distribution up to an additive shift, see \cite{PitCheWyl2021} for details.

To compute the value of QCV one may use the representation based on the probability density or quantile functions. More specifically, recalling that $X$ is absolutely continuous with the density $f_\alpha(\cdot)$ and using~\eqref{eq:cond_var} with $p=Q_{\alpha}(x)$, we get
\begin{align}
\sigma^2_{\alpha}(a,b)&=\frac{1}{b-a}\int_{Q_{\alpha}(a)}^{Q_{\alpha}(b)} x^2 f_{\alpha}(x)dx-\left(\frac{1}{b-a}\int_{Q_{\alpha}(a)}^{Q_{\alpha}(b)} x f_{\alpha}(x)dx\right)^2 =\frac{1}{b-a}\int_{a}^{b} (Q_{\alpha}(p))^2 dp-\left(\frac{1}{b-a}\int_{a}^{b} Q_{\alpha}(p)dp\right)^2.\label{eq:cond_var_quant}
\end{align}
As expected, the explicit formula for $\sigma^2_{\alpha}(a,b)$ could be derived only for the Gaussian ($\alpha=2$) and the Cauchy ($\alpha=1$) distributions, see \cite{PitCheWyl2021}. Namely, for the Gaussian case, i.e. $X\sim S\alpha S$ with $\alpha=2$, we get
 \begin{equation}\label{eq:normalconvar}
      \sigma^2_2(a,b) =2 \left(\frac{\Phi^{-1}(a) \phi(\Phi^{-1}(a)) - \Phi^{-1}(b)\phi(\Phi^{-1}(b))}{b-a} - \frac{\left( \phi \left(\Phi^{-1}(a) \right) - \phi \left(\Phi^{-1}(b) \right)\right) ^2}{(b-a)^2} \right),
 \end{equation}
 where $\phi(\cdot)$ and $\Phi(\cdot)$ are standard Gaussian PDF and CDF, respectively, while for the Cauchy case, i.e. $X\sim S\alpha S$ with $\alpha=1$, we get 
 \begin{equation}\label{eq:cauchyconvar}
     \sigma^2_1(a,b) = \frac{1}{b-a}\left(\frac{F_1^{-1}(b) - F_1^{-1}(a)}{D(a,b)} - 1\right) - \frac{1}{4D^2(a,b)(b-a)^2} \log^2 \left( \frac{1+F_1^{-1}(b)^2}{1+F_1^{-1}(a)^2} \right),
 \end{equation}
where $D(a,b) := \tan ^{-1} \left(F_1^{-1}(b)\right) - \tan^{-1} \left(F_1^{-1}(a)\right)$.

\begin{figure}[htp!]
\begin{center}
\includegraphics[width=0.39\textwidth]{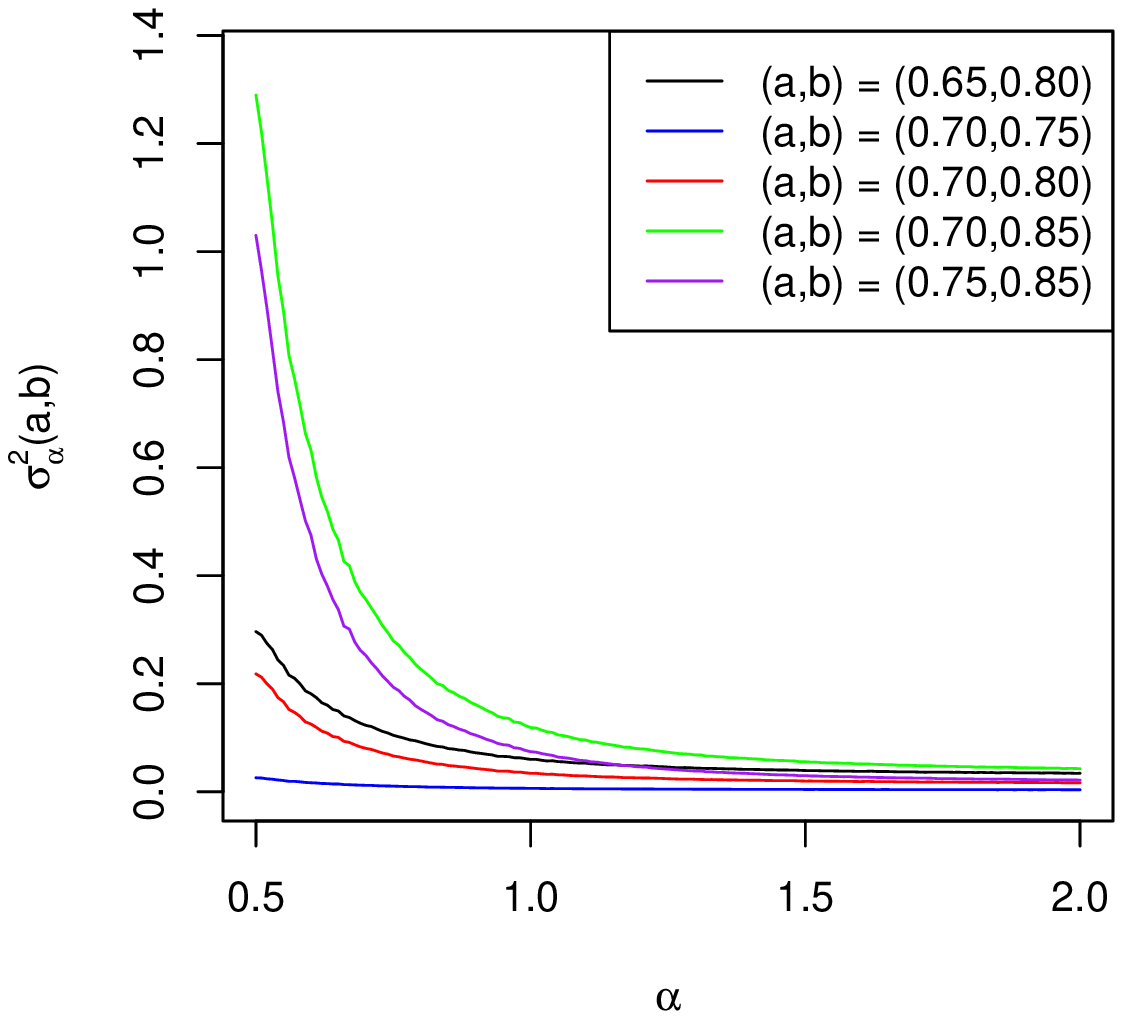}
\includegraphics[width=0.39\textwidth]{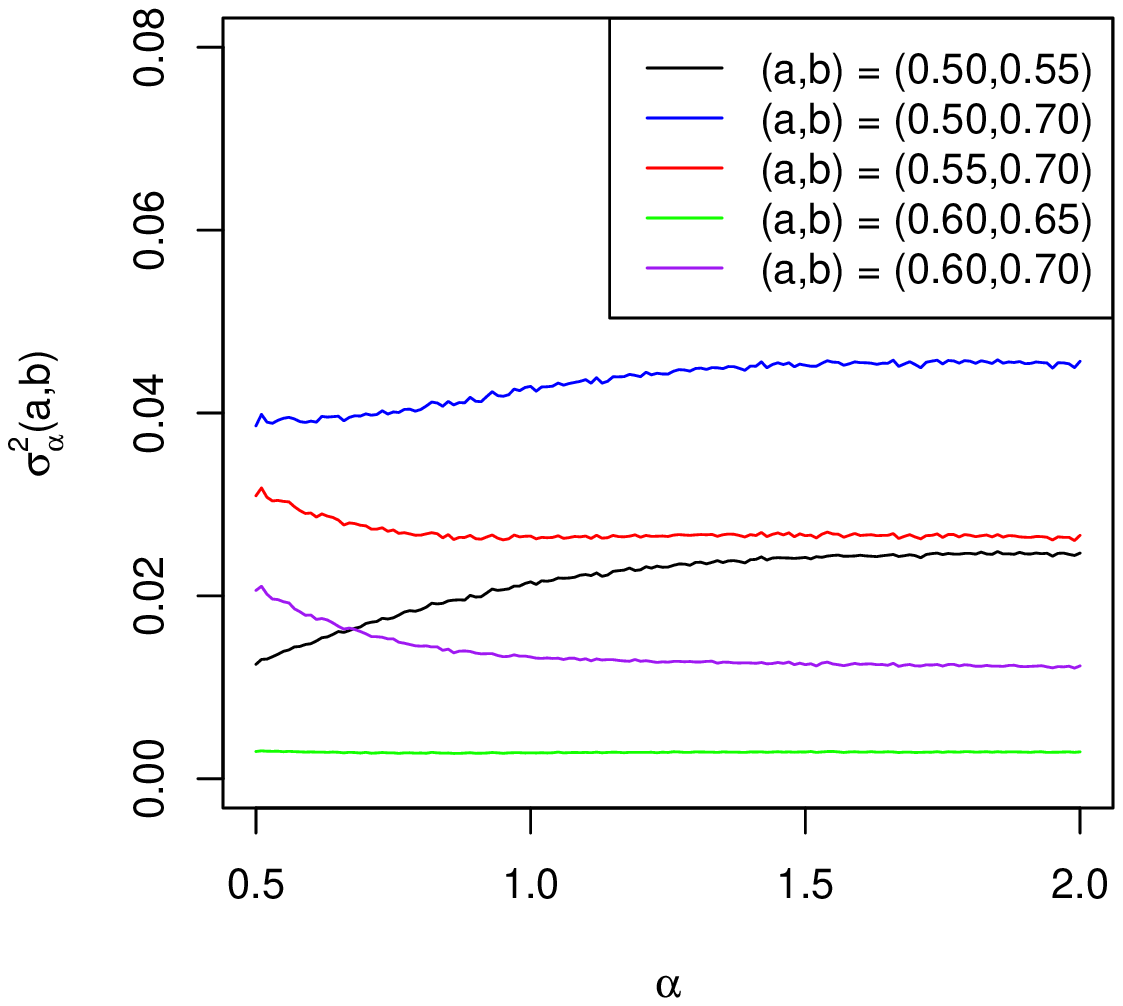}
\end{center}
\caption{Plots of the conditional variances $\sigma^2_{\alpha}(a,b)$ as functions of $\alpha$ with various $a$ and $b$. The curves are based on the Monte Carlo simulations with 1 000 000 replications each. As seen on the left panel, if $a$ and $b$ satisfy $0.65\leq a< b<1$, then we get that the map $\alpha \to\sigma^2_{\alpha}(a,b)$ is decreasing. However, as seen on the right panel, if $a<0.65$, this property may not hold.}
\label{fig:variances}
\end{figure}

Recalling that the smaller the tail index $\alpha$, the heavier the tails, we expect that the quantile conditional variance on an appropriately chosen tail set should be monotone with respect to $\alpha$. Consequently, QCV could be used to measure the heaviness of the tail of the symmetric $\alpha$-stable distributions. This intuition is formalized in Theorem~\ref{th:mono_variances}; see also Figure~\ref{fig:variances} for a numerical illustration.

\begin{theorem}\label{th:mono_variances}
Let $\alpha_1,\alpha_2\in (0,2]$ be such that $\alpha_1<\alpha_2$. Then, there exists $p_1\in (1/2,1)$ such that, for any $a,b\in [p_1,1)$ satisfying $a<b$, we get
\[
\sigma^2_{\alpha_1}(a,b)>\sigma^2_{\alpha_2}(a,b).
\]

\end{theorem}
\begin{proof}
Let us fix $\alpha_1,\alpha_2\in (0,2]$ satisfying $\alpha_1<\alpha_2$. Also, as in Proposition~\ref{pr:1}, assume $\alpha_2<2$. Note that, recalling~\eqref{eq:cond_var_quant}, for any $a,b\in [1/2,1)$ satisfying $a<b$, we get
\begin{align}\label{eq:th:mono_variances:1}
\sigma^2_{\alpha_1}(a,b)-\sigma^2_{\alpha_2}(a,b)&=\frac{1}{b-a}\int_{a}^{b} (Q_{\alpha_1}(p)-Q_{\alpha_2}(p))(Q_{\alpha_1}(p)+Q_{\alpha_2}(p)) dp\nonumber\\
&\phantom{=}-\left(\frac{1}{b-a}\int_{a}^{b} (Q_{\alpha_1}(p)-Q_{\alpha_2}(p)) dp\right)\left( \frac{1}{b-a}\int_{a}^{b} (Q_{\alpha_1}(p)+Q_{\alpha_2}(p))dp\right).
\end{align}
Let us define $g(\cdot):=Q_{\alpha_1}(\cdot)-Q_{\alpha_2}(\cdot)$ and $h(\cdot):=Q_{\alpha_1}(\cdot)+Q_{\alpha_2}(\cdot)$, and note that~\eqref{eq:th:mono_variances:1} could be expressed as
\[
\sigma^2_{\alpha_1}(a,b)-\sigma^2_{\alpha_2}(a,b)=\frac{1}{b-a}\int_{a}^{b} g(p) h(p) d p- \left(\frac{1}{b-a} \int_{a}^{b} g(p) d p \right)\left( \frac{1}{b-a} \int_{a}^{b} h(p) d p \right).
\]
Then, it is enough to show that $g(\cdot)$ and $h(\cdot)$ are (strictly) increasing. Indeed, using the classic Chebyshev integral inequality (see e.g. Theorem 8 in Section 2.5 of~\cite{Mit1970}), from the monotonicity of $g(\cdot)$ and $h(\cdot)$, we get $\frac{1}{b-a} \int_{a}^{b} g(x) h(x) d x > \left(\frac{1}{b-a} \int_{a}^{b} g(x) d x \right)\left( \frac{1}{b-a} \int_{a}^{b} h(x) d x \right)$, which shows $\sigma^2_{\alpha_1}(a,b)-\sigma^2_{\alpha_2}(a,b)>0$. Note that $h(\cdot)$ is increasing as the sum of the increasing functions $Q_{\alpha_1}(\cdot)$ and $Q_{\alpha_2}(\cdot)$. Thus, to conclude the proof it is enough to show that $g(\cdot)$ is increasing. Recalling that $Q_{\alpha}(\cdot)$ is the inverse of $F_{\alpha}(\cdot)$, we get
\[
\frac{d g}{dp}(p)=\frac{1}{f_{\alpha_1}(Q_{\alpha_1}(p))}-\frac{1}{f_{\alpha_2}(Q_{\alpha_2}(p))}, \quad p\in (1/2,1).
\]
Thus, for any $p\in (1/2,1)$, we get $\frac{d g}{dp}(p)>0$  if and only if
$
f_{\alpha_1}(Q_{\alpha_1}(p))<f_{\alpha_2}(Q_{\alpha_2}(p))$. Consequently, it is enough to show that
\begin{equation}\label{eq:pr:mono_variances:1}
\lim_{p\to 1^-} \frac{f_{\alpha_2}(Q_{\alpha_2}(p))}{f_{\alpha_1}(Q_{\alpha_1}(p))}=\infty.
\end{equation}
Indeed, identity \eqref{eq:pr:mono_variances:1} implies that, starting from some $p_1\in (1/2,1)$, we get $\frac{dg}{dp}(p)>0$, and consequently  the map $p\mapsto g(p)$ is (strictly) increasing on $[p_1,1)$.
To show~\eqref{eq:pr:mono_variances:1}, we use Theorem~\ref{th:approx_Nolan} and Proposition~\ref{pr:quantile_tail}. Namely, we get
\begin{align*}
\lim_{p\to 1^-} \frac{f_{\alpha_2}(Q_{\alpha_2}(p))}{f_{\alpha_1}(Q_{\alpha_1}(p))}&= \lim_{p\to 1^-} \frac{f_{\alpha_2}(Q_{\alpha_2}(p))}{\alpha_2 {c}_{\alpha_2}(Q_{\alpha_2}(p))^{-(\alpha_2 +1)}}\cdot \frac{\alpha_1 {c}_{\alpha_1}(Q_{\alpha_1}(p))^{-(\alpha_1 +1)}}{f_{\alpha_1}(Q_{\alpha_1}(p))}\cdot\frac{\alpha_2 {c}_{\alpha_2}(Q_{\alpha_2}(p))^{-(\alpha_2 +1)}}{\alpha_1 {c}_{\alpha_1}(Q_{\alpha_1}(p))^{-(\alpha_1 +1)}}\\
&=\frac{\alpha_2{c}_{\alpha_2}}{\alpha_1 {c}_{\alpha_1}}\lim_{p\to 1^-} \frac{(Q_{\alpha_2}(p))^{-(\alpha_2 +1)}}{(Q_{\alpha_1}(p))^{-(\alpha_1 +1)}} \\
&=\frac{\alpha_2{c}_{\alpha_2}}{\alpha_1 {c}_{\alpha_1}}\lim_{p\to 1^-} \left(\frac{Q_{\alpha_2}(p)}{\bar{c}_{\alpha_2}(1-p)^{-1/\alpha_2}}\right)^{-(\alpha_2 +1)}\cdot \left(\frac{\bar{c}_{\alpha_1}(1-p)^{-1/\alpha_1}}{Q_{\alpha_1}(p)}\right)^{-(\alpha_1 +1)}\cdot \frac{\left(\bar{c}_{\alpha_2}(1-p)^{-1/\alpha_2}\right)^{-(\alpha_2+1)}}{\left(\bar{c}_{\alpha_1}(1-p)^{-1/\alpha_1}\right)^{-(\alpha_1+1)}}\\
&=\frac{\alpha_2{c}_{\alpha_2}(\bar{c}_{\alpha_1})^{\alpha_1+1}}{\alpha_1 {c}_{\alpha_2}(\bar{c}_{\alpha_2})^{\alpha_2+1}} \lim_{p\to 1^-} (1-p)^{1/\alpha_2-1/\alpha_1}.
\end{align*}
Recalling that $\alpha_1<\alpha_2$, we get $ \lim_{p\to 1^-} (1-p)^{1/\alpha_2-1/\alpha_1} \to \infty $, which concludes the proof.
\end{proof}

\begin{remark}\label{rem:constant.variance}
From simulations, we get that the constant $p_1\in (1/2,1)$ from Theorem~\ref{th:mono_variances} can be chosen independently of $\alpha_1$ and $\alpha_2$, and it satisfies the inequality $p_1\leq 0.65$.  On the other hand, if the quantile level is smaller than $0.65$, then the monotonicity condition from Theorem~\ref{th:mono_variances} may be violated. Indeed, a numerical check shows that $\sigma^2_{0.5}(0.25,0.75)\approx 0.28$ and  $\sigma^2_{1}(0.25,0.75)\approx 0.275$, while $\sigma^2_{1.5}(0.25,0.75)\approx 0.285$. For transparency, these observations are illustrated in Figure~\ref{fig:variances}. The left panel shows exemplary choices of $0.65\leq a<b<1$ for which we can see that the map $\alpha \mapsto \sigma^2_{\alpha}(a,b)$ is decreasing while the right panel shows several other cases in which this property is violated.
\end{remark}

 %%%%%%%%%%%%%%%%%%%%%%%%%%%%%%%%%%%%%%%%%%%%%%%%%%%%%%%%%

\subsection{Extracting the tail index using a quantile conditional variance ratio}\label{S:extracting.tail}

From Theorem~\ref{th:mono_variances} combined with Remark~\ref{rem:constant.variance} we learn that for $X\sim S\alpha S$ and adequately chosen tail quantile split values $0<a<b<1$, the function $G_1(\alpha):=\sigma^2_\alpha(a,b)$ should be monotone. Moreover, one would expect that for a central quantile split $(d,1-d)$, where  $0<d<0.5$ is relatively close to 0.5,  the change of function $G_2(\alpha):=\sigma^2_\alpha(d,1-d)$ is smaller when compared to $G_1$ in a sense that for any $\alpha\in (0,2]$ we get
\begin{equation}\label{eq:derivative.ineq}
\left|\frac{\partial}{\partial\alpha}G_1(\alpha)\right|> \left|\frac{\partial}{\partial\alpha}G_2(\alpha)\right|.
\end{equation}
Thus, given the split $(a,b,d)$, we can define the variance ratio function $N\colon (0,2]\to \bR_{+}$,
\begin{equation}\label{eq:N.statistic}
N(\alpha):=2\frac{G_1(\alpha)}{G_2(\alpha)}=2\frac{\sigma^2_\alpha(a,b)}{\sigma^2_\alpha(d,1-d)}=\frac{\sigma^2_\alpha(a,b)+\sigma^2_\alpha(1-b,1-a)}{\sigma^2_\alpha(d,1-d)};
\end{equation}
note that the last equality follows from the fact that for the symmetric $\alpha$-stable distribution we always get $\sigma^2_\alpha (a,b)=\sigma^2_\alpha (1-b,1-a)$. The function $N(\cdot)$ is expected to be monotone with respect to $\alpha$. Moreover, noting that the  $N(\cdot)$ is invariant with respect to the affine transformations of the underlying random variable $X$, one could use values of $N(\cdot)$ to extract the value of the tail index $\alpha$ within the class of symmetric $\alpha$-stable random variables $X\sim S(\alpha,0,c,\mu)$. 

\begin{figure}[htp!]
\begin{center}
\includegraphics[width=0.31\textwidth]{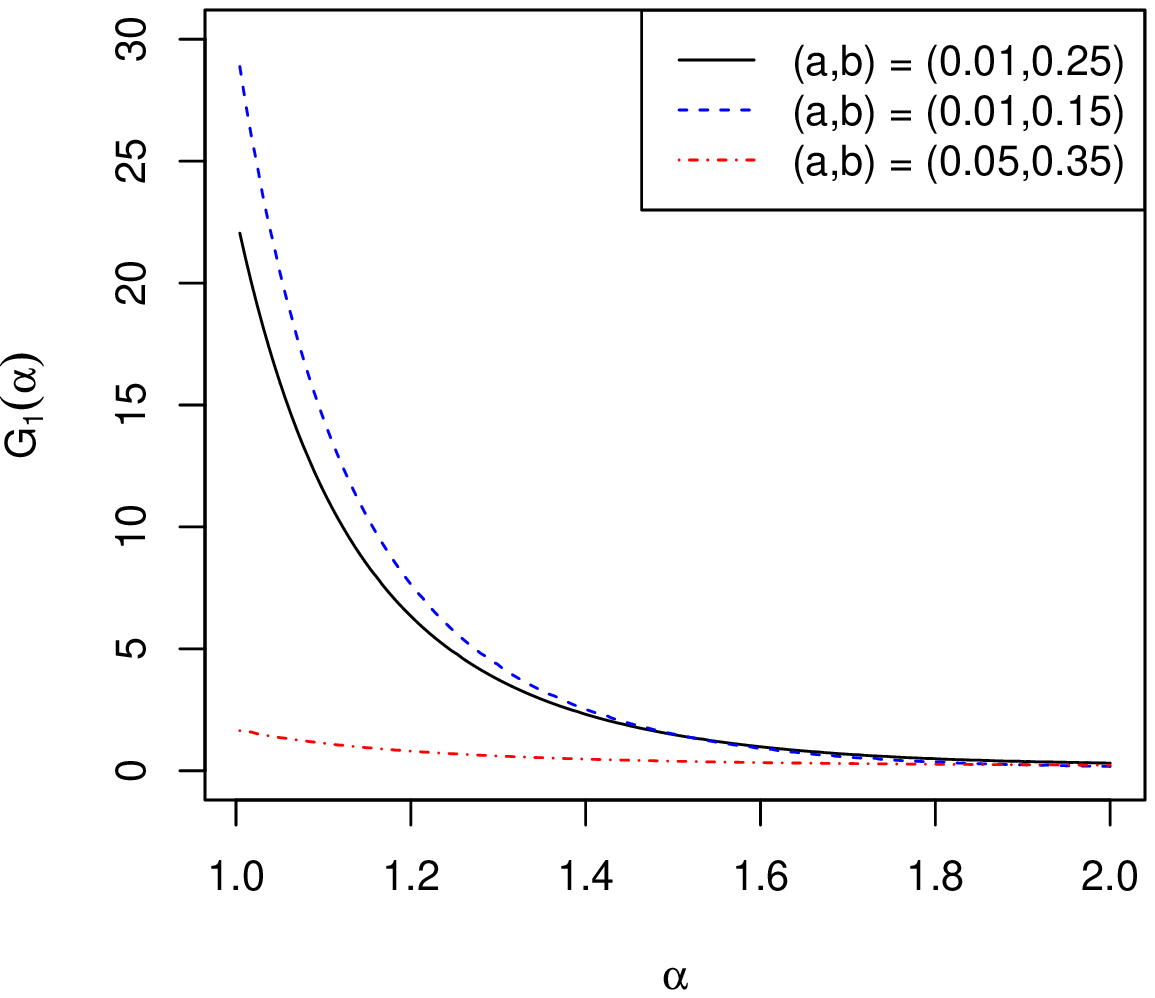}
\includegraphics[width=0.31\textwidth]{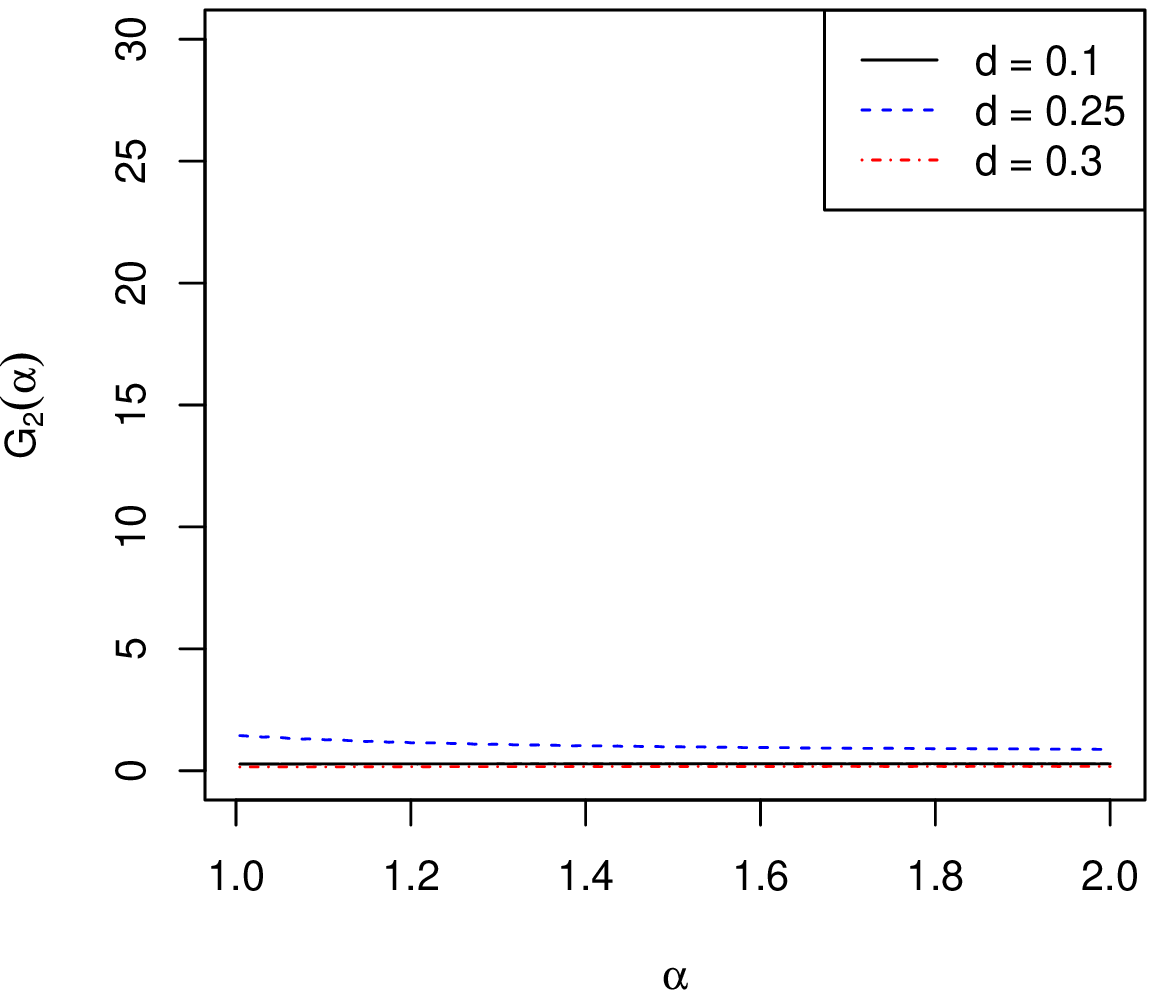}
\includegraphics[width=0.31\textwidth]{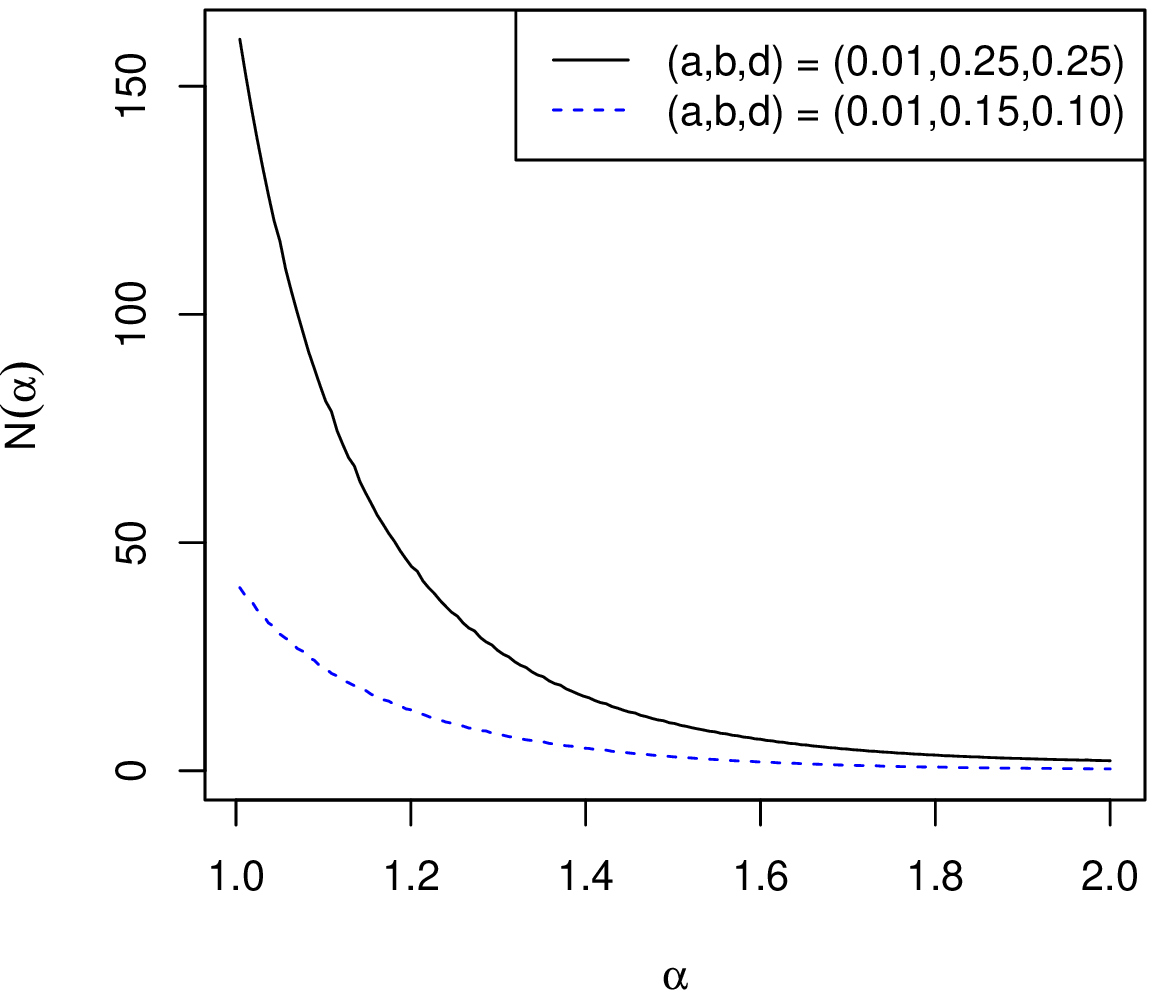}
\end{center}
\caption{Dynamics of $G_1(\alpha)=\sigma^2_\alpha(a,b)$, $G_2(\alpha)=\sigma^2_\alpha(d,1-d)$, and $N(\alpha)=2G_1(\alpha)/G_2(\alpha)$ for various choices of $(a,b,d)$. On the left panel we present the values of the function $G_1(\cdot)$ for $(a,b)\in \{(0.01,0.25),(0.01,0.15),(0.05,0.35)\}$. The middle panel present the value of $G_2(\cdot)$ for $d\in \{0.1,0.25,0.30\}$. The left panel presents the ratio values of $N(\cdot)$ for $(a,b,d)\in\{(0.01,0.25,0.25),(0.01,0.15,0.1)\}$. As one can observe, the function $N(\cdot)$ is indeed monotone with respect to $\alpha$ for adequately chosen values of $(a,b,d)$, which is expected due to \eqref{eq:derivative.ineq}.}
\label{F:S.bijection}
\end{figure}

While the direct proof of the monotonicity of~\eqref{eq:N.statistic} is challenging due to the theoretical difficulties stated in Remark~\ref{rem:po.ind} and Remark~\ref{rem:constant.variance}, one could numerically verify the monotonicity of~\eqref{eq:N.statistic} for an adequate choice of $(a,b,d)$ e.g. by approximating the related integrals by suitable numerical integration procedures. This is illustrated in Figure~\ref{F:S.bijection}. Having the strict monotonicity of $N(\cdot)$, we can easily recover the value of $\alpha$ given the value of $N(\alpha)$ by using the inverse transform $N^{-1}(\cdot)$. This simple intuition will be used to construct a fitting statistic in Section~\ref{S:cond.variance.fit}.

\section{Conditional variance as fitting statistics}\label{S:cond.variance.fit}

In this section we show how to use the variance ratio function $N(\cdot)$ introduced in \eqref{eq:N.statistic} for tail index parameter fitting. First, let us introduce the statistical framework and recall the basic properties of the sample QCV estimator. Given $n\in\bN$, i.i.d sample $(X_1,...X_n)$, and quantile split $0 < a < b < 1$, the sample QCV estimator is given by 
\begin{align}\label{eq:estimQCV}
 \hat{\sigma}^2_X(a,b) := \frac{1}{[nb] - [na]} \sum_{i=[na]+1}^{[nb]} (X_{(i)} - \hat{\mu}_X(a,b))^2,
\end{align}
where $\hat{\mu}_X(a,b)$ denotes the conditional sample mean, $X_{(i)}$ corresponds to the $i$th order statistic of the sample, and  $[x] := \max\{k \in \mathbb{Z}: k \leq x \}$ is the integer part of $x\in\bR$.  It should be noted that \eqref{eq:estimQCV} can be easily calculated as the sample variance of a sub-sample in which the conditioning is based on order statistics. Moreover, from \cite{JelPit2018} and \cite{PitCheWyl2021} we know that estimator $\hat{\sigma}^2_X(a,b)$ is consistent and $\sqrt{n}\cdot \hat{\sigma}^2_X(a,b)$ tends to the Gaussian distribution as $n\to\infty$.

Second, we introduce and discuss properties of the variance ratio estimator. Given the split $(a,b,d)$ for $0<a<b<1$ and $0<d<0.5$, we introduce a sample version of \eqref{eq:N.statistic} by setting
\begin{align}\label{eq:S:general2}
     \hat N :=
\frac{\hat \sigma^2_X (a,b)  + \hat \sigma^2_X (1-b,1-a)}{\hat \sigma^2_X (d,1 - d)}.
\end{align}

The consistency and asymptotic normality of the statistic given in  \eqref{eq:estimQCV} imply consistency and asymptotic normality of \eqref{eq:S:general2}, see Theorem 1 in \cite{JelPit2018} for the idea of the proof. Moreover, since statistic \eqref{eq:estimQCV} is location invariant and scale proportional, we get that statistic given in \eqref{eq:S:general2} is both location and scale invariant. In general, the statistic \eqref{eq:S:general2} may be interpreted as the ratio between tail and central dispersion.

Finally, from Section~\ref{S:extracting.tail} we know that for appropriate choices of $(a,b,d)$, the function $\alpha \to N(\alpha)$ defined in \eqref{eq:N.statistic} is a~bijection. Thus, using consistency of $\hat N$, we can take  $N^{-1}(\cdot)$ to fit the tail index parameter given the value of $\hat N$. Namely, using the standard statistical plug-in procedure, we set
\begin{align}\label{eq:estimAlpha}
\hat{\alpha}:= N^{-1}(\hat{N}).
\end{align}

In the following, we propose two specific choices of the splits $(a,b,d)$ that guarantee the bijection property. Namely, we consider the splits $(a_1,b_1,d_1):=(0.015,0.25,0.25)$ and $(a_2,b_2,d_2):=(0.01,0.17,0.1)$ that correspond to variance ratios $N_1(\alpha):= 2\sigma_\alpha^2(0.015,0.25) / \sigma_\alpha^2(0.25,0.75)$ and $N_2(\alpha):= 2\sigma_\alpha^2(0.01,0.17) / \sigma_\alpha^2(0.1,0.9)$, with sample estimators
\begin{equation}\label{eq:S1S2}
     \hat N_1 :=\frac{\hat\sigma^2_X (0.015,0.25)  + \hat\sigma^2_X (0.75,0.985)}{\hat\sigma^2_X (0.25,0.75)}\quad\textrm{and}\quad
     \hat N_2 :=
\frac{\hat \sigma^2_X (0.01,0.17)  + \hat \sigma^2_X (0.83,0.99)}{\hat \sigma^2_X (0.1,0.9)},
\end{equation}
and output estimators of the tail-index $\alpha$ are given by
\begin{equation}\label{eq:hat_alpha}
\hat \alpha_1:=N_1^{-1}(\hat N_1)\quad \textrm{and}\quad \hat \alpha_2:=N_2^{-1}(\hat N_2).
\end{equation}
For brevity, we often refer to $\hat\alpha_1$ and $\hat\alpha_2$ as {\it QCV tail index $N_1$ estimator } and {\it QCV tail index $N_2$ estimator}, respectively, or simply as {\it $N_1$ estimator} and {\it $N_2$ estimator}, respectively. The first estimator is introduced to provide a generic fit for any $\alpha\in [1,2]$ which covers most practical application of the $\alpha$-stable distribution, while the second one is tailored for near-Gaussian cases, where $\alpha$ is relatively close to 2. The choice of the split values in both cases was based on numerical experiments. Namely, we decided to choose the parameters which minimise the average root mean square error (RMSE) between true and estimated values of $\alpha$ for different sample sizes. In the first case, the average was taken on tail index interval $[1,2]$ while in the second case we used the narrowed interval $[1.85,2]$.

\begin{figure}[htp!]
\begin{center}
\includegraphics[width=0.30\textwidth, angle = 270]{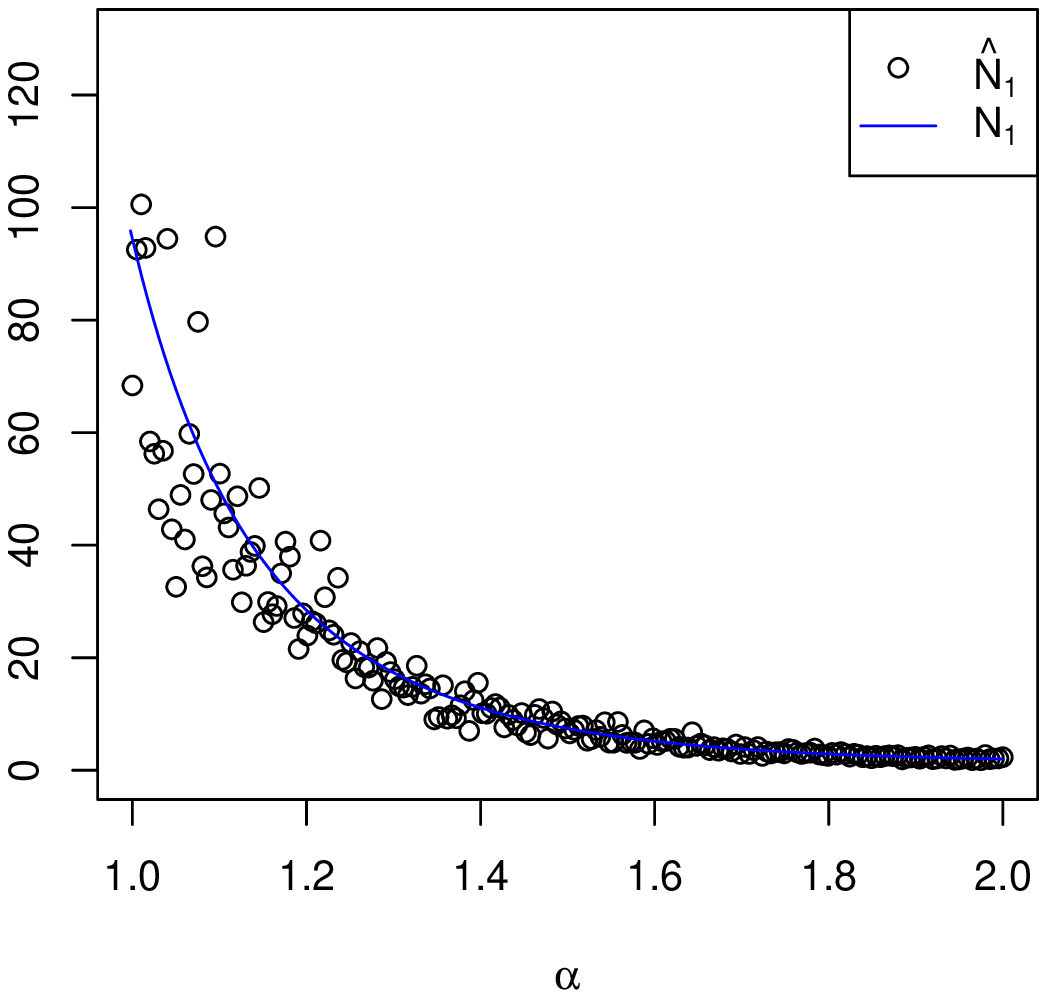}
\includegraphics[width=0.30\textwidth,angle = 270]{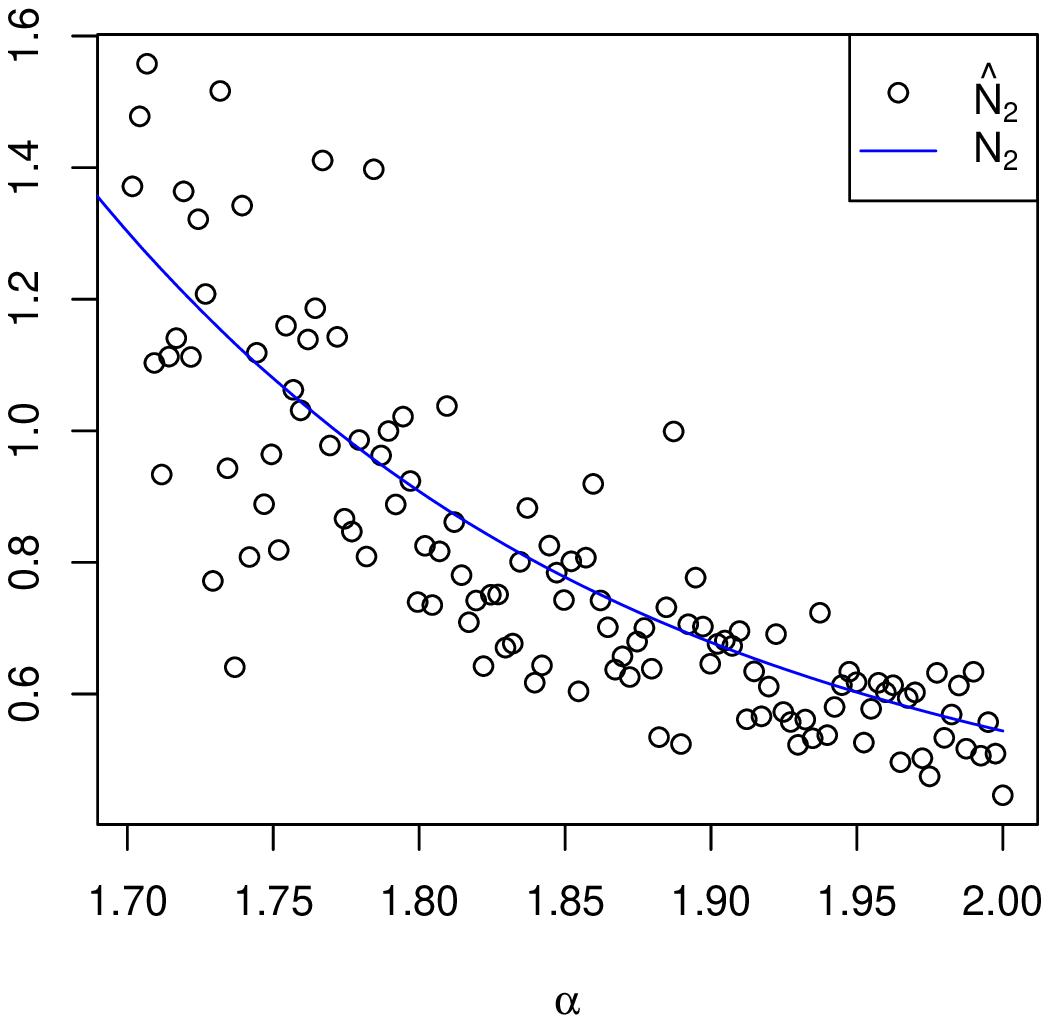}

\end{center}
\caption{Conditional variance ratios $N_1(\alpha)$ and $N_2(\alpha)$ compared with their estimates, i.e. values of $\hat N_1(\alpha)$ and $\hat N_2(\alpha)$ obtained for various values of $\alpha$. The left panel shows $N_1(\alpha)$ for $\alpha \in [1,2]$ while the right panel shows $N_2(\alpha)$ for $\alpha \in [1.7,2]$. The results are based on 400 strong Monte Carlo samples for $\alpha$ dispersed uniformly on the underlying interval and for $n=1000$. The curves $\alpha\to N_1(\alpha)$ and $\alpha\to N_2(\alpha)$ are based on the trapezoidal integration procedure applied to~\eqref{eq:cond_var_quant}.}
\label{fig:N_12_numerical_value}
\end{figure}

Let us now provide a more extensive comment on the estimation procedure. First, note that the specific values of $N_1(\alpha)$ and $N_2(\alpha)$ are not available in the closed-form. In fact, these quantities are defined as the ratios of the appropriate QCVs, which can be computed explicitly only in specific cases; see the discussion following~\eqref{eq:cond_var_quant}. However, the values of $N_1(\alpha)$ and $N_2(\alpha)$ can be approximated using various numerical integration procedures. In this paper we use the standard trapezoidal rule approximation applied to the $\alpha$-stable distribution PDF. For completeness, in Figure~\ref{fig:N_12_numerical_value} we compare the approximated values for $N_1$ and $N_2$ with sample-based values of $\hat N_1$ and $\hat N_2$, for $n=1000$. The obtained results indirectly confirm reasonable accuracy of the approximation procedure.

\begin{figure}[htp!]
\begin{center}
\includegraphics[width=0.30\textwidth, angle = 270]{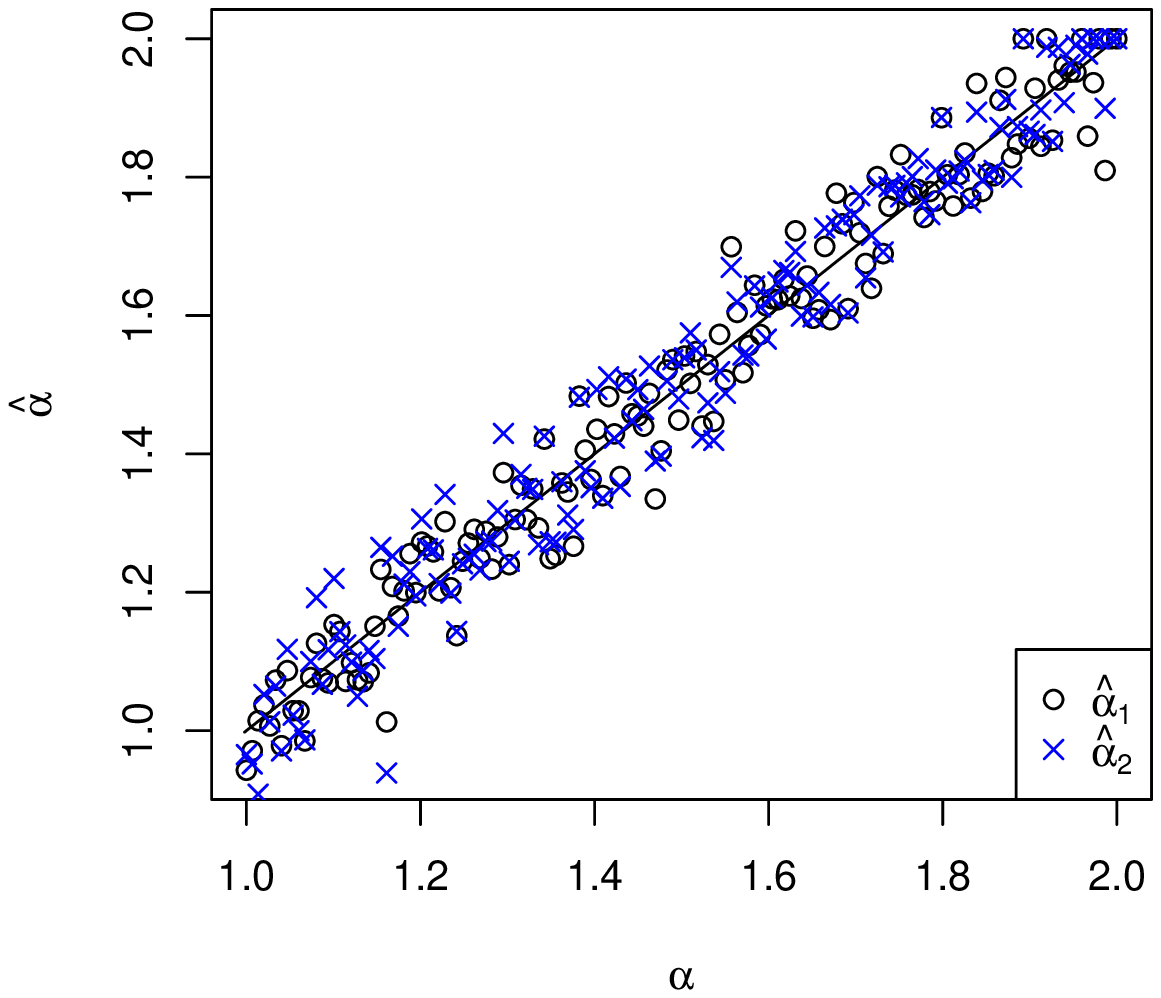}
\includegraphics[width=0.30\textwidth, angle = 270]{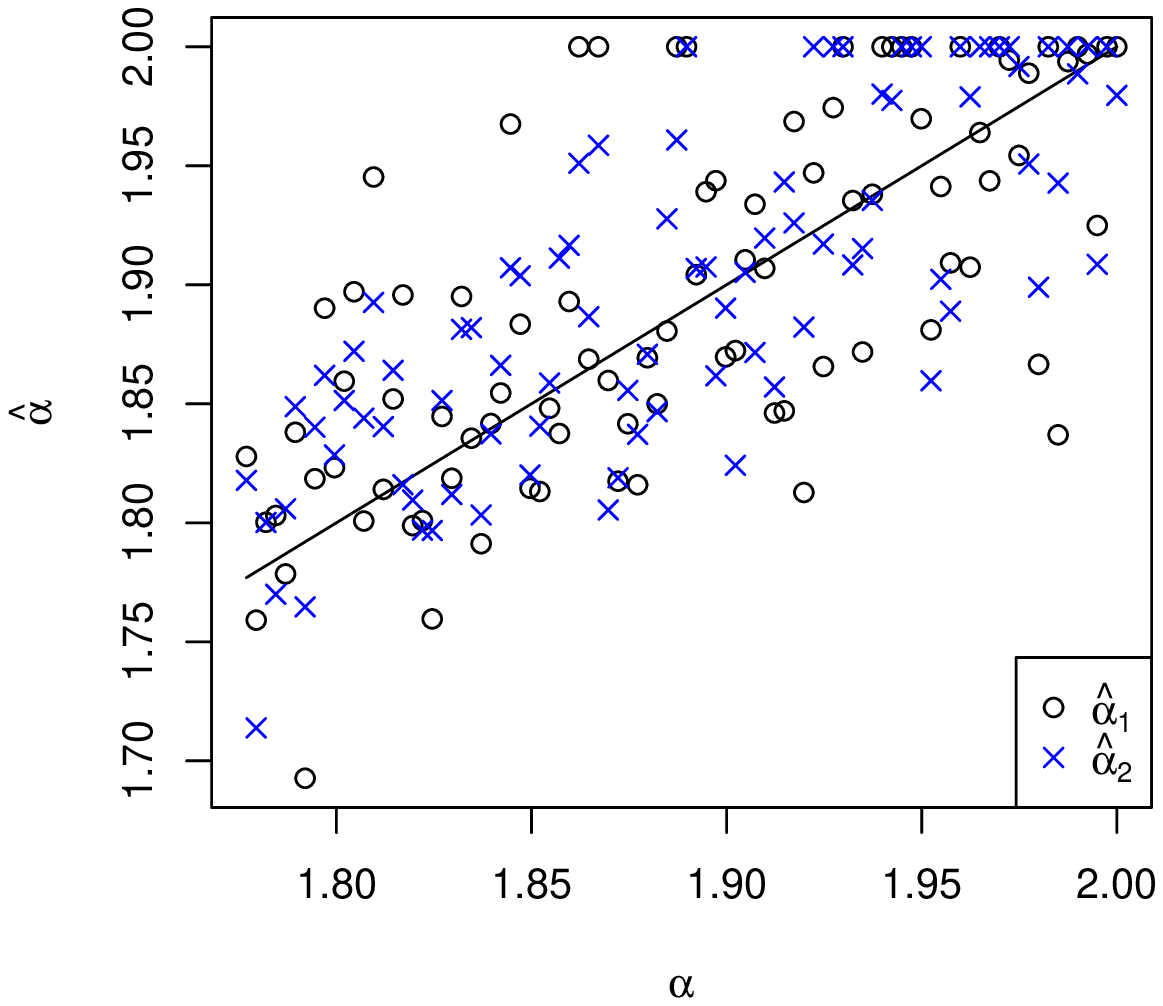}
\end{center}
\caption{Tail index $\alpha$ compared with estimators $\hat \alpha_1$ and $\hat \alpha_2$, for various values of true $\alpha$ and sample size $n=1000$. The left panel shows the full domain $\alpha \in [1,2]$ based on 400 elements grid, while the right panel shows the sub-domain $\alpha \in [1.8,2]$. The results are based of 400 strong Monte Carlo samples for $\alpha$ dispersed uniformly on the underlying interval and for $n=1000$. Note that estimates $\hat\alpha_1$ and $\hat \alpha_2$ are obtained using the inverse of $N_1$ and $N_2$, see~\eqref{eq:hat_alpha} and Figure~\ref{fig:N_12_numerical_value} for details.}
\label{fig:hat_alpha_vs_alpha}
\end{figure}

Next, in Figure~\ref{fig:hat_alpha_vs_alpha} we compare true values of $\alpha$ with the estimates
$\hat{\alpha}_1$ and $\hat{\alpha}_2$ obtained via \eqref{eq:hat_alpha} for multiple Monte Carlo simulations with samples of size $n=1000$. This serves as a first sanity check indicating that the  procedures proposed correctly estimate the underlying parameter. Also, the right panel in Figure~\ref{fig:hat_alpha_vs_alpha} suggests that, for $\alpha$ close to $2$, the estimator $\hat\alpha_2$ outperforms $\hat\alpha_1$ as the estimates exhibit smaller dispersion. This is consistent with a the rationale behind the construction of $\hat\alpha_2$, i.e. the fact that this estimator was intended to capture the near-Gaussian cases of the $\alpha$-stable distributions.

\begin{figure}[htp!]
\begin{center}
\includegraphics[width=0.30\textwidth, angle =270]{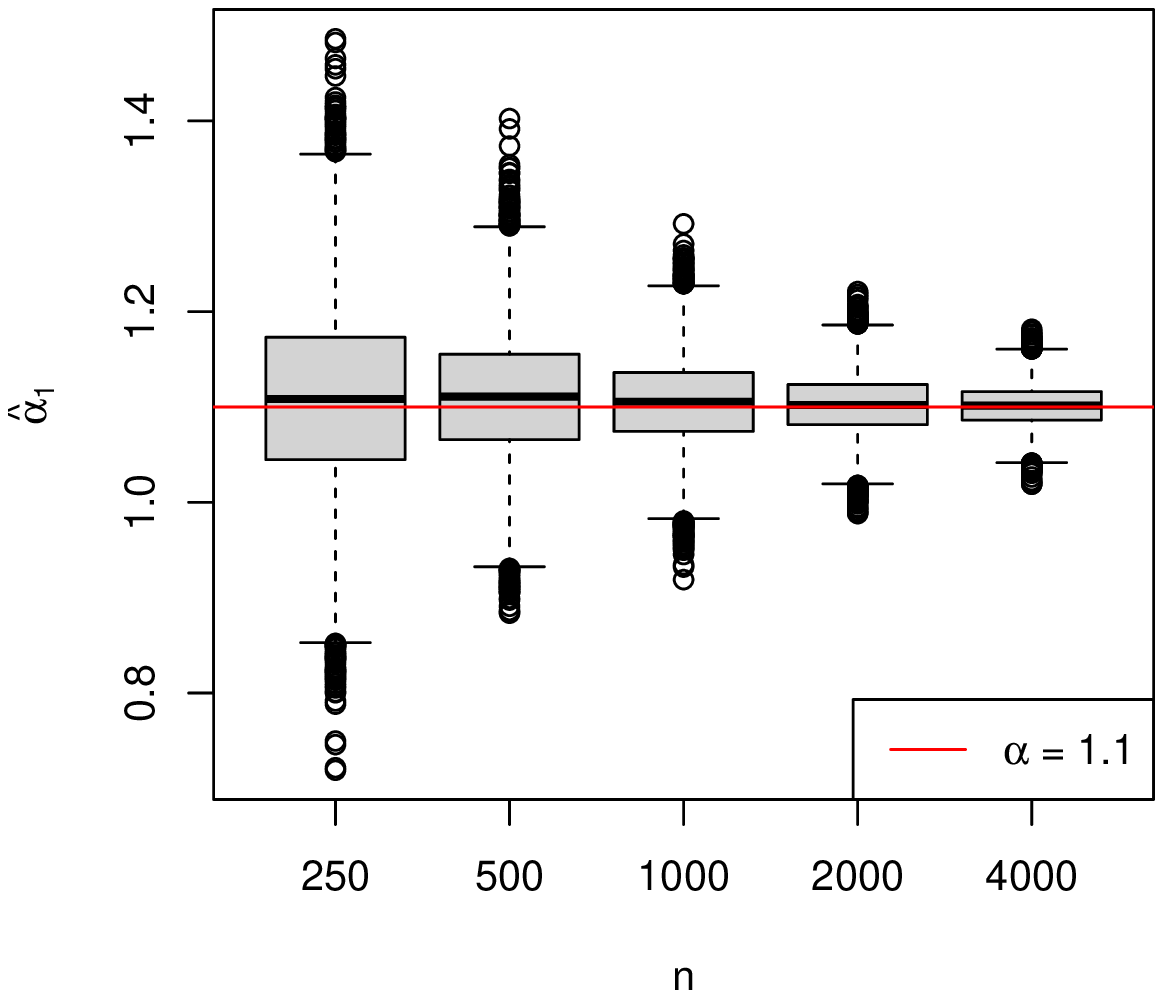}
\includegraphics[width=0.30\textwidth, angle =270]{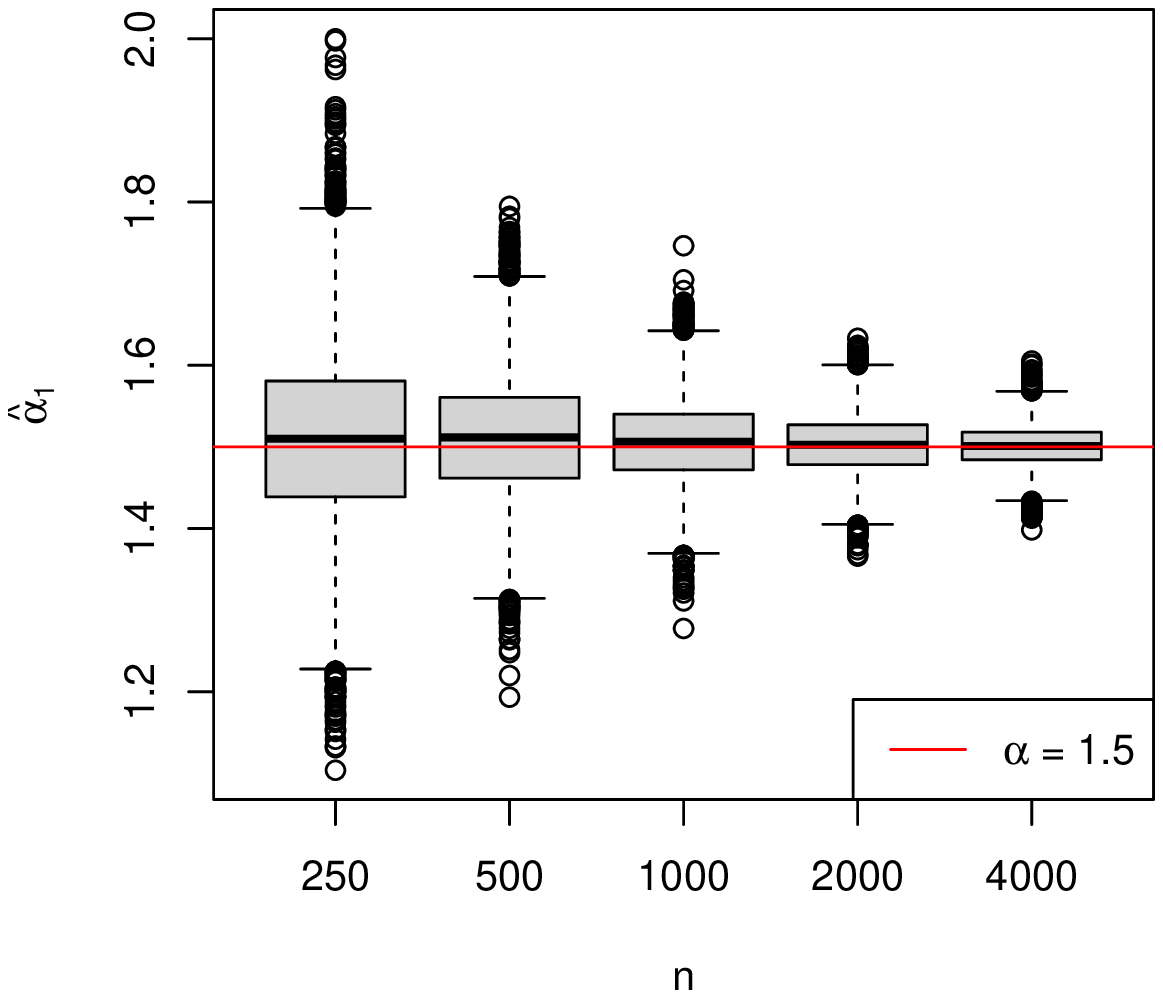}
\includegraphics[width=0.30\textwidth, angle =270]{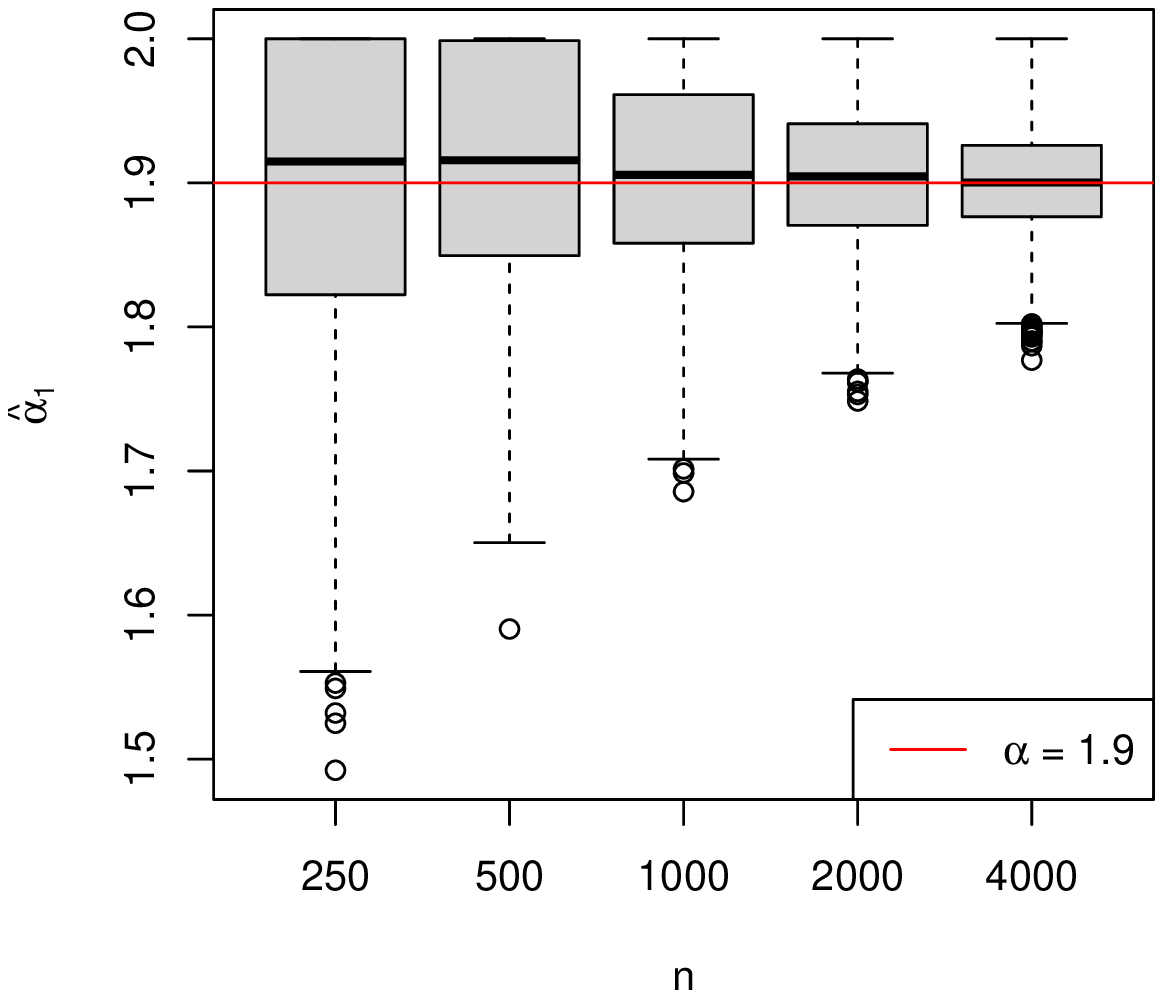}
\end{center}
\caption{Empirical box-plots of $\hat\alpha_1$ for sample size $n\in \{250,500,1000,2000,4000\}$ and tail-index $\alpha\in \{1.1,1.5,1.9\}$. Each box-plot is based on 10\,000 strong Monte Carlo samples of size $n$ drawn from the relevant $S\alpha S$ distribution.}
\label{fig:hat_alpha_vs_alph2}
\end{figure}

\begin{figure}[htp!]
\begin{center}
\includegraphics[width=0.31\textwidth, angle = 270]{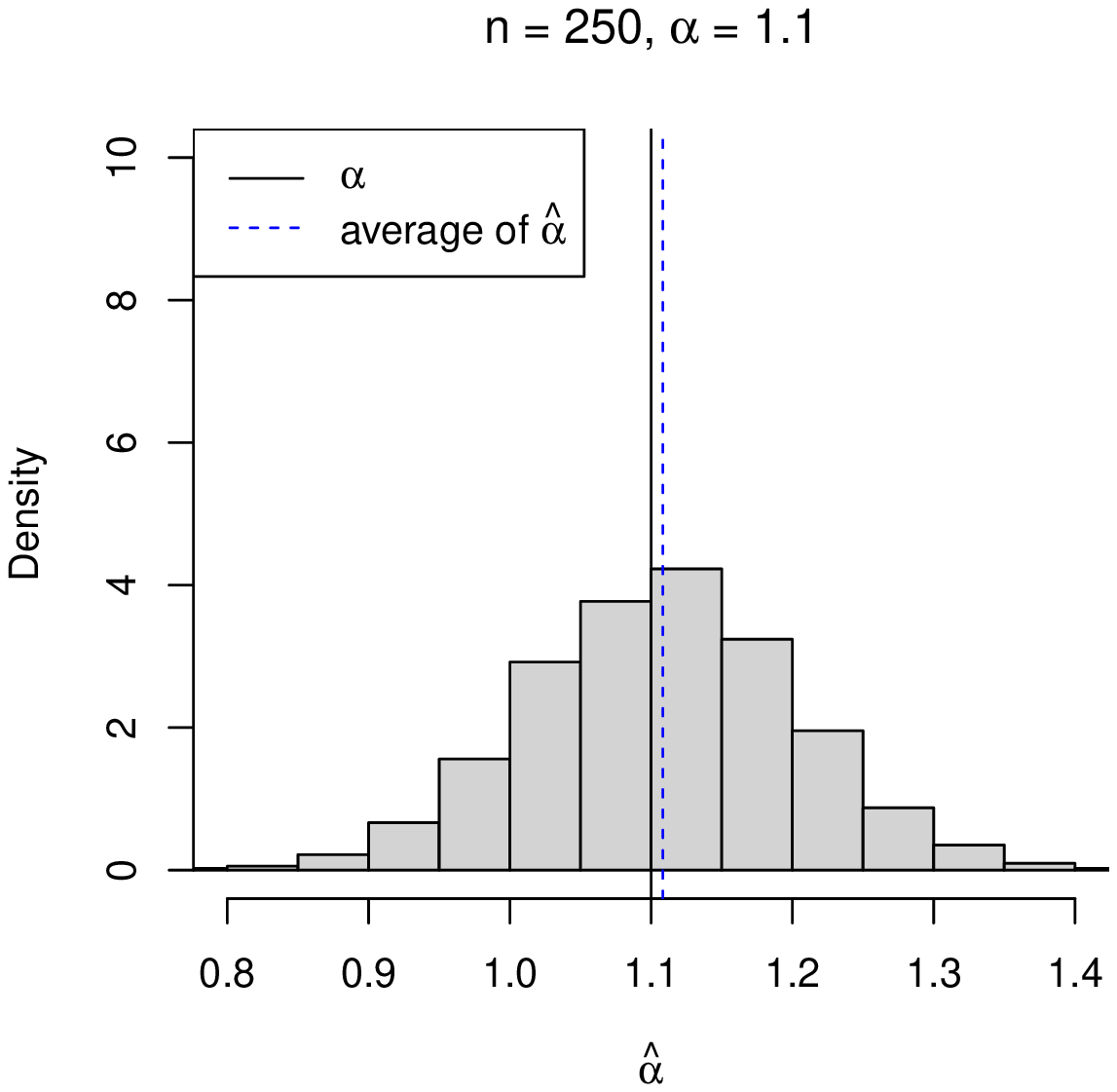}
\includegraphics[width=0.31\textwidth, angle = 270]{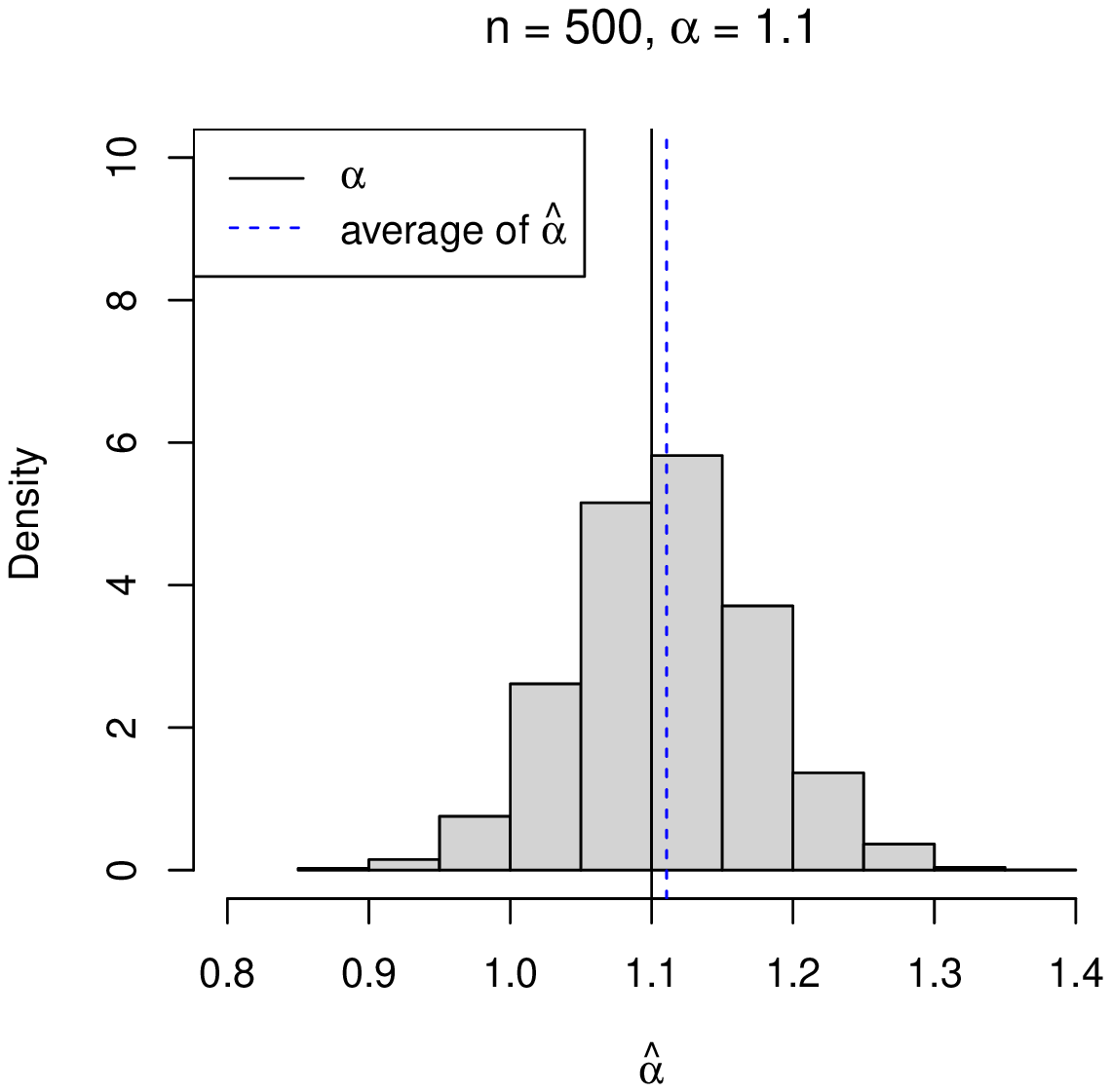}
\includegraphics[width=0.31\textwidth, angle = 270]{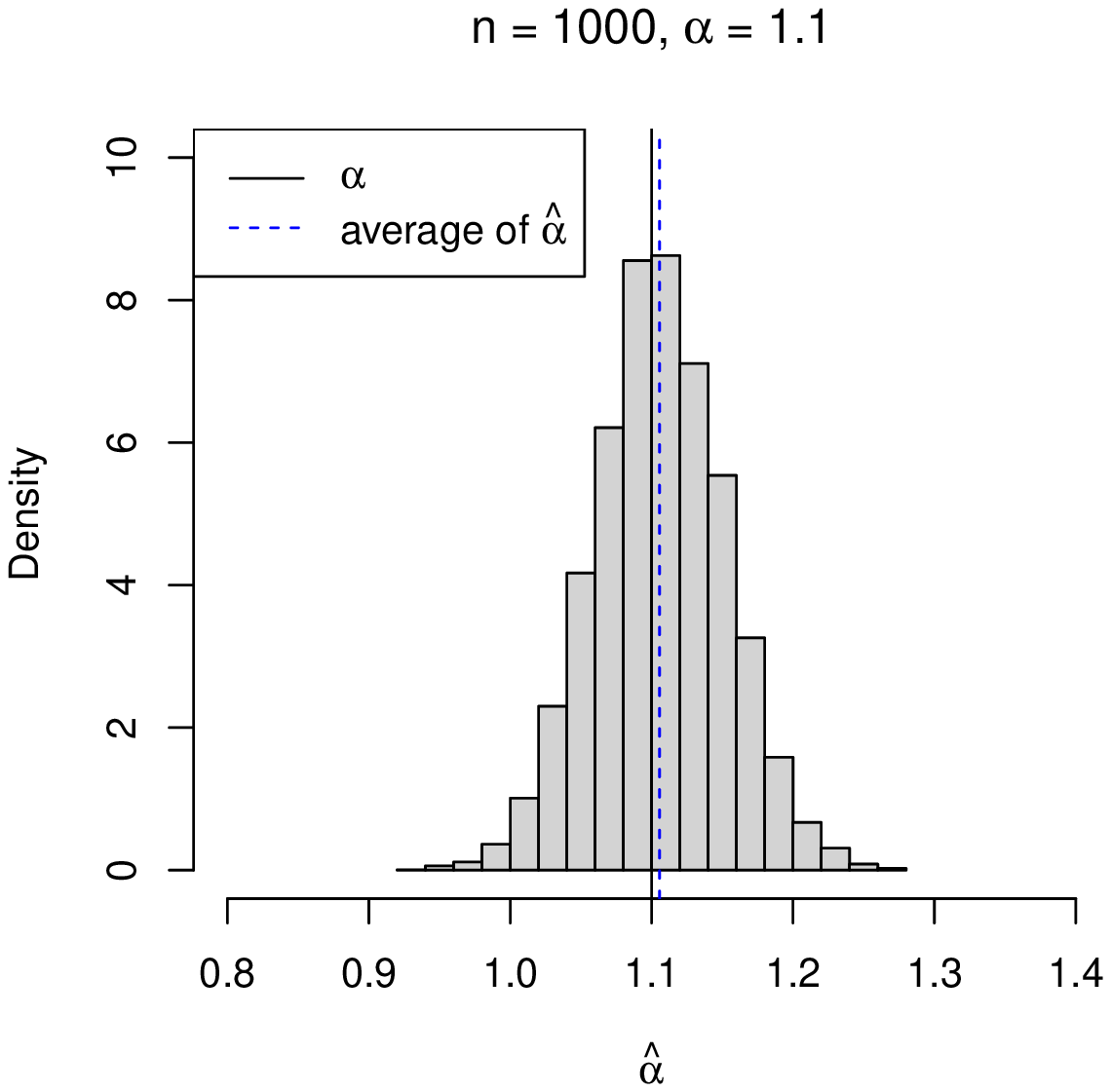}

\includegraphics[width=0.31\textwidth, angle = 270]{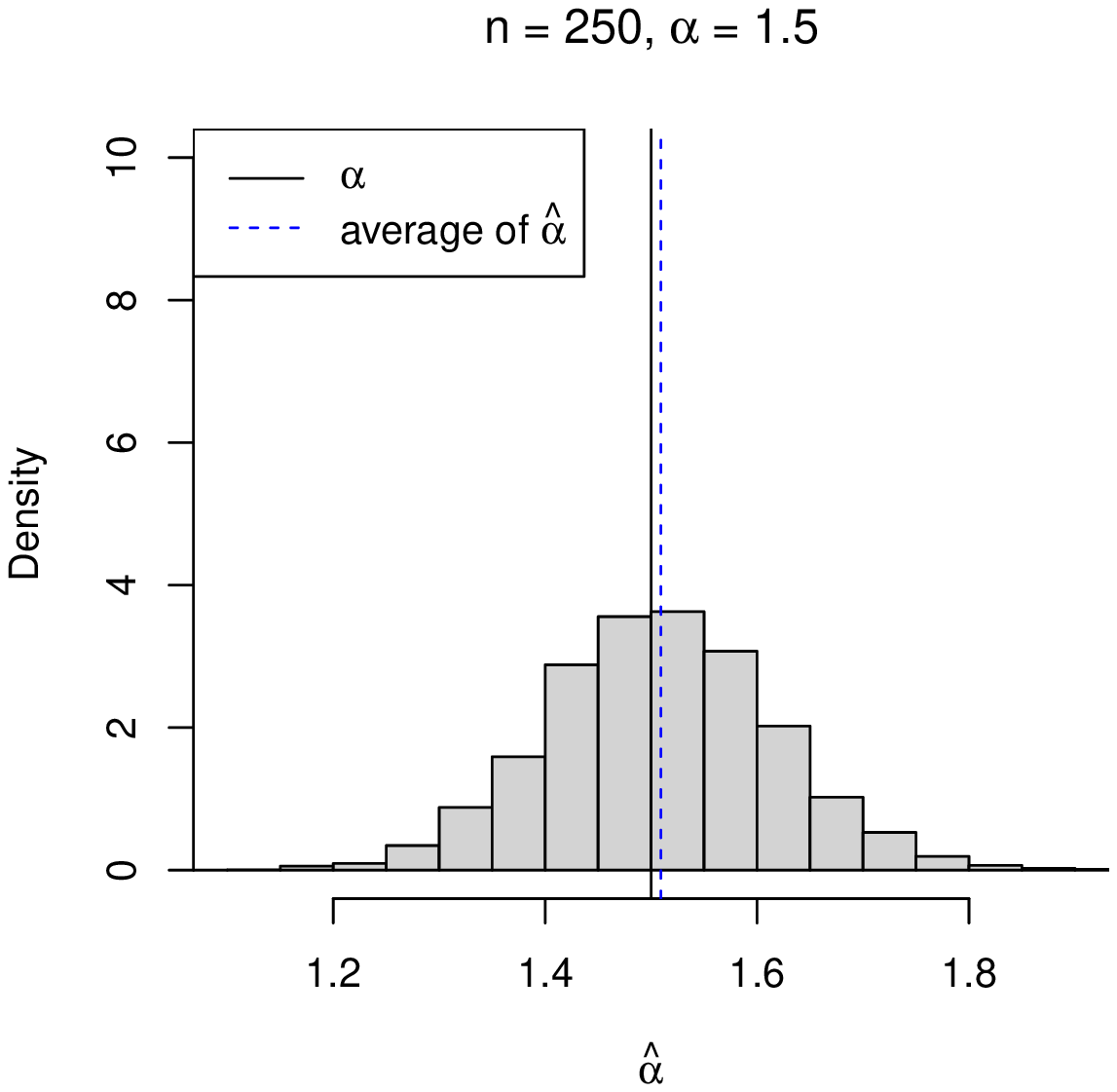}
\includegraphics[width=0.31\textwidth, angle = 270]{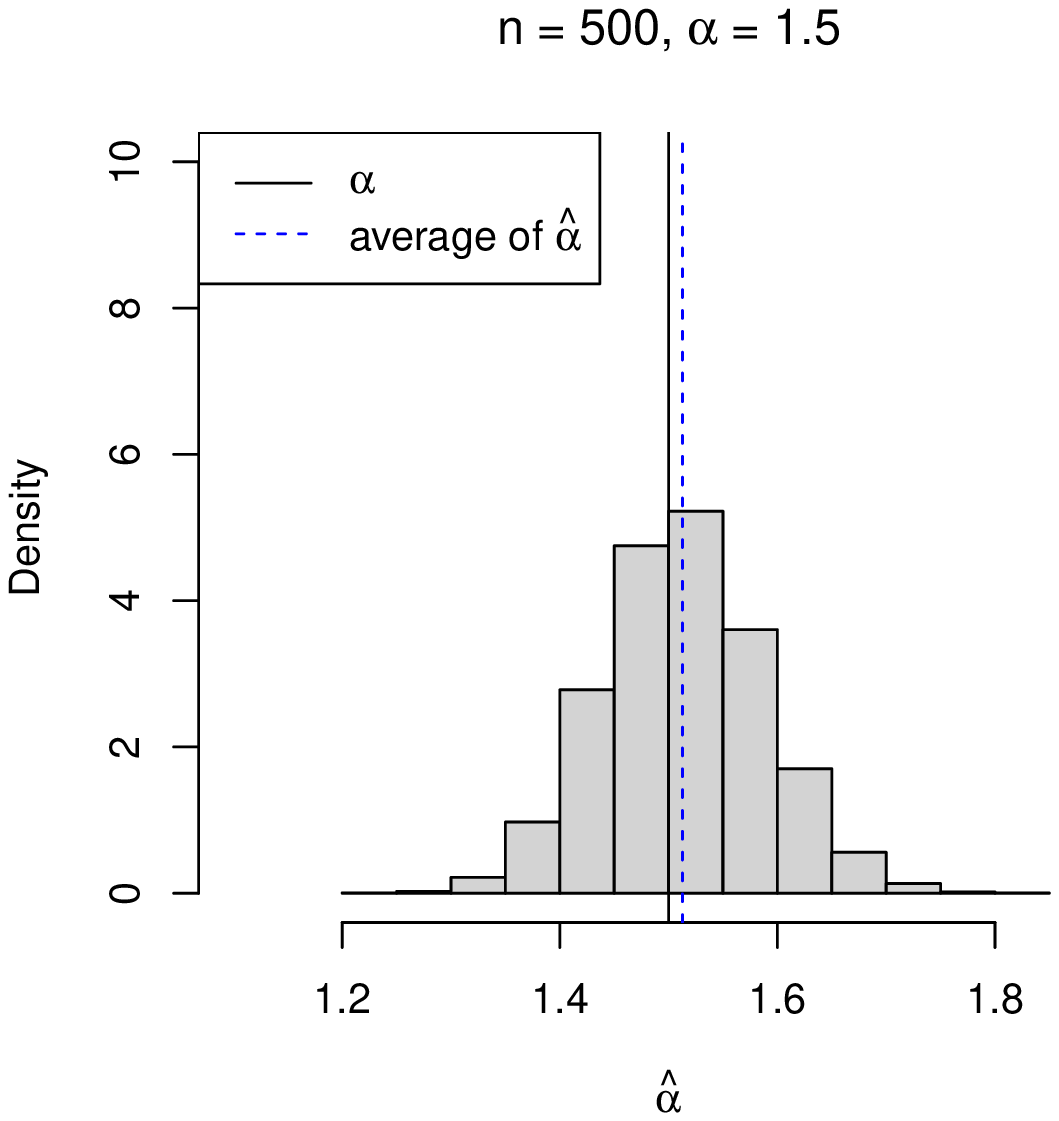}
\includegraphics[width=0.31\textwidth, angle = 270]{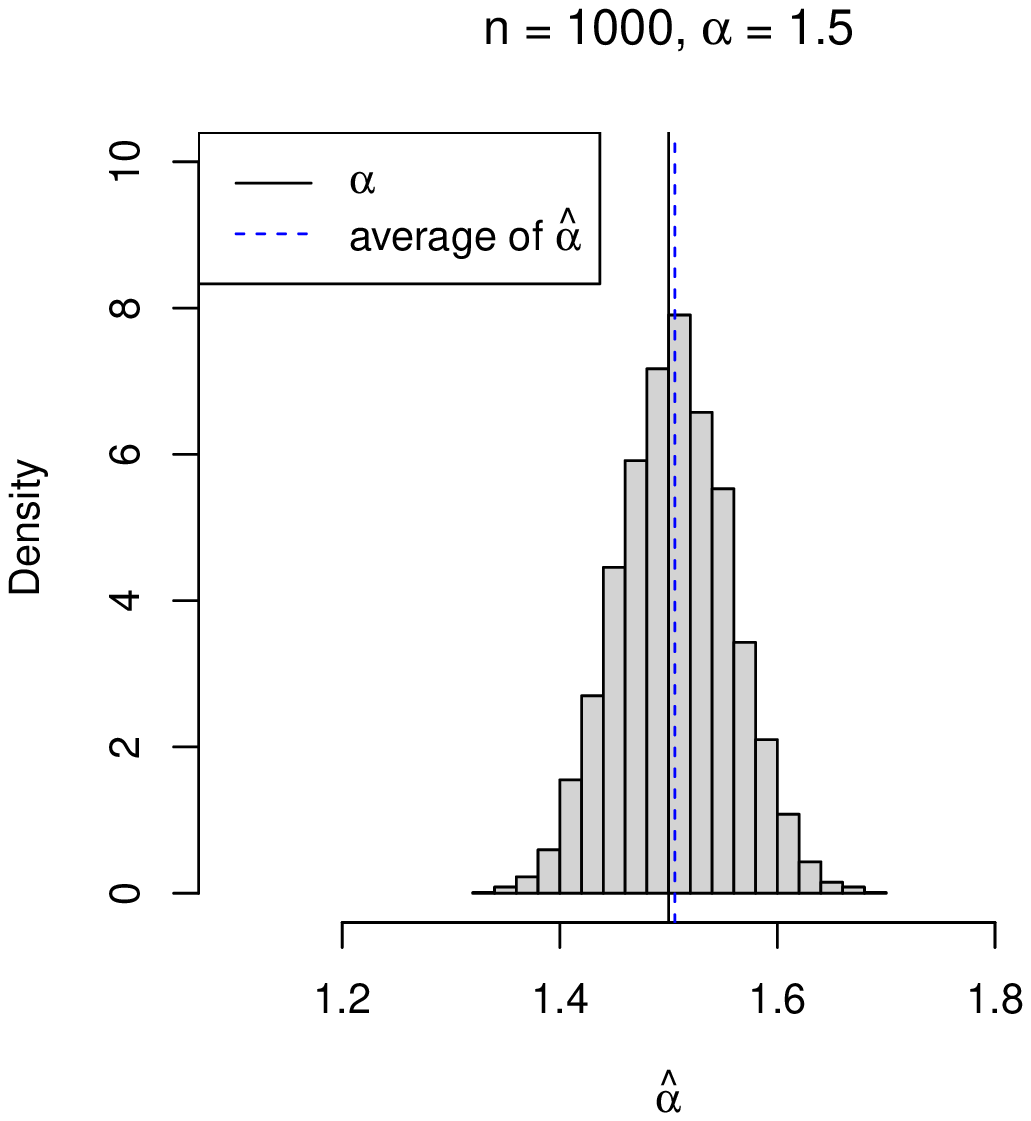}

\includegraphics[width=0.31\textwidth, angle = 270]{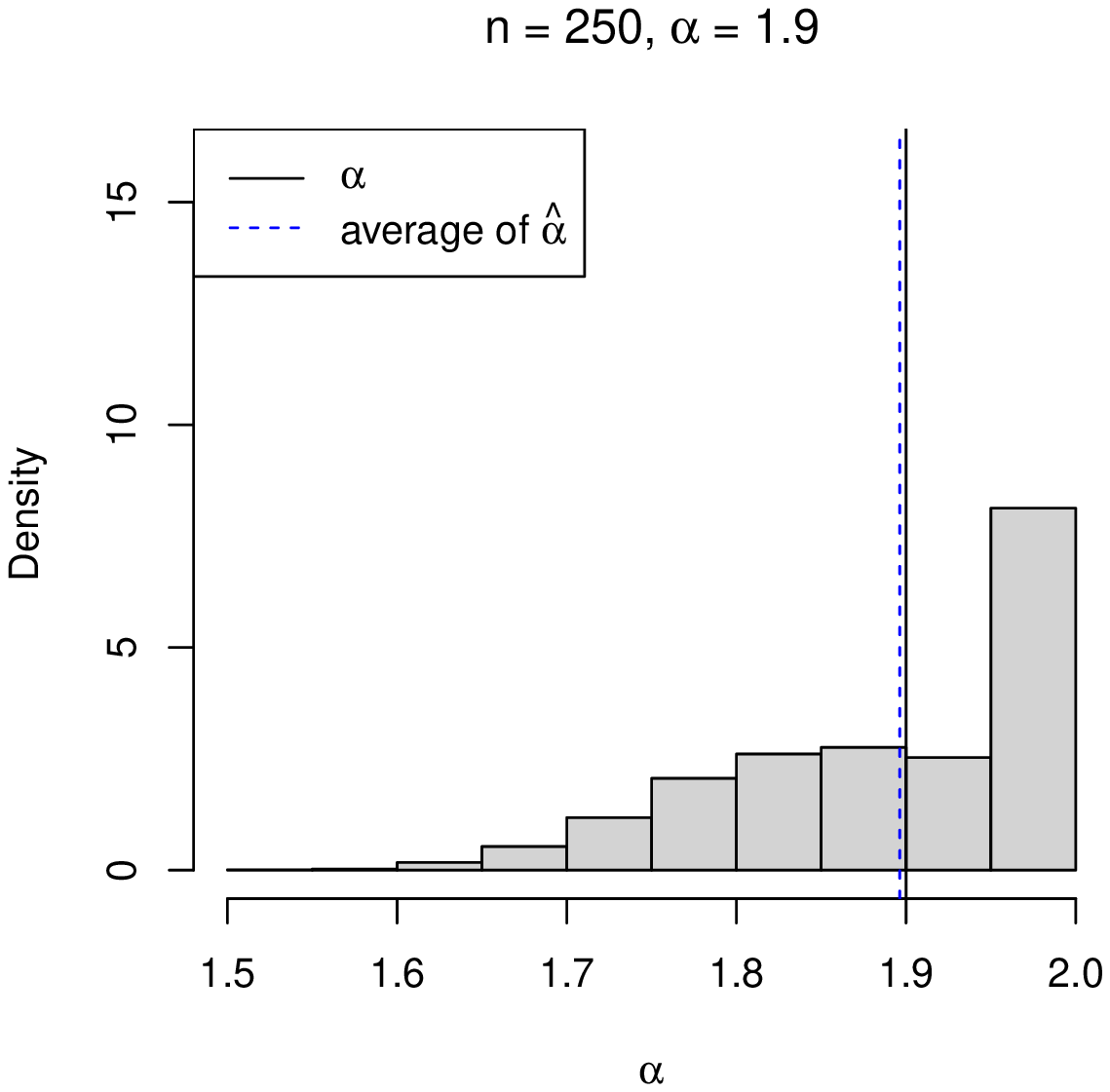}
\includegraphics[width=0.31\textwidth, angle = 270]{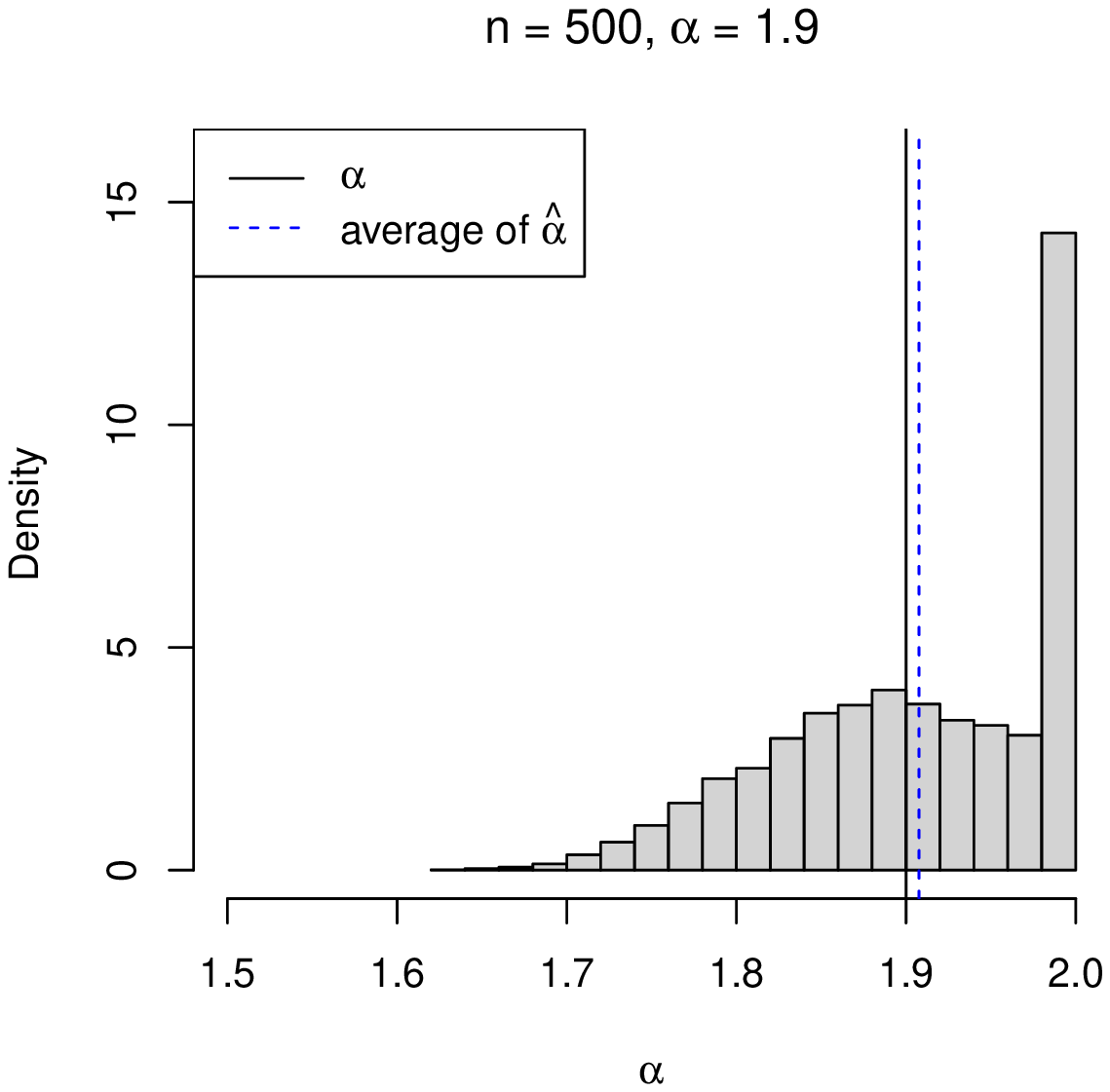}
\includegraphics[width=0.31\textwidth, angle = 270]{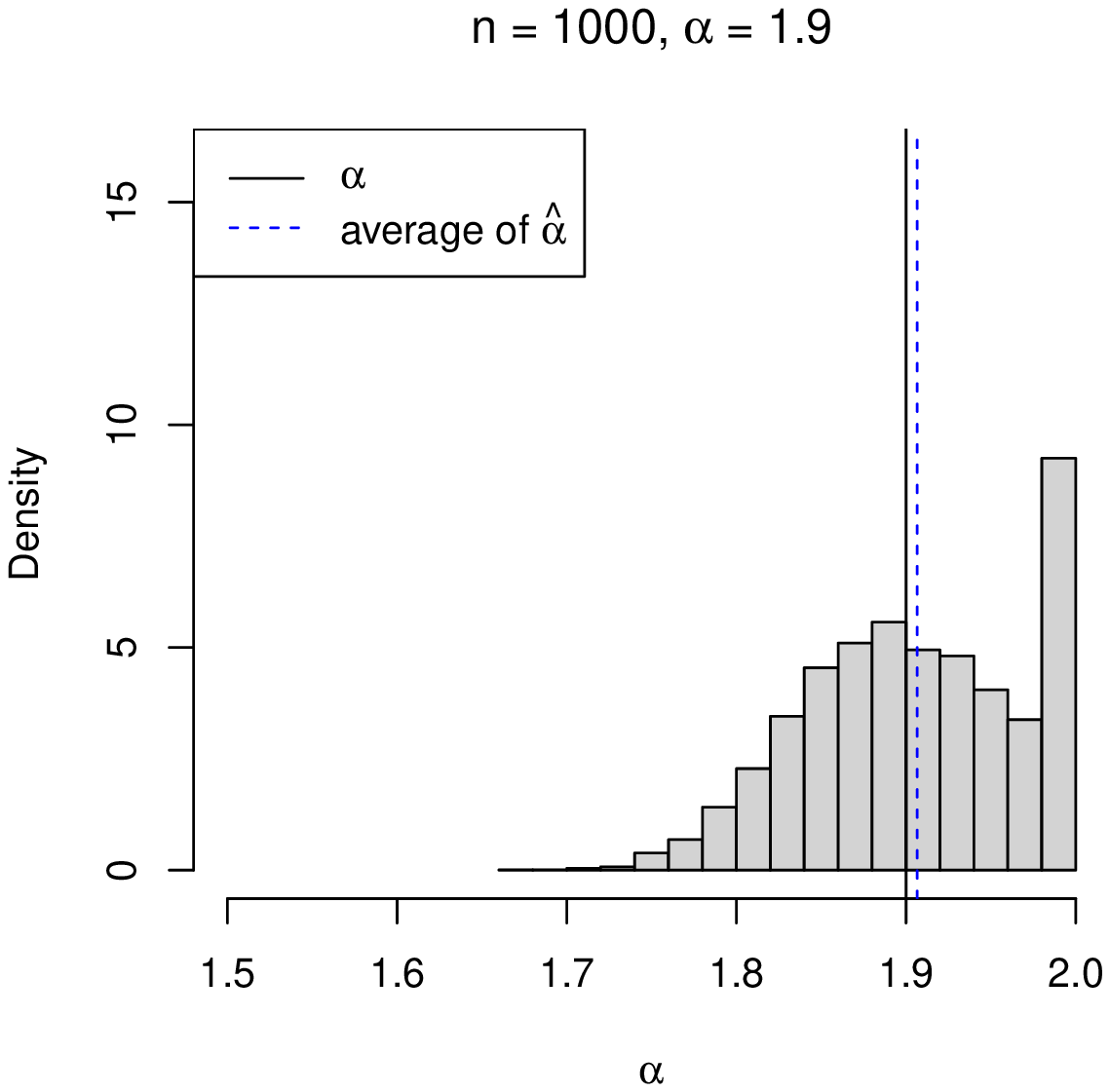}
\end{center}
\caption{Empirical distribution of $\hat{\alpha}_1$ for $\alpha \in \{1.1,1.5,1.9\}$ and the sample sizes $n \in \{250,500,1000\}$. Each plot is based on 10\,000 strong  Monte Carlo samples. Vertical lines confronts the true value of $\alpha$ with the average of $\hat{\alpha}_1$.}
\label{fig:hat_alpha_distribution}
\end{figure}

To illustrate the consistency of the proposed estimators, in Figure~\ref{fig:hat_alpha_vs_alph2} we show Monte Carlo based box-plots for $\hat\alpha_1$ estimator for various pre-fixed values of $\alpha$ and different sample sizes. As expected, the estimated value tends to the true value with decreasing error rate. The statistical properties of the proposed estimators are further investigated in Figure~\ref{fig:hat_alpha_distribution} in which, we present the full Monte Carlo distribution of $\hat\alpha_1$. As expected, the plot shows that the average value of the estimate is close to the true value of $\alpha$ and the variance of the estimator decreases as the sample size increases. Also, the distribution of $\hat\alpha_1$ for moderate values of $\alpha$ is close to the Gaussian distribution. However, since the tail index cannot exceed 2, the distribution of $\hat\alpha_1$ for $\alpha$ close to $2$ is naturally asymmetric; asymmetry disappears in the limit when the sample size tends to infinity. The results for $\hat{\alpha}_2$ are similar and are omitted for brevity.

To sum up, we conclude that the estimators proposed in~\eqref{eq:hat_alpha} behave as expected and enjoy multiple useful properties, such as Gaussian-like limiting distribution and small bias. In the next section, we confront the proposed estimators with other benchmark procedures to illustrate that the framework introduced in this paper is competitive and, in many instances, outperforms some commonly used estimation methods.

%%%%%%%%%%%%%%%%%%%%%%%%%%%%%%%%%%%%%%%%%%%%%%%%%%%%%%%%%
\section{Simulation study and performance evaluation}\label{simul}

In this section we study the performance of the  estimators $\hat\alpha_1$ and $\hat\alpha_2$ defined in~\eqref{eq:hat_alpha} and confront it with the performance of various benchmark alternatives. The evaluation process is Monte Carlo based, where the estimated values for different tail indices $\alpha\in [1,2]$ and different sample sizes $n\in \{250,500,1000\}$ are confronted with each other. In each case, we set strong Monte Carlo size to $k = 100\, 000$. Given reference tail-index $\alpha$, the reference sample size $n$, and the reference estimation technique, we use a performance metric given by the root mean squared error (RMSE) defined as
\begin{align}\label{rmse}
RMSE := \sqrt{\frac{1}{k} \sum_{i=1}^{k} (\alpha - \hat{\alpha}^i)^2},
\end{align}
where  $\alpha$ is the true value of the parameter and $\hat{\alpha}^i$ is the value estimated on the $i$th sample of size $n$, for $i=1, \ldots, k$. 

The remaining part of this section is organized as follows. In Section~\ref{S:otherEst}, we briefly describe several benchmark estimation procedures discussed in the literature and, in Section~\ref{S:comparison}, we compare them with the estimators defined in~\eqref{eq:hat_alpha}. Next, in Section~\ref{S:resistance}, we show that the estimators presented in~\eqref{eq:hat_alpha} are in fact robust with respect to asymmetry, so that our estimation technique could also be efficiently applied for $\beta\neq 0$, which indicates asymmetry in data.

\subsection{Benchmark estimation procedures}\label{S:otherEst}

In the literature, one can find various approaches used for the estimation of the stability index for the $\alpha$-stable distribution. They can be classified into three main categories, i.e. quantile methods, characteristic function-based algorithms, and  maximum likelihood techniques. In this section, we present the three most common approaches that can be considered as representatives of these classes. We focus on the estimation of the stability parameter $\alpha$ as this corresponds to the main topic of this paper. However, note that some of the methods discussed could also be used for an estimation of the other parameters of generic stable distributions.

The quantile-based approach for the estimation of $\alpha$ parameter in the symmetric case  was first  proposed in \cite{fama}, where authors applied a simple idea based on the observation that the quantiles of an appropriate large level  (e.g. $95\%$) of the symmetric $\alpha$-stable distribution decrease monotonically for $\alpha\in [1,2]$.  This idea was then extended by \cite{est3_new} for the general
(non-symmetric) case.  In a nutshell, to estimate $\alpha$, McCulloch proposes to compute the following statistic 
\begin{equation}\label{eq:McCulloch_nu}
\nu(\alpha):=\frac{Q_{\alpha}^{-1}(0.95)-Q_{\alpha}^{-1}(0.05)}{Q_{\alpha}^{-1}(0.75)-Q_{\alpha}^{-1}(0.25)},    
\end{equation}
where $Q_{\alpha}$ is the quantile function of the symmetric $\alpha$-stable distribution. Then, denoting by $\hat\nu$ the sample estimator of~\eqref{eq:McCulloch_nu}, we look for the parameter $\alpha$ which gives the same sample and theoretical values, i.e. we define the McCulloch Estimator (MCH) as
\begin{equation}\label{eq:MCH.def}
    \hat\alpha_{MCH}:=\nu^{-1}(\hat\nu).
\end{equation}
It should be noted that the value of $\hat{\alpha}_{MCH}$  cannot be determined analytically as the theoretical quantiles of the symmetric $\alpha$-stable distribution are not given in an explicit form. Thus, it was proposed to apply the Monte Carlo simulations and was provided the tabulated values of $\nu(\alpha)$.  In \cite{est3_new} it was proven that if the sample size is large enough, then the method gives reliable estimates of the underlying parameters, see also Chapter 4 in~\cite{nolan_book} for further discussion. For other generalizations of the quantile-based approach see e.g.. \cite{dominicy,Huixia,Maynon,leitch}.

The second technique is the so-called regression method based on the characteristic function. This approach is a natural choice for the estimation of $\alpha$-stable distribution's parameters. The idea of this approach was first introduced by \cite{press}. In a nutshell, using~\eqref{eq:fchar} with $(\beta,c,\mu)=(0,1,0)$, we get
\[
\log (-\log \phi(u))=\alpha \log |u|, \quad u\in \mathbb{R}.
\] 
Thus, the parameter $\alpha$ can be recovered as the regression coefficient in the regression of $\log (-\log \phi(u))$ against $\log|u|$ for $u$ from some predetermined set $\{u_1,\ldots, u_k\}$. In practice, given a sample $\{X_1,X_2,\ldots,X_n\}$, the characteristic function $\phi(u)$ is estimated by its sample version
$
\hat{\phi}(u):=\frac{1}{n}\sum_{j=1}^{n}\exp\left(iuX_j\right)$,  $ u\in \mathbb{R},
$
and the Regression Estimator (REG) is given as a solution to the least squares problem, i.e.
\begin{equation}\label{eq:REG.def}
\hat\alpha_{REG}:= \argmin_{\alpha} \left( \sum_{i=1}^k\left(\log (-\log \hat\phi(u_i))-\alpha \log|u_i|\right)^2\right);
\end{equation}
see e.g.~\cite{weronr} for details. Usually, the convergence of Press's method to the population values depends on the choice of estimation points, whose selection is problematic. Thus, \cite{est4_new} proposed a much more accurate method which starts with an initial estimate of the parameters and proceeds iteratively until some prespecified convergence criterion is satisfied. This technique is now considered as the classical regression-type approach for the estimation of the $\alpha$-stable distribution's parameters.  In the literature one can find various extensions and improved versions of this approach; see e.g. \cite{kogon,HASSANNEJAD,arad,Paulson} for details. 

The last benchmark method presented in this paper is based on the maximum likelihood approach. For a sample $\mathbf{X}:=\{X_1,X_2,\cdots,X_n\}$, we calculate the log-likelihood function 
\begin{equation*}
L(\mathbf{X};\alpha)=\sum_{j=1}^n\log\left(f_{\alpha}(X_j)\right),
\end{equation*}
where $f_{\alpha}(\cdot)$ is the PDF of  the symmetric $\alpha$-stable distribution. Then, the Maximum Likelihood Estimator (MLE) is given as the maximiser of $ L(\mathbf{x};\alpha)$, i.e.
\begin{equation}\label{eq:MLE.def}
\hat{\alpha}_{MLE}:=\argmax_{\alpha} L(\mathbf{X};\alpha).
\end{equation}
It is worth highlighting that, usually, the $\alpha$-stable PDF is not available in the explicit form and it needs to be approximated numerically. Thus, the maximum-likelihood-based approaches discussed in the literature differ in the choice of the approximating algorithm, see e.g. \cite{dumouchel,Nolan2001,MITTNIK,Matsui,Brorsen}.

%%%%%%%%%%%%%%%%%%%%%%%%%%%%%%%%%%%%%%%%%
\subsection{QCV estimators performance assessment}\label{S:comparison}

In this section we confront the RMSE performance of $N_1$, $N_2$, MCH, and REG estimators defined in \eqref{eq:hat_alpha}, \eqref{eq:MCH.def}, and \eqref{eq:REG.def}. While the remaining maximum likelihood method (MLE) requires big amount of time-consuming calculations, our preliminary check indicates that this method is outperformed by all procedures in almost all instances. For completeness, in Appendix~\ref{S:app_figures} we present the comparison of all benchmark methods, including MLE, for reduced number of Monte Carlo simulations, i.e. for $k_1:=1000$. See Table~\ref{tab:RMSE_MLE} therein for the summary of the results.

All computations are performed in {\bf R 4.0.4} and library \textit{stabledist} is used for simulations. The results for the benchmark estimators (MCH, REG, MLE) are based on our own implementation of the  classical algorithms adjusted to the stable symmetric case.

The RMSE results for all considered estimators computed for 100 000 strong Monte Carlo runs for various values of $\alpha$ and $n\in \{250,500,1000\}$ are presented in Table~\ref{table:MC}. 

\begin{table}[ht!]
\begin{center}
\scalebox{0.92}{
\begin{tabular}{|r|rrrr|}
\multicolumn{5}{c}{$n=250$}\\
  \hline
 $\alpha$ & $N_1$ & $N_2$ & MCH & REG \\ 
  \hline
1.0 & {\bf 0.091} & 0.124 & 0.092 & {\bf 0.091} \\ 
  1.1 & {\bf 0.096} & 0.126 & 0.098 & {\bf 0.096} \\ 
  1.2 & {\bf 0.100} & 0.126 & 0.104 & {\bf 0.100} \\ 
  1.3 & {\bf 0.103} & 0.125 & 0.112 & 0.104 \\ 
  1.4 & {\bf 0.104} & 0.121 & 0.121 & 0.108 \\ 
  1.5 & {\bf 0.107} & 0.117 & 0.134 & 0.109 \\ 
  1.6 & {\bf 0.110} & 0.112 & 0.149 & {\bf 0.110} \\ 
  1.7 & 0.115 & {\bf 0.107} & 0.157 & {\bf 0.107} \\ 
  1.8 & 0.116 & 0.101 & 0.152 & {\bf 0.098} \\ 
  1.9 & 0.104 & 0.084 & 0.139 & {\bf 0.077} \\ 
  2.0 & 0.103 & 0.072 & 0.143 & {\bf 0.041} \\ 
   \hline
\end{tabular}
\,\,\,\,\,\,\,\,\,\,\,\,
\begin{tabular}{|r|rrrr|}
\multicolumn{5}{c}{$n=500$}\\
  \hline
 $\alpha$ & $N_1$ & $N_2$ & MCH & REG \\ 
  \hline
1.0 & {\bf0.063} & 0.087 & 0.064 & 0.064 \\ 
  1.1 & {\bf 0.066} & 0.089 & 0.068 & 0.067 \\ 
  1.2 & {\bf 0.069} & 0.088 & 0.072 & 0.070 \\ 
  1.3 & {\bf 0.071} & 0.087 & 0.077 & 0.072 \\ 
  1.4 & {\bf 0.072} & 0.084 & 0.083 & 0.075 \\ 
  1.5 & {\bf 0.073} & 0.080 & 0.091 & 0.077 \\ 
  1.6 & {\bf 0.075} & 0.076 & 0.105 & 0.077 \\ 
  1.7 & 0.080 & {\bf 0.072} & 0.119 & 0.076 \\ 
  1.8 & 0.086 & {\bf 0.071} & 0.122 & {\bf 0.071} \\ 
  1.9 & 0.083 & 0.066 & 0.111 & {\bf 0.060} \\ 
  2.0 & 0.079 & 0.058 & 0.110 & {\bf 0.030} \\ 
   \hline
\end{tabular}
\,\,\,\,\,\,\,\,\,\,\,\,
\begin{tabular}{|r|rrrr|}
\multicolumn{5}{c}{$n=1000$}\\
  \hline
$\alpha$ & $N_1$ & $N_2$ & MCH & REG \\ 
  \hline
1.0 & {\bf 0.043} & 0.058 & 0.045 & 0.045 \\ 
  1.1 & {\bf 0.046} & 0.060 & 0.048 & 0.047 \\ 
  1.2 & {\bf 0.048} & 0.060 & 0.051 & 0.049 \\ 
  1.3 & {\bf 0.050} & 0.059 & 0.054 & 0.051 \\ 
  1.4 & {\bf 0.050} & 0.058 & 0.058 & 0.053 \\ 
  1.5 & {\bf 0.051} & 0.055 & 0.063 & 0.054 \\ 
  1.6 & {\bf 0.051} & 0.052 & 0.072 & 0.054 \\ 
  1.7 & 0.054 & {\bf 0.049} & 0.085 & 0.053 \\ 
  1.8 & 0.061 & {\bf 0.048} & 0.095 & 0.051 \\ 
  1.9 & 0.067 & 0.052 & 0.089 & {\bf 0.044} \\ 
  2.0 & 0.065 & 0.054 & 0.083 & {\bf 0.022} \\ 
   \hline
\end{tabular}}
\end{center}
\caption{RMSE for $N_1$, $N_2$, MCH, and REG estimators based on 100 000 strong Monte Carlo samples for $n \in \{250, 500, 1000\}$ and $\alpha \in \{1, 1.1, \ldots, 2\}$.  The best performance is marked in bold.}
\label{table:MC}
\end{table}

We see that for any choice of $\alpha$ and $n$, the estimator based on $N_1$ outperforms the McCulloch method. Also, for  $\alpha$ in small to medium range, the $N_1$ method shows the best results among all compared procedures. For $\alpha$ close to 2, the regression and $N_2$ methods provide the best results; the closer we are to the Gaussian case, the better the regression method performs. This shows that while $N_1$ is a good overall fit statistic, $N_2$ is good for the tail-index estimation when  $\alpha$ is close to $2$.

Next, we want to check whether the QCV method meaningfully complements the existing benchmarks methods and brings some new statistical information. If this is the case, one might refine the existing methodology e.g. by using {\it ensemble learning} with the QCV estimator, see e.g.~\cite{Pol2006} and references therein. We decided to focus on comparison between QCV estimators and REG estimator since they exhibited the best performance in Table~\ref{table:MC}.

\begin{figure}[htp!]
\begin{center}
\includegraphics[width=0.30\textwidth]{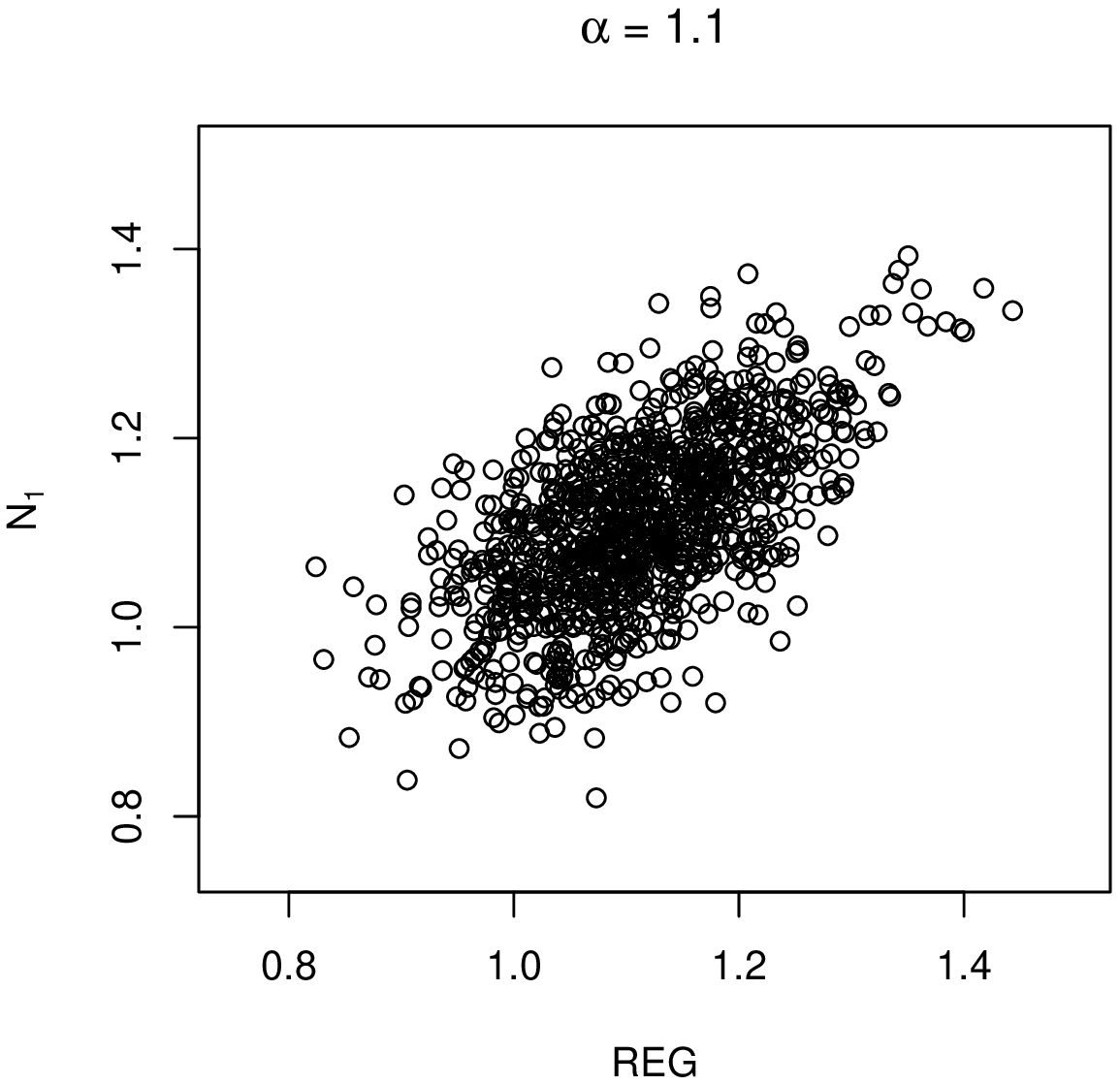}
\includegraphics[width=0.30\textwidth]{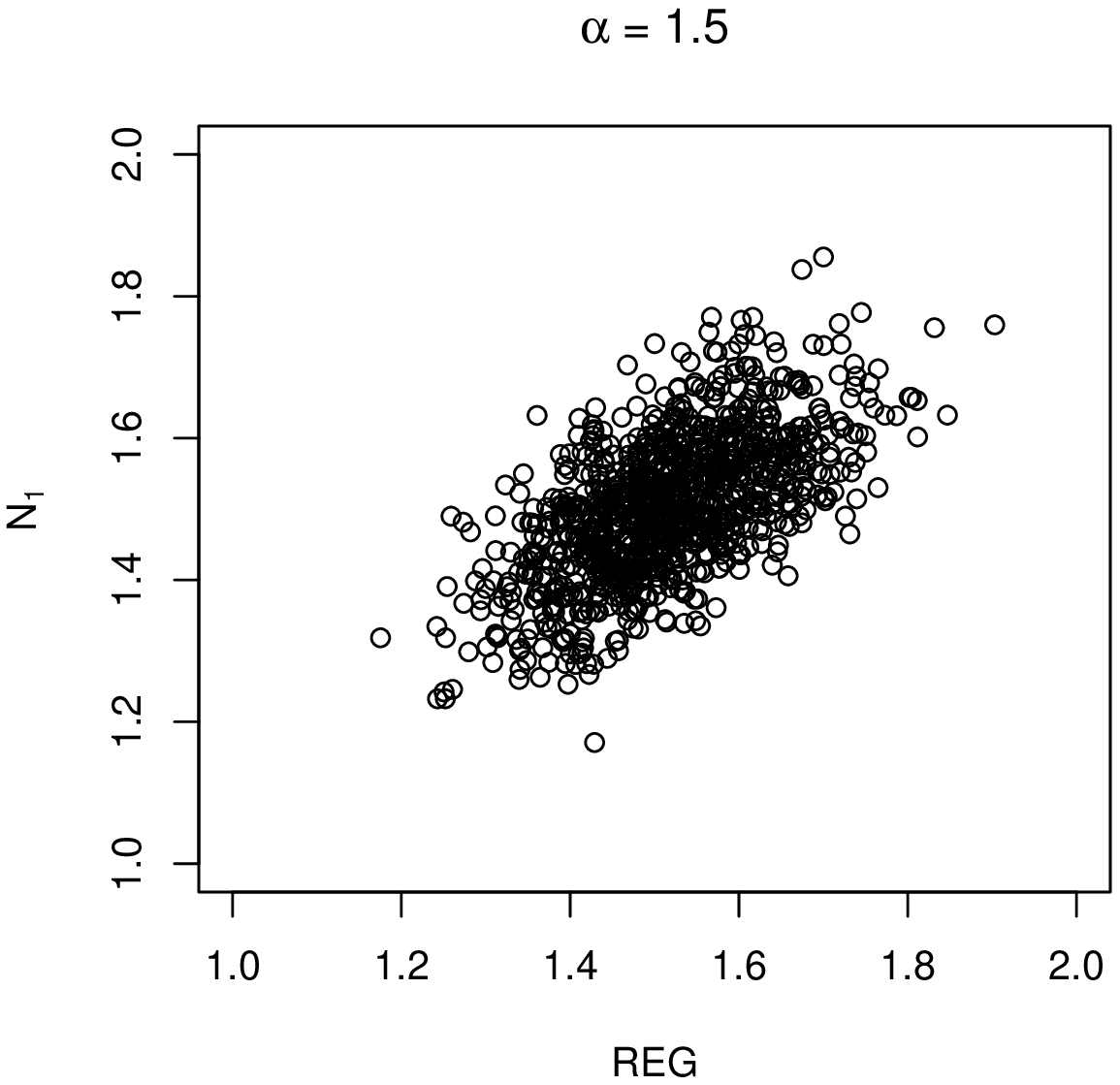}
\includegraphics[width=0.30\textwidth]{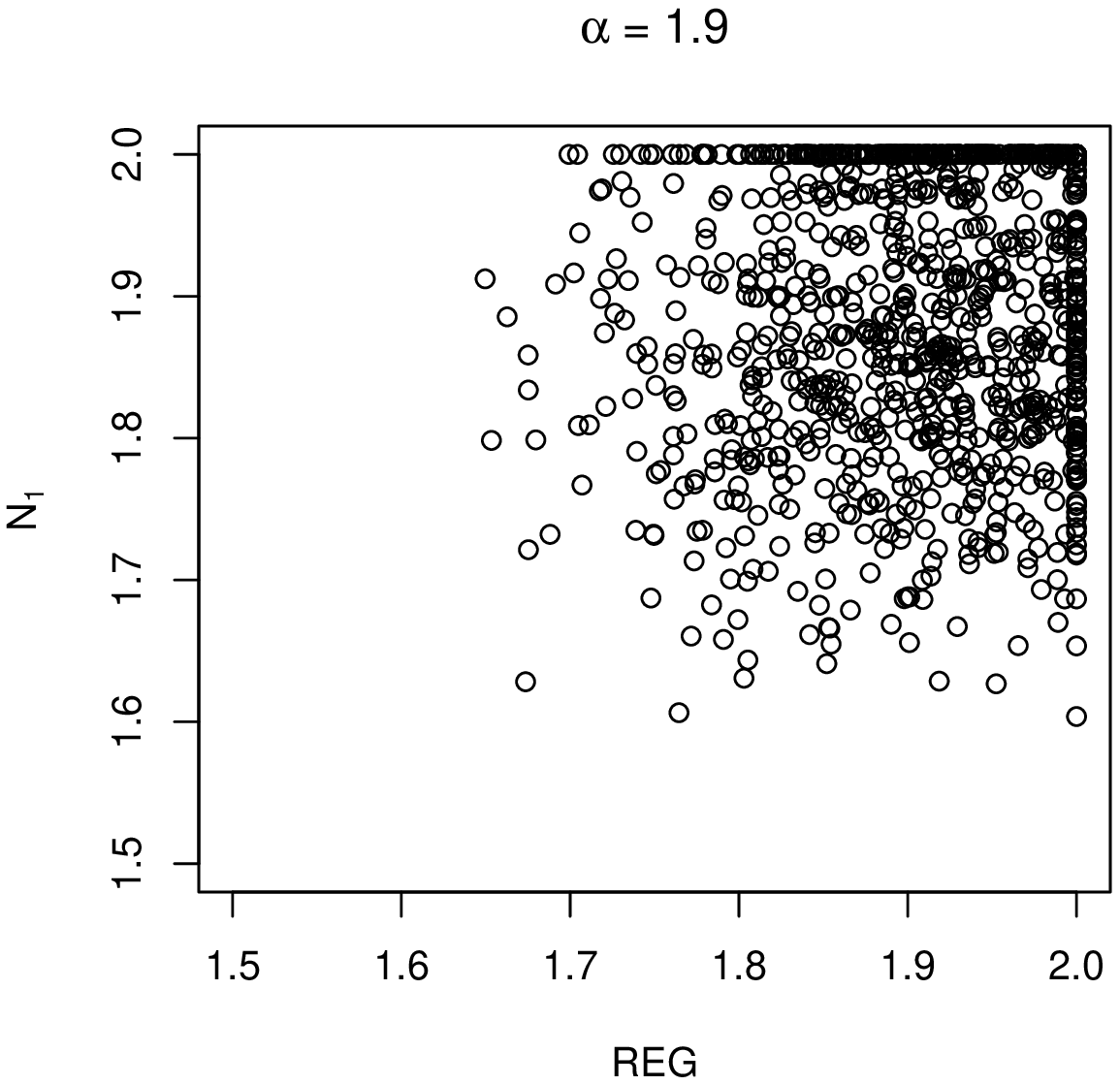}
\end{center}
\caption{REG and $N_1$ estimated values for exemplary values of $\alpha\in \{1.1,1.5,1.9\}$ and $n=250$. For each $\alpha$, we present the results for 1\,000 strong Monte Carlo simulations.}
\label{fig:CorN1}
\end{figure}

First, we decided to confront the estimated REG and $N_1$ values. Because $N_1$ is based on a different characteristics than the regression method, one would expect certain level of independence between the estimated values. This is indeed the case as shown in Figure~\ref{fig:CorN1}, in which the results for sample size $n=250$ and $\alpha\in \{1.1,1.5,1.9\}$ are presented. Similar conclusions are true for other sample sizes, and for $N_2$ statistics.

Second, we want to sanity check whether we can refine the estimation procedure by combining the information encoded in the REG and QCV  estimators. Given a sample, and estimated values $\hat\alpha_{1}$, $\hat\alpha_2$ and $\hat\alpha_{\textrm{REG}}$,  the simplest way to check this is to consider the averaged estimators
\[
\hat\alpha_{M_1}:=\frac{\hat\alpha_{1}+\hat\alpha_{\textrm{REG}}}{2}\quad\textrm{and}\quad \hat\alpha_{M_2}:=\frac{\hat\alpha_{2}+\hat\alpha_{\textrm{REG}}}{2},
\]
which we refer to simply as {\it $M_1$ estimator} and {\it $M_2$ estimator}. In particular, we can confront the performance of REG, $M_1$, and $M_2$ method using RMSE metric in a similar way it was done in Table~\ref{table:MC}. The results of this test are presented in Table~\ref{table:MC.22}.

\begin{table}[ht!]
\begin{center}
\scalebox{0.92}{
\begin{tabular}{|r|rrr|}
\multicolumn{4}{c}{$n=250$}\\
  \hline
 $\alpha$ & REG & $M_1$ & $M_2$ \\ 
  \hline
  1.0 & 0.091 & {\bf 0.079} & 0.089 \\ 
  1.1  & 0.096 & {\bf 0.085} & 0.095 \\ 
  1.2  & 0.100 & {\bf 0.090} & 0.099 \\ 
  1.3  & 0.104 & {\bf 0.094} & 0.103 \\ 
  1.4  & 0.108 & {\bf 0.096} & 0.104 \\ 
  1.5  & 0.109 & {\bf 0.097} & 0.103 \\ 
  1.6  & 0.110 & {\bf 0.097} & 0.100 \\ 
  1.7  & 0.107 & {\bf 0.094} & {\bf 0.094} \\ 
  1.8  & 0.098 & 0.087 & {\bf 0.084} \\ 
  1.9  & 0.077 & 0.069 & {\bf 0.063} \\ 
  2.0  & {\bf 0.041} & 0.061 & 0.047 \\ 
   \hline
\end{tabular}
\,\,\,\,\,\,\,\,\,\,\,\,
\begin{tabular}{|r|rrr|}
\multicolumn{4}{c}{$n=500$}\\
\hline
 $\alpha$ & REG & $M_1$ & $M_2$ \\ 
  \hline
1.0 & 0.064 & {\bf 0.055} & 0.062 \\ 
  1.1  & 0.067 & {\bf 0.060} & 0.067 \\ 
  1.2  & 0.070 & {\bf 0.063} & 0.070 \\ 
  1.3  & 0.073 & {\bf 0.066} & 0.072 \\ 
  1.4  & 0.075 & {\bf 0.067} & 0.073 \\ 
  1.5  & 0.077 & {\bf 0.068} & 0.072 \\ 
  1.6  & 0.077 & {\bf 0.067} & 0.069 \\ 
  1.7  & 0.076 & 0.066 & {\bf 0.065} \\ 
  1.8  & 0.071 & 0.064 & {\bf 0.060} \\ 
  1.9  & 0.060 & 0.054 & {\bf 0.049} \\ 
  2.0    & {\bf 0.030} & 0.047 & 0.037 \\ 
   \hline
\end{tabular}
\,\,\,\,\,\,\,\,\,\,\,\,
\begin{tabular}{|r|rrr|}
\multicolumn{4}{c}{$n=1000$}\\
\hline
$\alpha$ & REG & $M_1$ & $M_2$ \\ 
  \hline
1.0  & 0.045 & {\bf 0.039} & 0.042 \\ 
  1.1  & 0.047 & {\bf 0.042} & 0.046 \\ 
  1.2  & 0.049 & {\bf 0.044} & 0.048 \\ 
  1.3  & 0.051 & {\bf 0.046} & 0.050 \\ 
  1.4  & 0.053 & {\bf 0.047} & 0.051 \\ 
  1.5  & 0.054 & {\bf 0.047} & 0.050 \\ 
  1.6  & 0.054 & {\bf 0.047} & 0.048 \\ 
  1.7 & 0.054 & 0.046 & {\bf 0.045} \\ 
  1.8  & 0.051 & 0.045 & {\bf 0.041} \\ 
  1.9  & 0.044 & 0.042 & {\bf 0.037} \\ 
  2.0  & {\bf 0.022} & 0.037 & 0.033 \\ 
   \hline
\end{tabular}}
\end{center}
\caption{RMSE for REG, $M_1$, and $M_2$ estimators based on 100 000 strong Monte Carlo samples for $n \in \{250, 500, 1000\}$ and $\alpha \in \{1, 1.1, \ldots, 2\}$.  The best performance is marked in bold.}
\label{table:MC.22}
\end{table}

One can note that in almost all instances the averaged estimators $M_1$ and $M_2$ outperforms the REG estimator. It confirms that the QCV estimator contains information that is not encoded in the REG estimator and could be used to refine existing approaches using e.g. ensemble learning. In fact, the simple averaging scheme already leads to a substantial reduction of RMSE.

\begin{figure}[htp!]
\begin{center}
\includegraphics[width=0.31\textwidth]{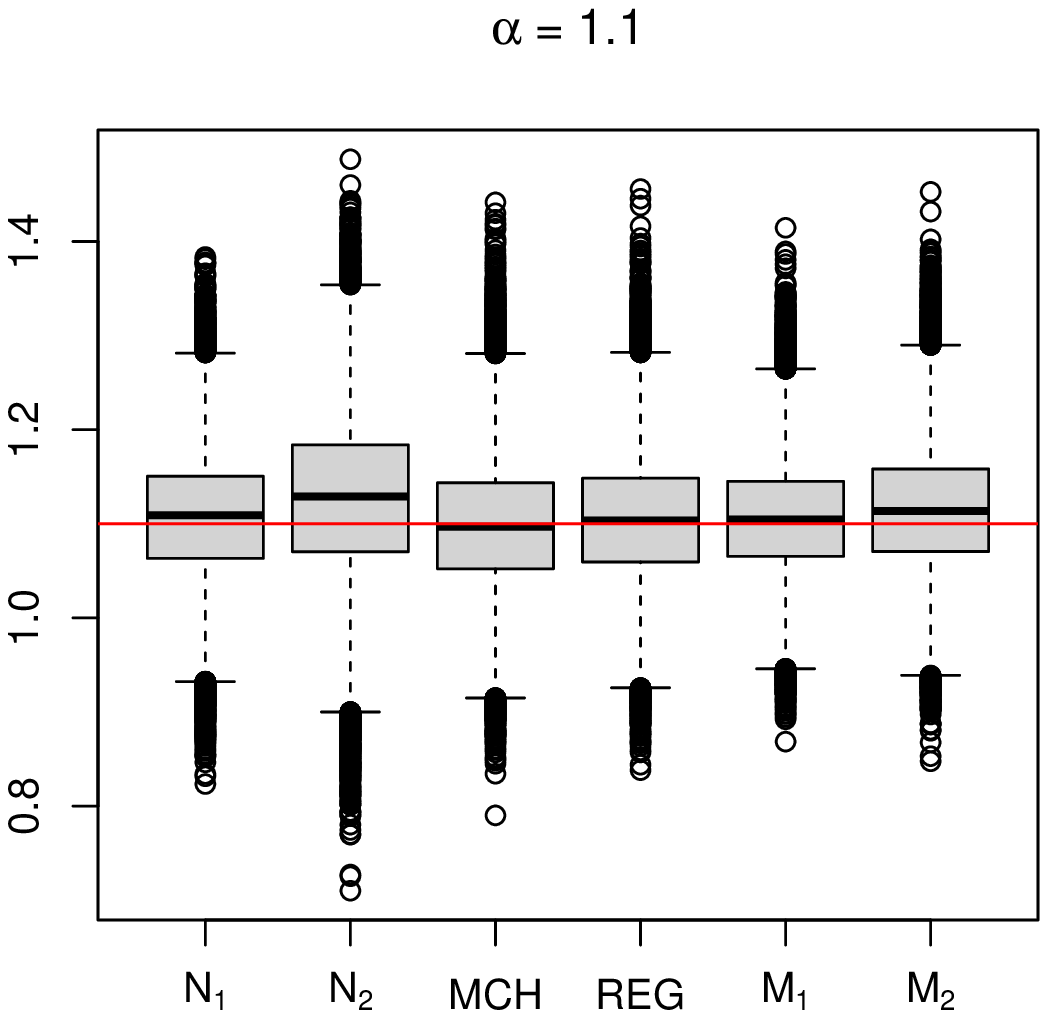}
\includegraphics[width=0.31\textwidth]{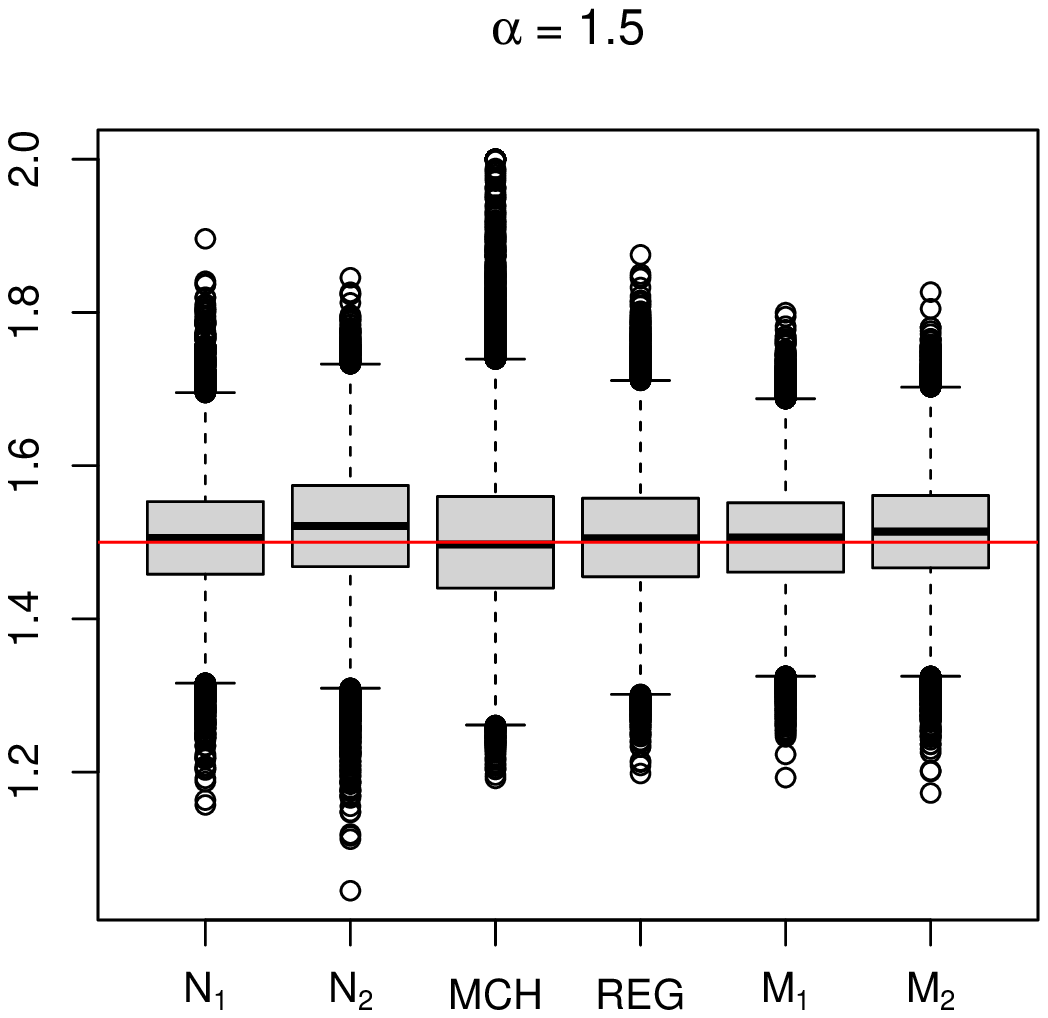}
\includegraphics[width=0.31\textwidth]{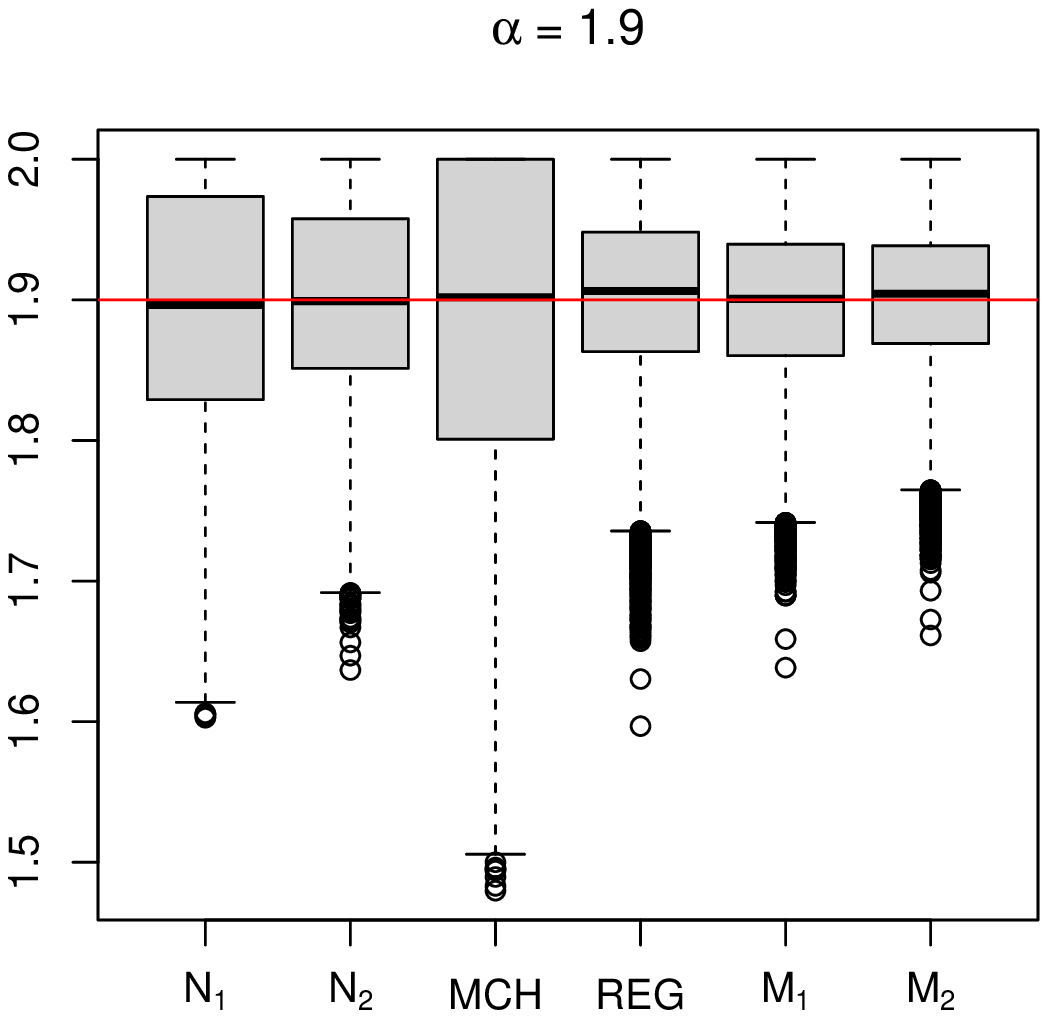}
\end{center}
\caption{Boxplots of 100 000 estimates for the considered methods with $n = 500$ and $\alpha \in \{1.1, 1.5, 1.9\}$. The horizontal line represents the true value of $\alpha$. }
\label{fig:boxplots}
\end{figure}

To provide a summary, in Figure~\ref{fig:boxplots} we present box-plots of estimated values for all considered estimators, for $\alpha\in \{1.1,1.5,1.9\}$ and $n = 500$. It can be seen that, in each case, the median of estimates for $N_1$, MCH, and REG are close to the true value of $\alpha$. Next, the dispersion of the results for $N_1$ is smaller than for MCH and comparable with REG. Still, for $\alpha=1.1$ and $\alpha=1.5$, the $N_1$ estimator outperforms the REG estimator in terms of RMSE as already shown in Table~\ref{table:MC}. Also, the right panel in Figure~\ref{fig:boxplots} confirms that the $N_2$ method is particularly effective for $\alpha$ close to 2. Finally, looking at $M_1$ and $M_2$, we see that the best results are obtained by combining REG and QCV approaches.

\FloatBarrier

\subsection{Asymmetry robustness}\label{S:resistance}
While the methodology in question is created for the symmetric \stab distribution family, in this section we demonstrate that it can also be applied for general $\alpha$-stable distribution, especially when the stability index is relatively close to $2$; recall that for $\alpha=2$, the \stab distribution is reduced to the Gaussian distribution which is independent of $\beta$ parameter. To demonstrate the robustness to the parameter change, in Table \ref{tab:MC_beta} we present the averages of the estimated  $\alpha$ parameters based on $10\,000$ Monte Carlo simulations for sample size $n=1000$ and various combinations of theoretical $\alpha$ and $\beta$ parameters from sets  $\{1.1,1.2,\ldots, 2\}$ and $\{0,0.1,\ldots,1\}$, respectively. The negative values of $\beta$ were not taken into account, as changing the sign of $\beta$ corresponds to the reflection of the random variable value across the Y axis. The results in Table~\ref{tab:MC_beta} show that, irregardless of $\beta$, the average estimates of $\alpha$ are relatively close to the theoretical values. This suggests that both $N_1$ and $N_2$ estimators can be considered as robust in reference to $\beta$.

\begin{figure}[htp!]
\begin{center}
\includegraphics[width=0.4\textwidth]{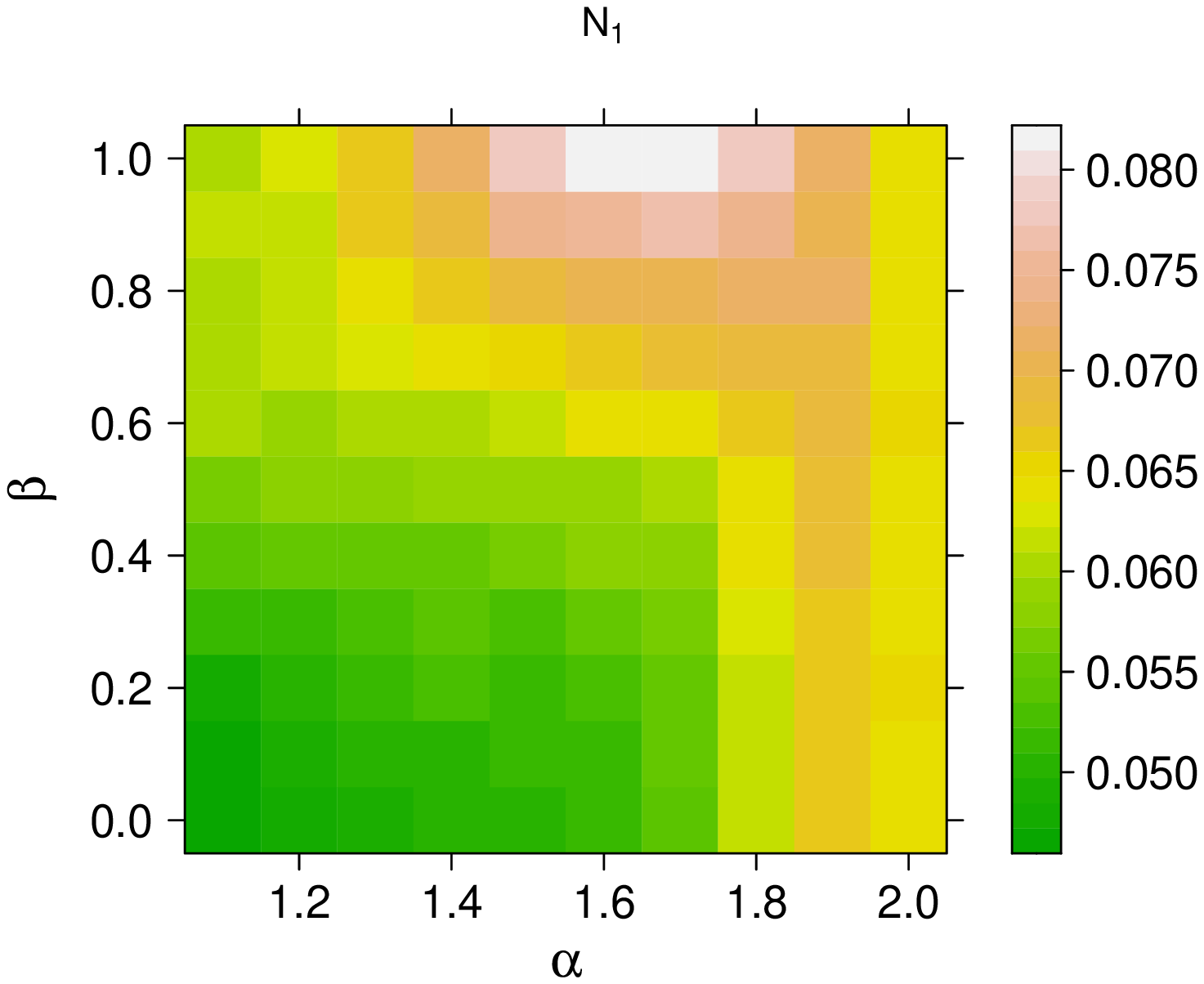}
\includegraphics[width=0.4\textwidth]{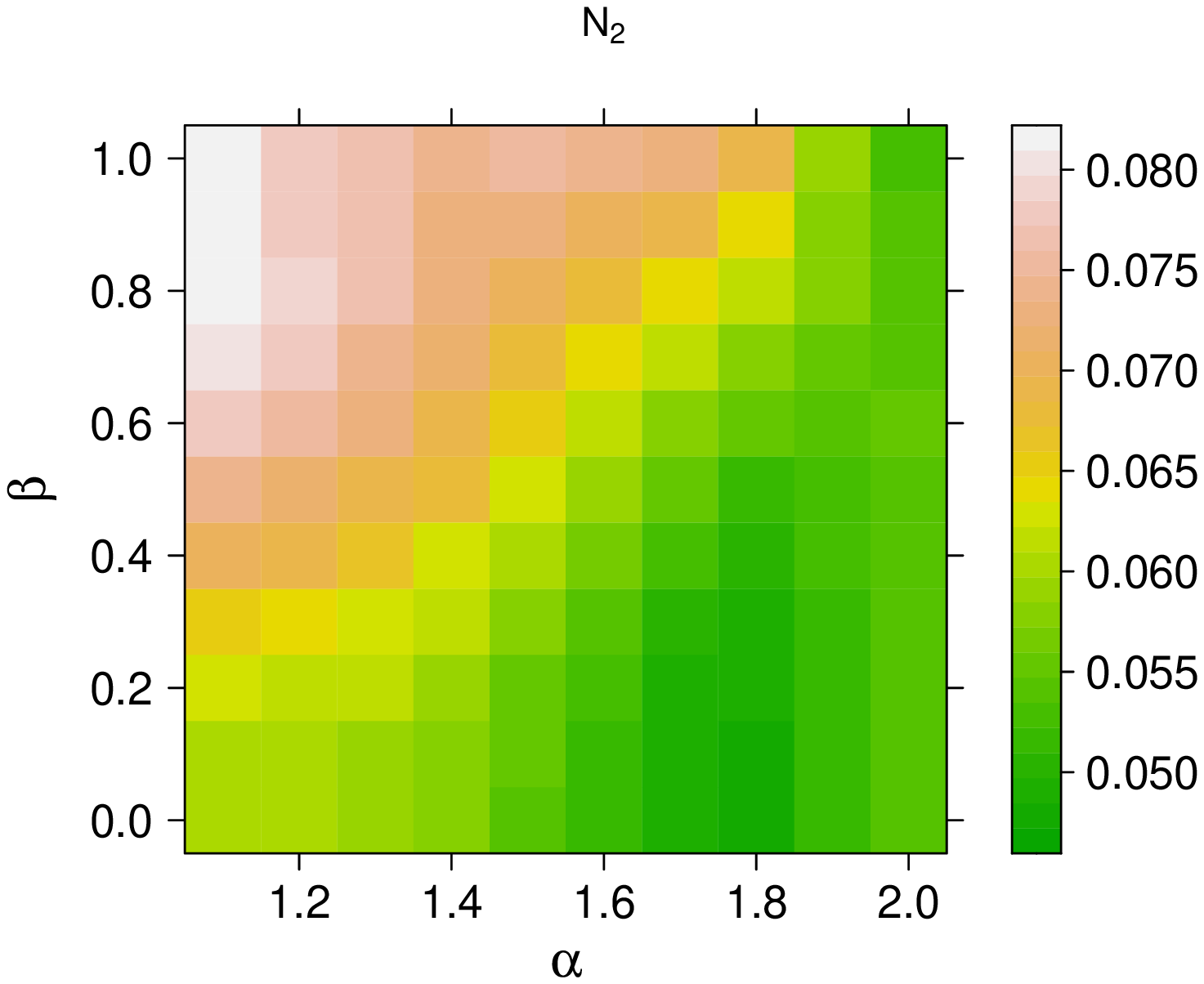}
\end{center}
\caption{RMSE of the estimated stability index for various combinations of  $\alpha \in \{1.1,1.2,\ldots,2\}$ and  $\beta \in \{0.0, 0.1,\ldots,1\}$; the number of strong Monte Carlo simulations is $10\, 000$ and the sample size is $n=1000$. The left panel shows the results for $N_1$, while the right panel shows the results for $N_2$.}\label{fig_res_2}
\end{figure}

\begin{table}[ht]
\centering
\scalebox{0.75}{
\begin{tabular}{|r|rrrrrrrrrrr|}
\multicolumn{12}{c}{$N_1$}\\
  \hline
$\alpha$ \textbackslash \, $\beta$ & 0.0 & 0.1 & 0.2 & 0.3 & 0.4 & 0.5 & 0.6 & 0.7 & 0.8 & 0.9 & 1.0 \\ 
\hline
1.1 & 1.105 & 1.106 & 1.108 & 1.110 & 1.113 & 1.114 & 1.115 & 1.115 & 1.113 & 1.112 & 1.110 \\ 
  1.2 & 1.206 & 1.206 & 1.208 & 1.208 & 1.209 & 1.210 & 1.208 & 1.207 & 1.205 & 1.200 & 1.197 \\ 
  1.3 & 1.307 & 1.307 & 1.307 & 1.306 & 1.306 & 1.304 & 1.302 & 1.299 & 1.294 & 1.289 & 1.283 \\ 
  1.4 & 1.406 & 1.406 & 1.404 & 1.404 & 1.402 & 1.400 & 1.396 & 1.390 & 1.384 & 1.379 & 1.371 \\ 
  1.5 & 1.507 & 1.505 & 1.504 & 1.503 & 1.499 & 1.496 & 1.490 & 1.484 & 1.476 & 1.469 & 1.461 \\ 
  1.6 & 1.606 & 1.605 & 1.604 & 1.601 & 1.598 & 1.594 & 1.587 & 1.580 & 1.573 & 1.564 & 1.556 \\ 
  1.7 & 1.709 & 1.707 & 1.706 & 1.702 & 1.699 & 1.693 & 1.687 & 1.681 & 1.675 & 1.667 & 1.658 \\ 
  1.8 & 1.812 & 1.810 & 1.810 & 1.806 & 1.803 & 1.801 & 1.794 & 1.791 & 1.785 & 1.780 & 1.772 \\ 
  1.9 & 1.907 & 1.907 & 1.906 & 1.906 & 1.905 & 1.903 & 1.904 & 1.901 & 1.897 & 1.896 & 1.895 \\ 
  2.0 & 1.971 & 1.971 & 1.970 & 1.971 & 1.971 & 1.971 & 1.970 & 1.971 & 1.971 & 1.971 & 1.971 \\ 
   \hline
\end{tabular}}

\centering
\scalebox{0.75}{
\begin{tabular}{|r|rrrrrrrrrrr|}
\multicolumn{12}{c}{$N_2$}\\

  \hline
$\alpha$ \textbackslash \, $\beta$ & 0.0 & 0.1 & 0.2 & 0.3 & 0.4 & 0.5 & 0.6 & 0.7 & 0.8 & 0.9 & 1.0 \\ 
  \hline
1.1 & 1.109 & 1.112 & 1.115 & 1.119 & 1.123 & 1.124 & 1.127 & 1.129 & 1.129 & 1.129 & 1.127 \\ 
  1.2 & 1.210 & 1.211 & 1.215 & 1.217 & 1.219 & 1.222 & 1.223 & 1.224 & 1.224 & 1.219 & 1.217 \\ 
  1.3 & 1.310 & 1.312 & 1.313 & 1.314 & 1.317 & 1.318 & 1.318 & 1.317 & 1.315 & 1.310 & 1.304 \\ 
  1.4 & 1.409 & 1.411 & 1.411 & 1.413 & 1.414 & 1.414 & 1.412 & 1.409 & 1.404 & 1.400 & 1.393 \\ 
  1.5 & 1.510 & 1.510 & 1.510 & 1.511 & 1.509 & 1.509 & 1.505 & 1.502 & 1.495 & 1.489 & 1.482 \\ 
  1.6 & 1.609 & 1.609 & 1.609 & 1.608 & 1.607 & 1.605 & 1.600 & 1.594 & 1.589 & 1.581 & 1.574 \\ 
  1.7 & 1.710 & 1.709 & 1.709 & 1.707 & 1.706 & 1.701 & 1.697 & 1.691 & 1.686 & 1.679 & 1.670 \\ 
  1.8 & 1.812 & 1.811 & 1.812 & 1.809 & 1.807 & 1.805 & 1.799 & 1.796 & 1.789 & 1.785 & 1.778\\ 
  1.9 & 1.912 & 1.912 & 1.912 & 1.911 & 1.911 & 1.909 & 1.909 & 1.906 & 1.902 & 1.901 & 1.900 \\ 
  2.0 & 1.981 & 1.981 & 1.981 & 1.981 & 1.981 & 1.981 & 1.980 & 1.981 & 1.981 & 1.981 & 1.982 \\ 
   \hline
\end{tabular}}

\caption{Averages of the estimated $\alpha$ parameter for  $10\,000$ Monte Carlo simulations of \stab distributed samples with $n=1000$ elements. In the simulations we considered all combinations of the $\alpha$ and $\beta$ parameters from the sets $\{1.1,1.2,\ldots, 2\}$ and $\{0,0.1,\ldots,1\}$, respectively. }
\label{tab:MC_beta}

\end{table}
In Figure \ref{fig_res_2}, we demonstrate the RMSE value of the estimated stability index for various combinations of  $\alpha \in \{1.1,1.2,\ldots,2\}$ and  $\beta \in \{0, 0.1,\ldots,1\}$; as before, the number of strong Monte Carlo simulations is equal to $10\,000$ and sample size equals $n=1000$. The results presented in the level plots indicate relatively small error changes when the methodology is applied to non-symmetric data, that is, for $\beta\neq 0$. This effect is visible especially for the parameter $\alpha$ close to $2$. The vertical stripes on the level plots for the $\alpha$ parameter close to $2$ have almost the same colors, which confirms that for this case the $N_1$ and $N_2$ estimators are robust for the $\beta$ parameter. Namely, we observe the same degree of error in Table~\ref{fig_res_2} but from Table~\ref{tab:MC_beta} we learn that the mean estimated value does not depend on $\beta$ which indicates robustness when combined together.

\begin{figure}[htp!]
\begin{center}
\includegraphics[width=0.4\textwidth]{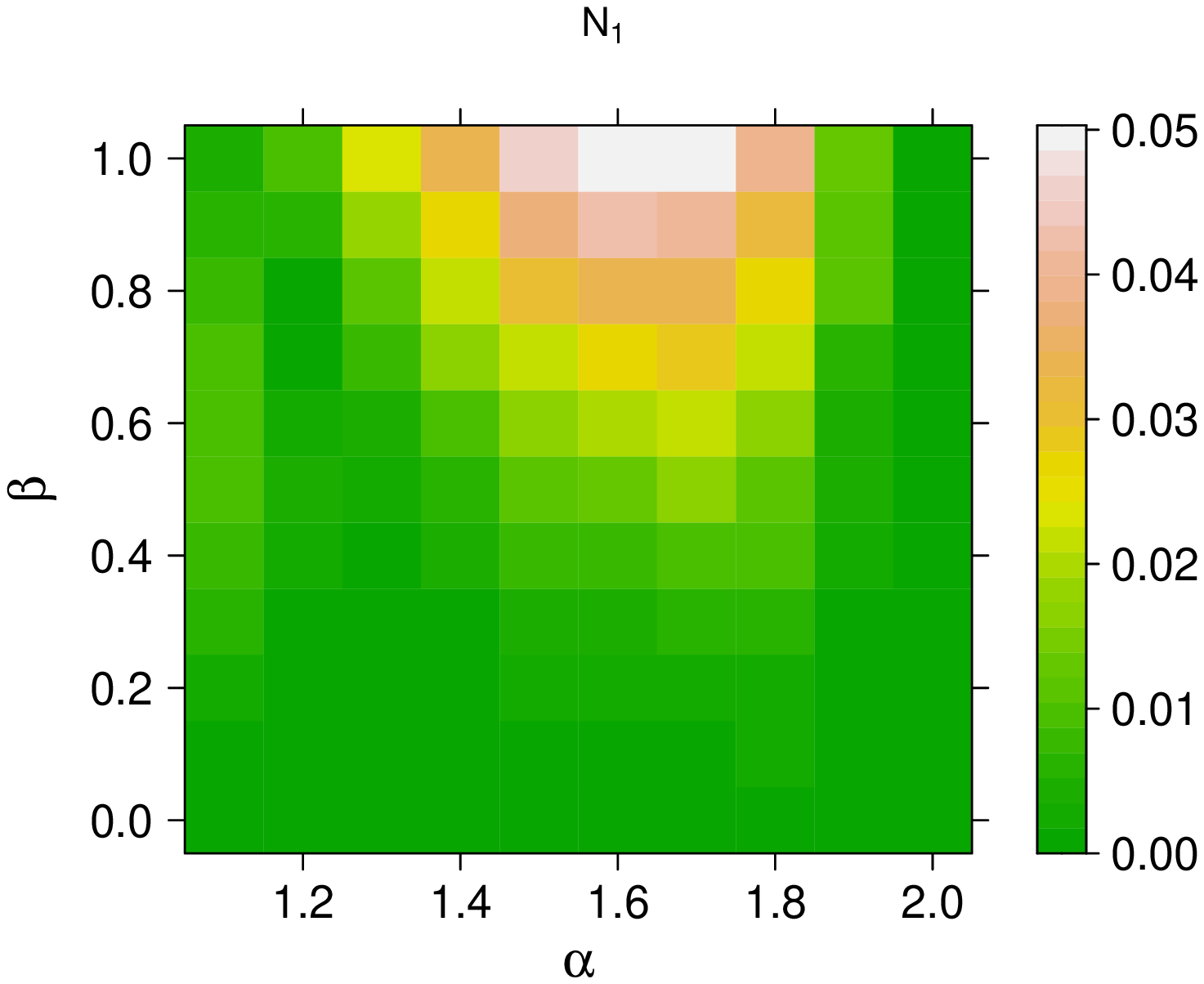}
\includegraphics[width=0.4\textwidth]{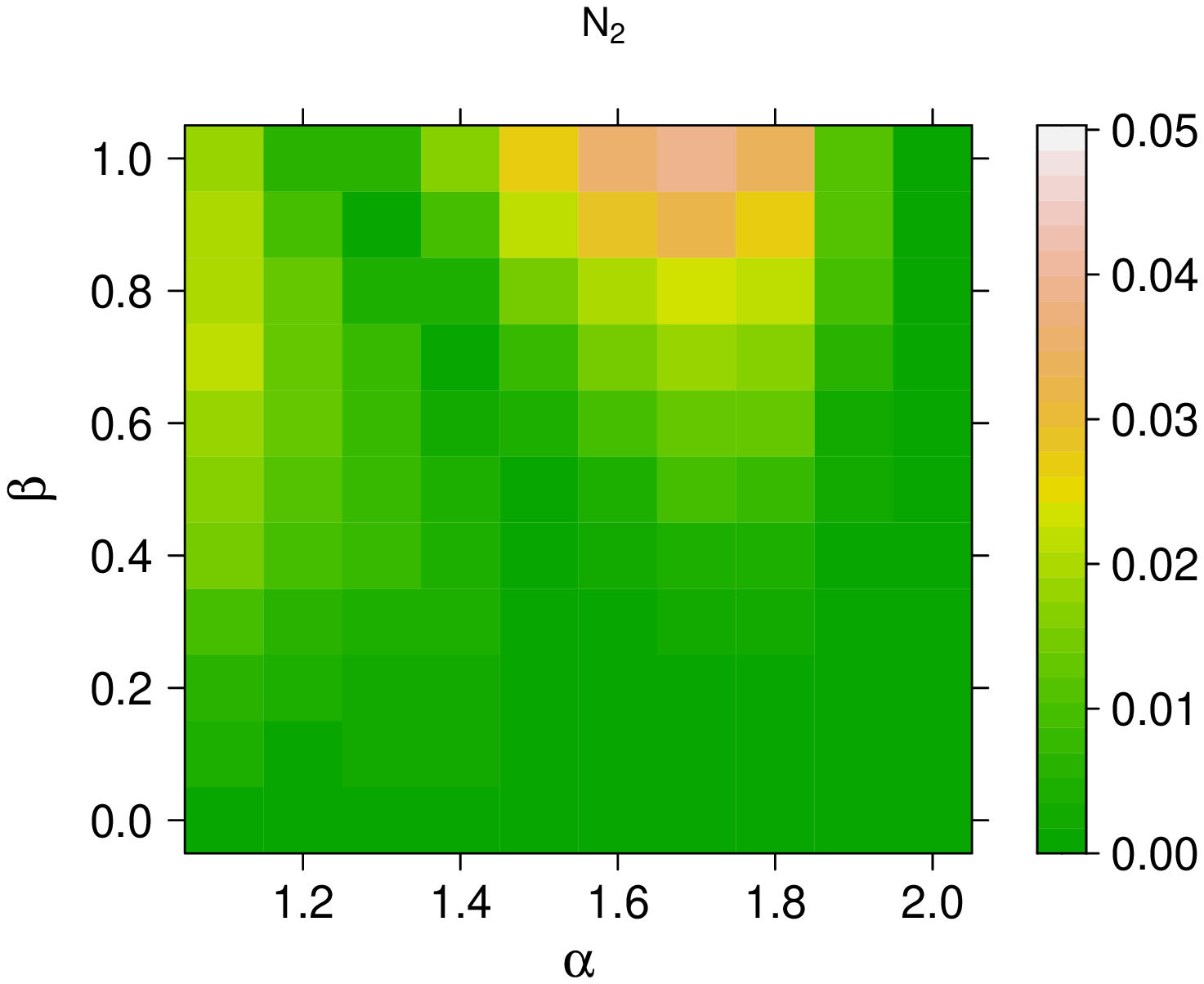}
\end{center}
\caption{Absolute differences between the averaged estimates of the stability parameter $\alpha$ in the symmetric and asymmetric case, see~\eqref{eq:diff.sym.nonsym}. The results for various combinations of  $\alpha \in \{1.1,1.2,\ldots,2\}$ and  $\beta \in \{0, 0.1,\ldots,1\}$ are presented. The number of strong Monte Carlo simulations is $10\, 000$ and the sample size is $n=1000$. The left panel shows the results for $N_1$, while the right panel shows the results for $N_2$.}\label{fig_res3}
\end{figure}

To further confirm the resistance of the proposed estimation methods to $\beta$ induced skewness, in Figure~\ref{fig_res3} we show the absolute differences between the averaged estimated values for fixed $\alpha$ and $\beta$, when confronted with symmetric alternative ($\beta=0$). More specifically, for each $\alpha \in \{1.1, 1.2, \ldots, 2\}$ and $\beta \in \{0, 0.1,\ldots,1\}$, we calculate
\begin{equation}\label{eq:diff.sym.nonsym}
\textstyle \left|\frac{1}{k}\sum_{i=1}^k \hat\alpha(\mathbf{X}_i^{(\alpha,\beta)})-\frac{1}{k}\sum_{i=1}^k \hat\alpha(\mathbf{X}_i^{(\alpha,0)})\right|,
\end{equation}
where $k=10\,000$ is the number of Monte Carlo samples, $\hat\alpha(\mathbf{X})$ is the estimate based on a sample $\mathbf{X}$, and $\mathbf{X}_i^{(\alpha,\beta)}$, $i=1, \ldots, M$, are i.i.d. $n=1000$-element samples from $S(\alpha, \beta, 1,0)$ distribution. The left panel shows the results for the estimator based on $N_1$, while the right panel shows the results for $N_2$. 
The results presented in Figure~\ref{fig_res3} demonstrate how $|\beta| \to 1$ affects the quality of the estimators. We observe that the estimated values of the stability index for the asymmetric data are relatively close to those obtained for the symmetric samples (i.e. when $\beta=0$), with maximal (average) difference equal to $0.05$ for sample size $n=1000$. This confirms that both $N_1$ and $N_2$ estimators are robust in reference to $\beta$ specification.  
\FloatBarrier

%%%%%%%%%%%%%%%%%%%%%%%%%%%%%%%%%%%%%%%%%%%%%%%%%%%%%%%%%

\section{Real data analysis}\label{real_data}

In this section we show how to apply our estimation methodology to real data. Namely, we investigate the data obtained in experiments on the controlled thermonuclear fusion from  the device ``Kharkiv Institute of Physics and Technology'', Kharkiv, Ukraine. The same data set was examined in  \cite{krzysiek0,krzysiek1,PitCheWyl2021}, where  detailed description of time series was provided; see \cite{beletskii} for the detailed description of the experimental set-up and the measurement procedure. Here, we only mention that the data describe the floating potential fluctuations (in volts) of plasma turbulence that is characterized by high levels of fluctuations of the electric field and particle density. In the fluctuations, the phenomenon called {\it L-H transition} has been observed. The {\it L-H transition} is a sudden transition from the low confinement mode (L mode) to a high confinement mode (H mode). During the transition, a regime shift could be observed, which might be associated with the change of the tail index.

\begin{figure}[htp!]
\begin{center}
\includegraphics[width=0.24\textwidth]{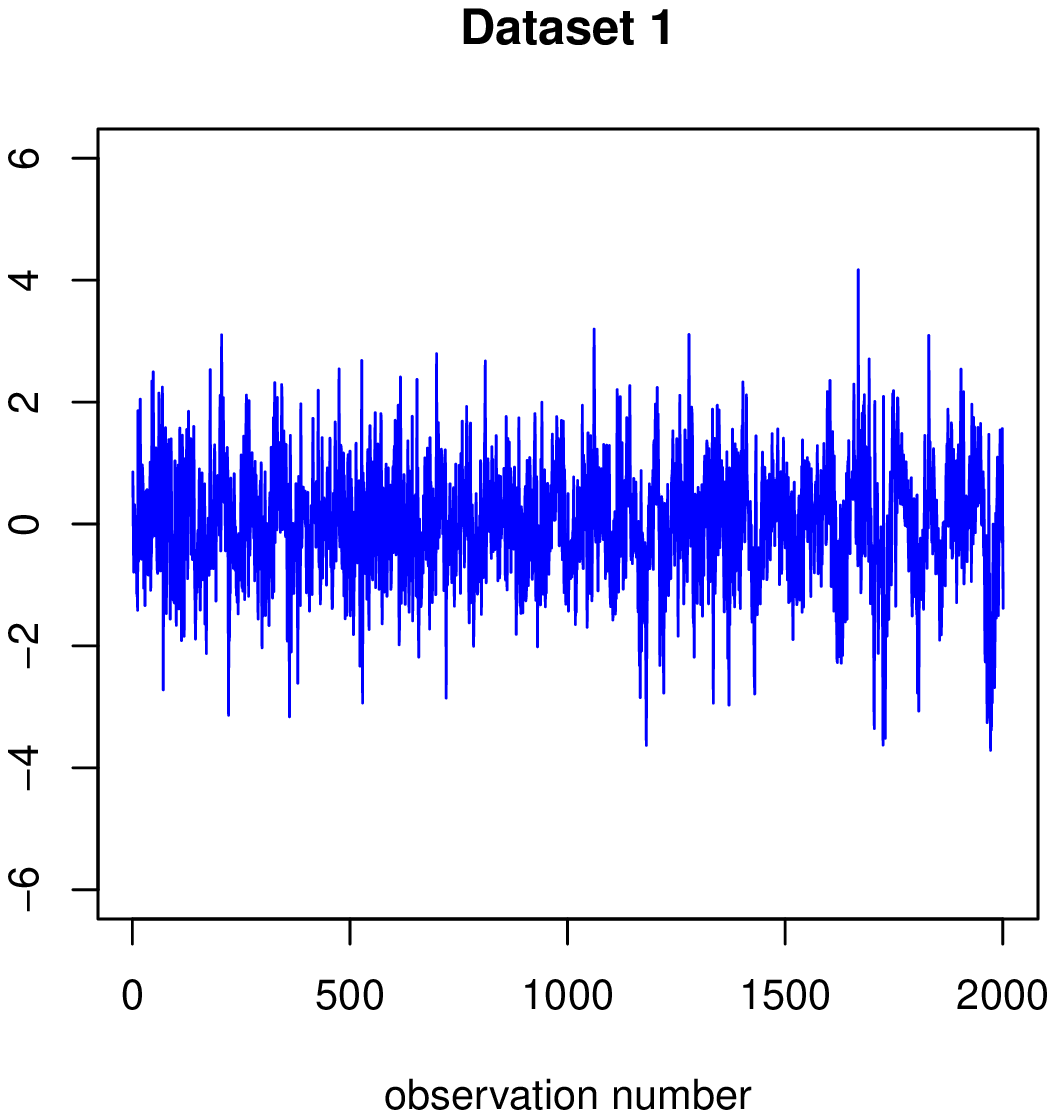}
\includegraphics[width=0.24\textwidth]{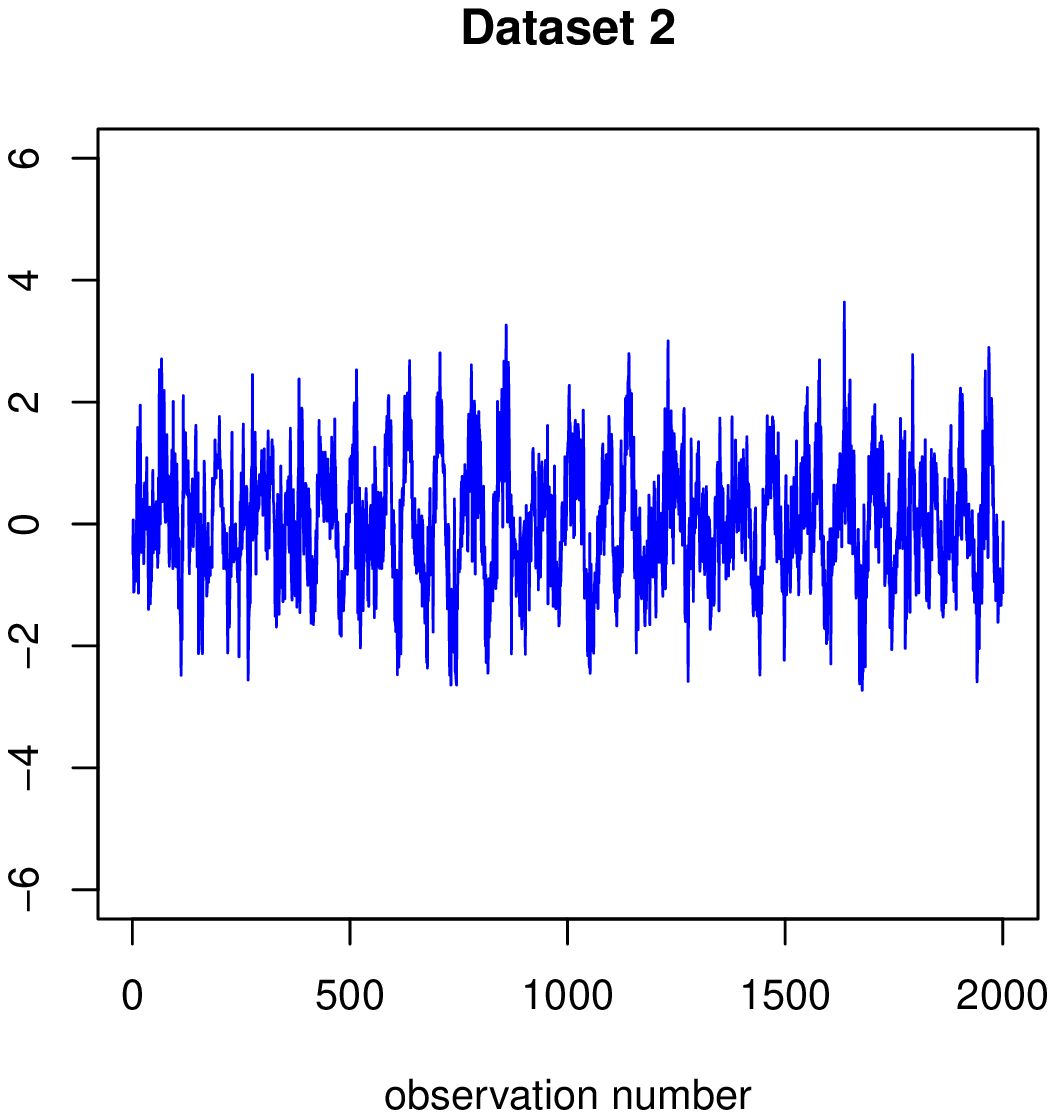}
\includegraphics[width=0.24\textwidth]{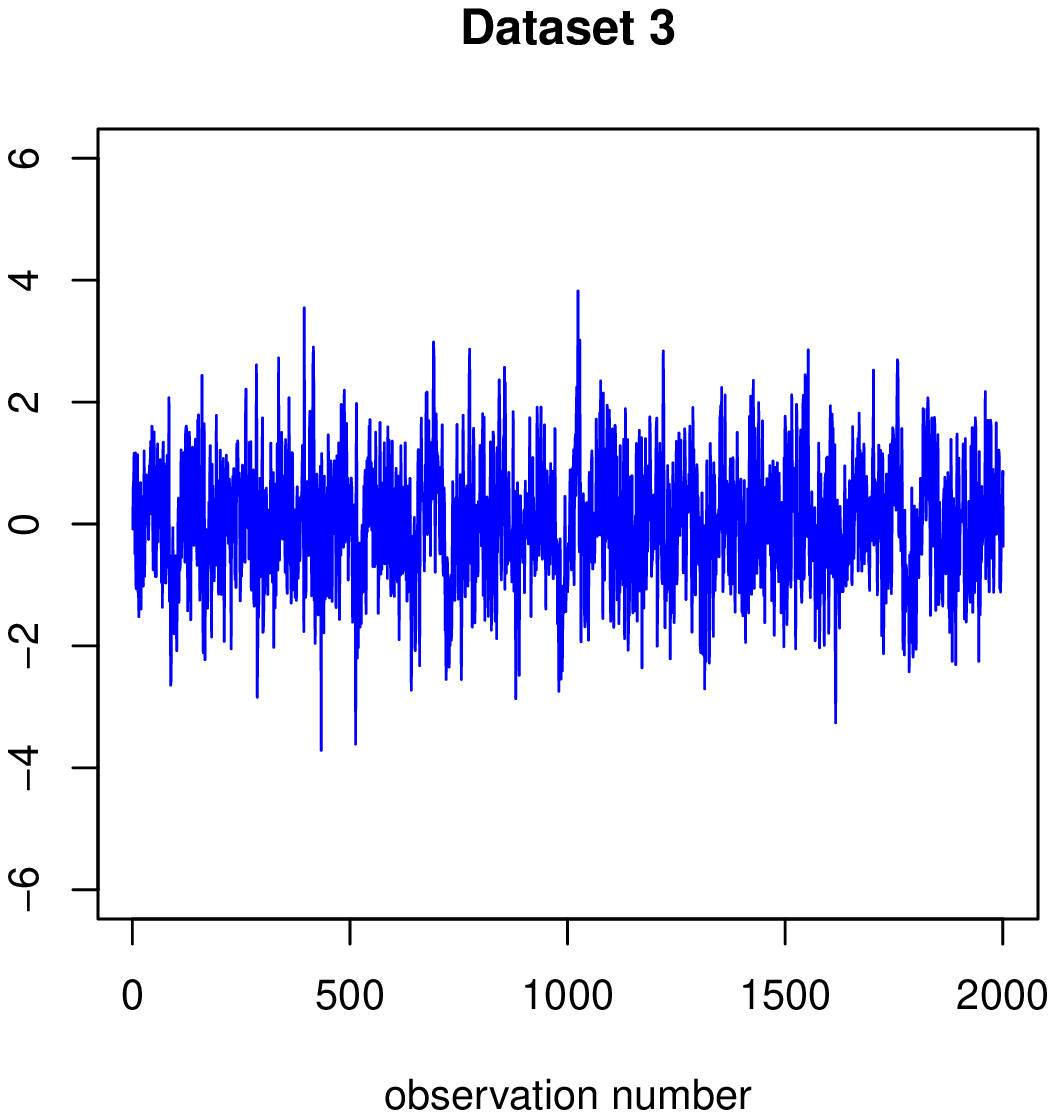}
\includegraphics[width=0.24\textwidth]{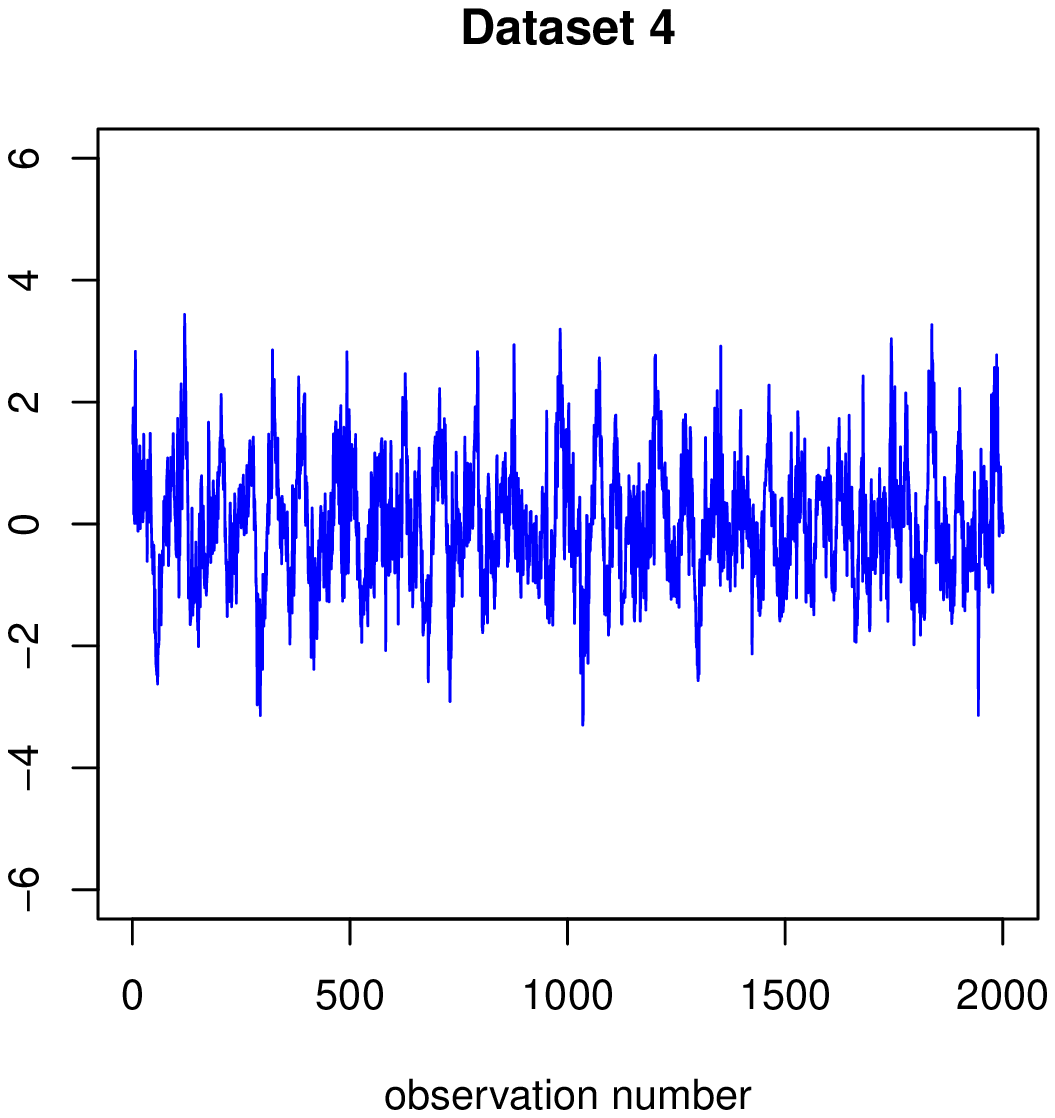}\\
\includegraphics[width=0.24\textwidth]{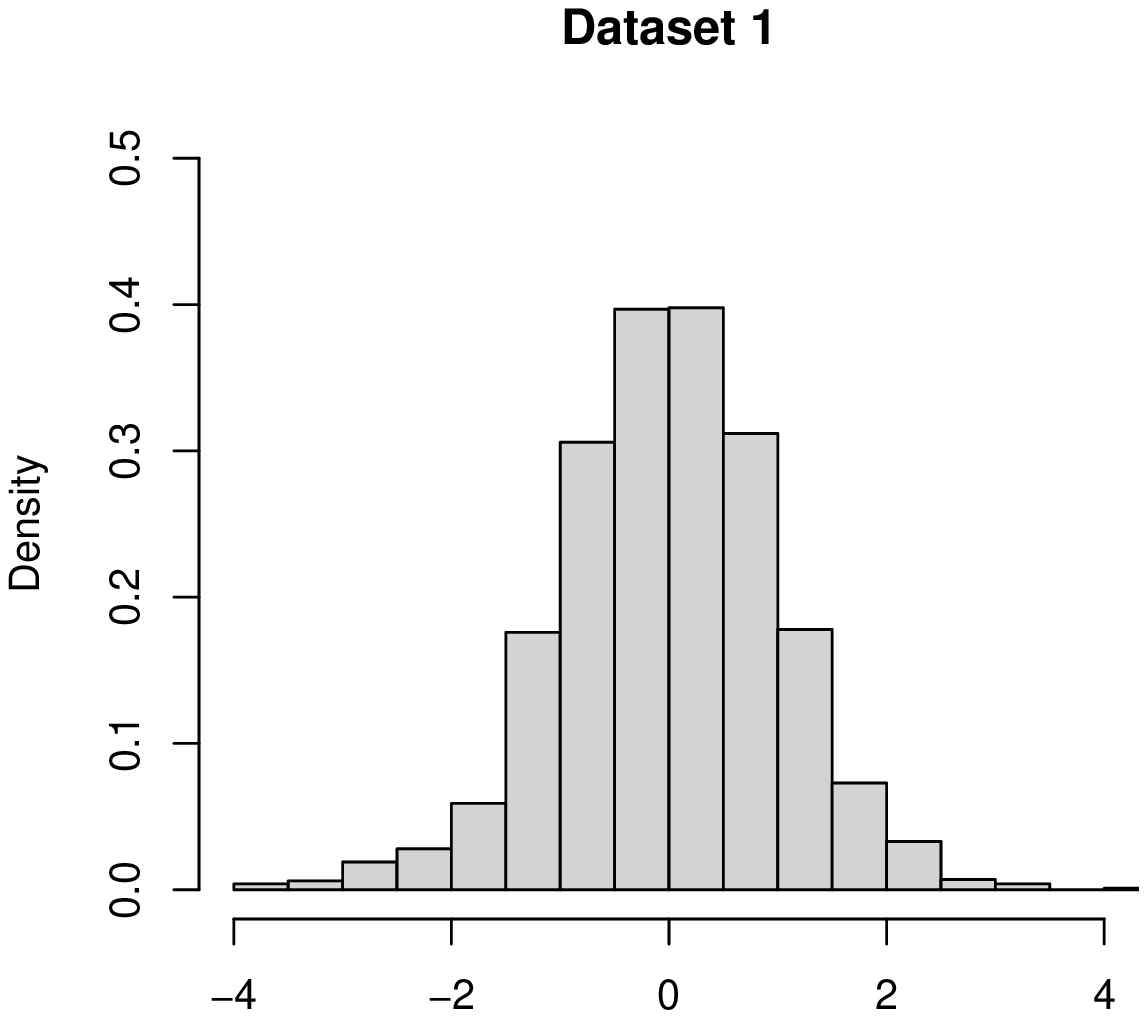}
\includegraphics[width=0.24\textwidth]{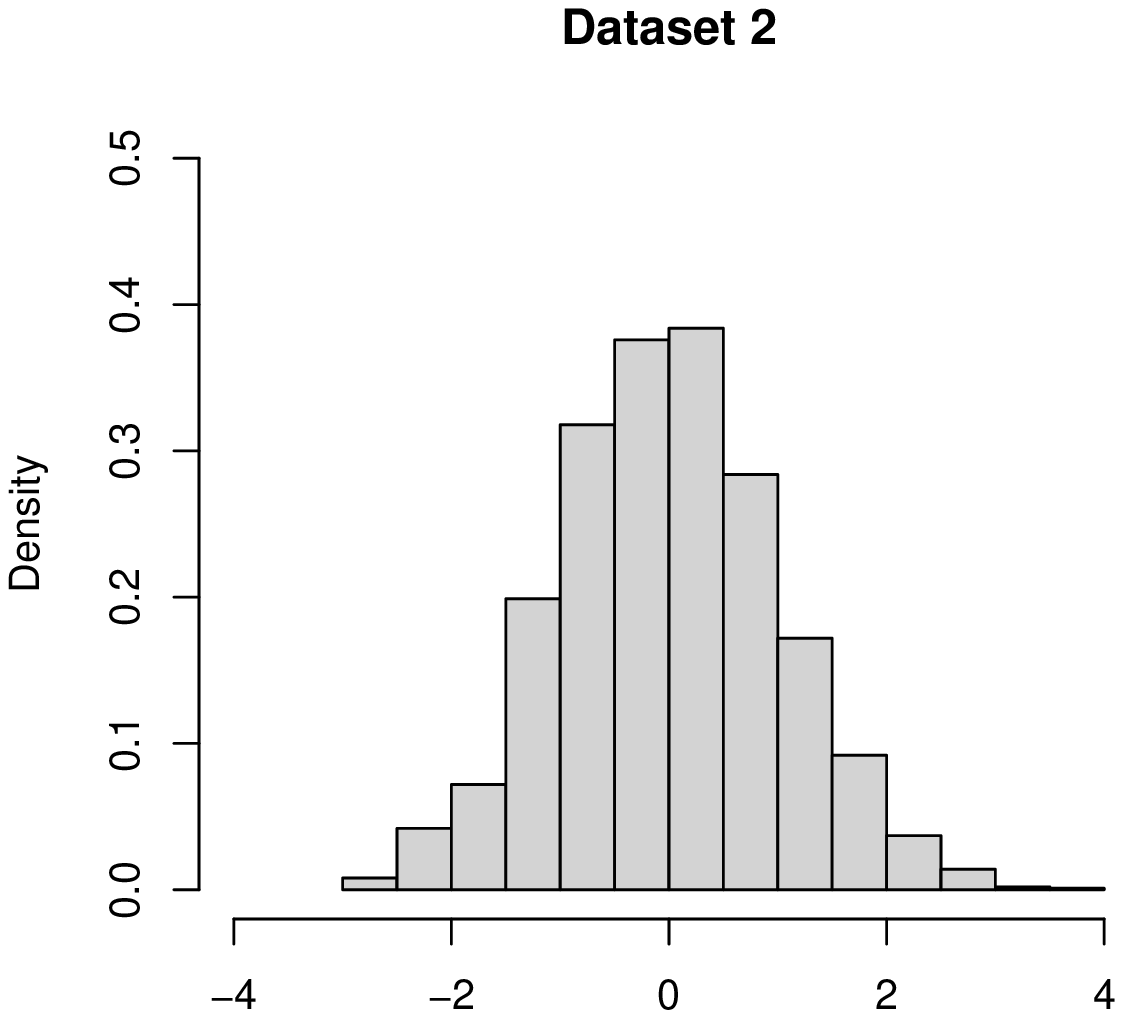}
\includegraphics[width=0.24\textwidth]{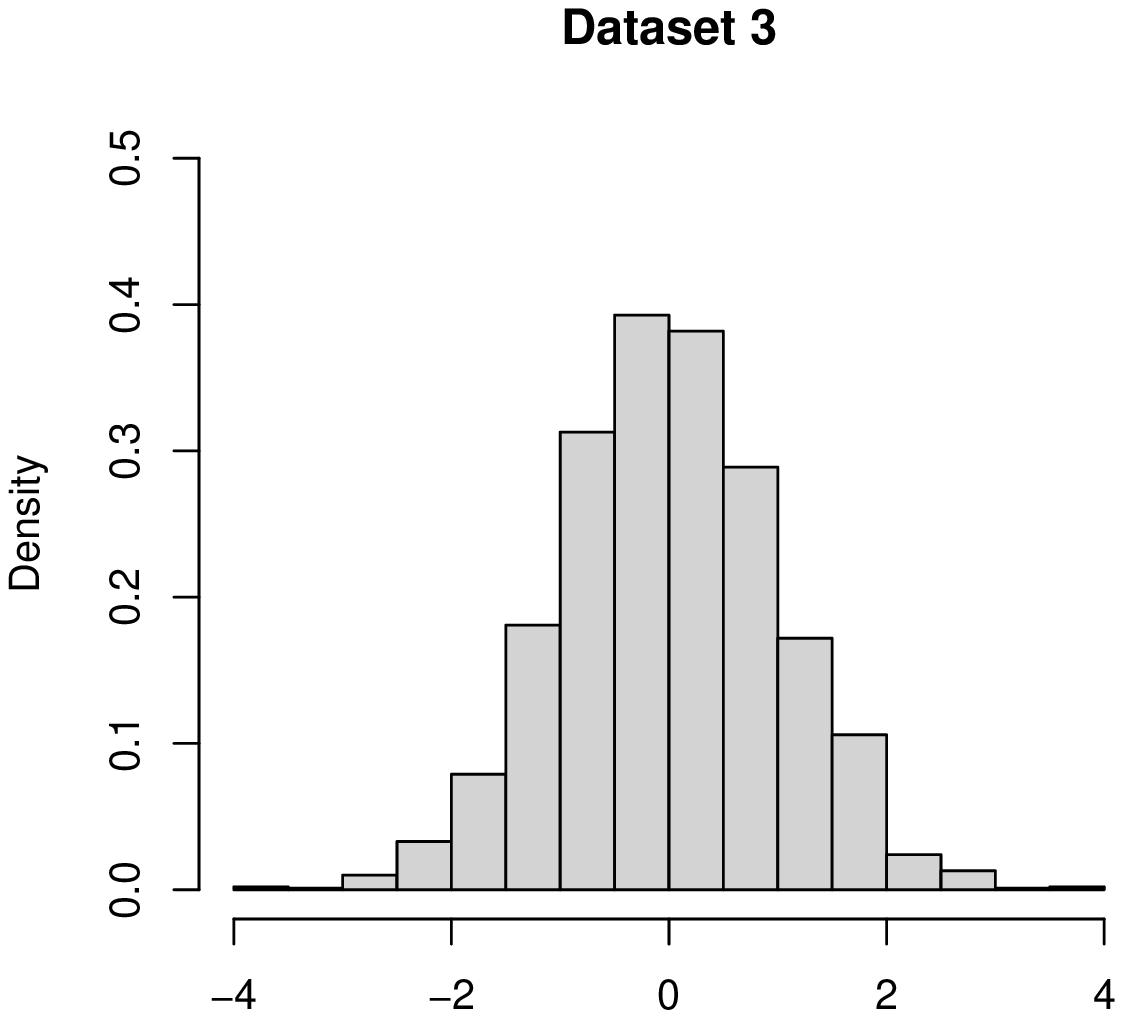}
\includegraphics[width=0.24\textwidth]{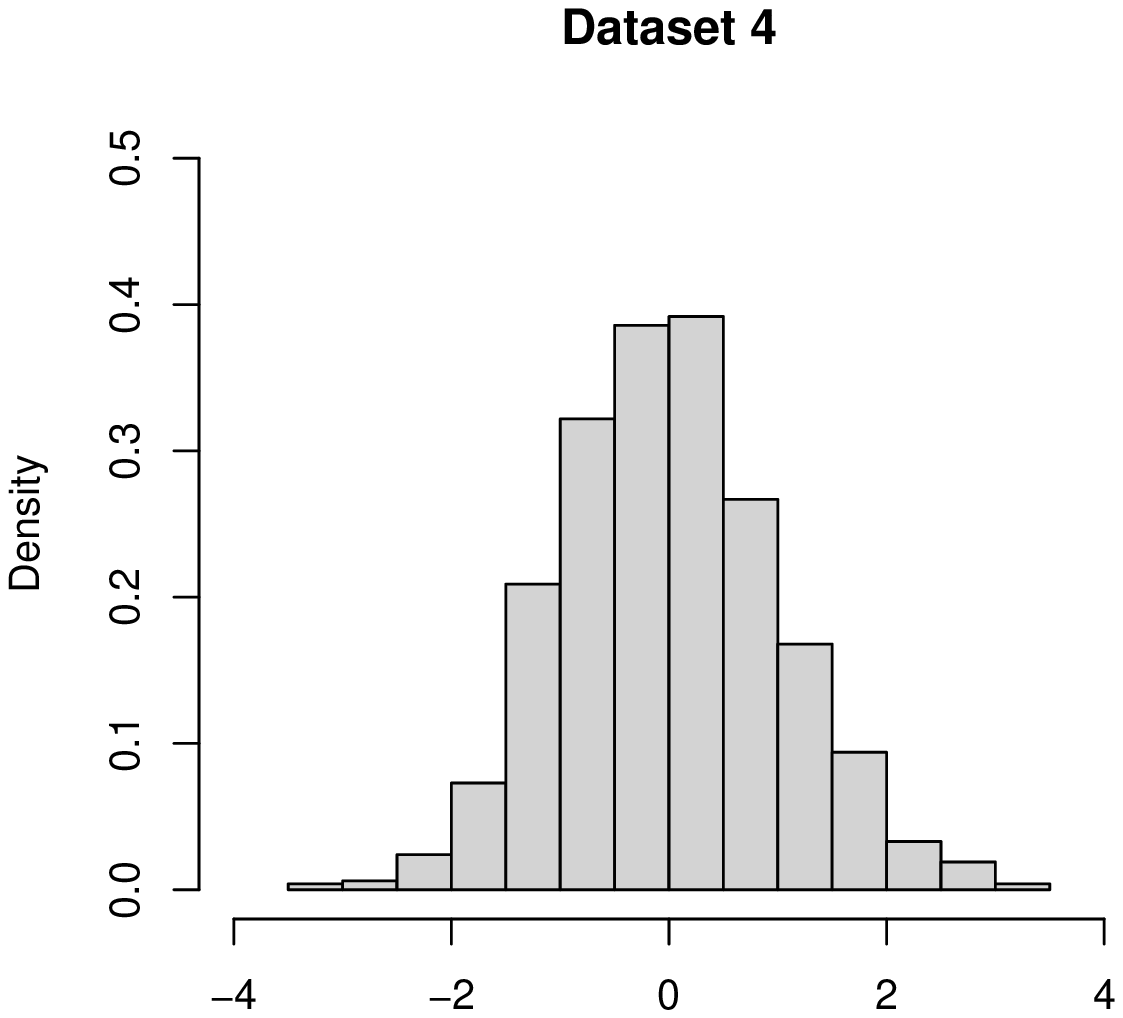}
\end{center}
\caption{Top panels: plasma data for torus radial position $r=9.5$ cm (Dataset 1 and Dataset 2) and $r=9.6$ cm (Dataset 3 and Dataset 4). The Dataset 1 and Dataset 3 describe the fluctuations before the L-H transition point.  Dataset 2 and Dataset 4 represent the fluctuation of the plasma after the L-H transition point. Bottom panels: histograms of analyzed datasets. }\label{Fig1_real}
\end{figure}

We examine four datasets denoted by Dataset 1, Dataset 2, Dataset 3, and Dataset 4. The  Dataset 1 and Dataset 2 describe the floating potential fluctuations for torus radial position $r = 9.5$ cm. Dataset 1 is related to the fluctuations before the transition point, Dataset 2 describes the fluctuation after the transition. Dataset 3 and Dataset 4 describe the potential fluctuations for torus radial position $r = 9.6$ cm. Dataset 3 is related to prior-transition fluctuations (L mode) while Dataset 4 is linked to posterior-transition fluctuations (H mode); see \cite{beletskii} for details. Each dataset contains $n=2000$ normalized observations, see Figure \ref{Fig1_real} for the data visualization (top panels) and histograms (bottom panel).

In \cite{krzysiek1}, a visual test for discriminating between light- and heavy-tailed distributed data was proposed, which pointed out differences between a prior transition point and a posterior transition point. Namely, it was shown that the distributions of Datasets 1 and Dataset 2, or Dataset 3 and Dataset 4, respectively, are slightly different. However, the differences were not visible on the level of the PDFs or CDFs (they were relatively close) and thus, more advanced techniques were applied to distinguish the corresponding distributions.  It was demonstrated that Dataset 1 can be described by the $\alpha$-stable distribution with $\alpha<2$ (yet close to 2), Dataset 2 and Dataset 3 can be modeled by the Gaussian distribution, while Dataset 4 belongs to the domain of attraction of the Gaussian law, however, it is non-Gaussian distributed.

The results presented in \cite{krzysiek1} were confirmed by \cite{PitCheWyl2021}, where a goodness-of-fit test for  the $\alpha$-stable distribution based on the conditional variance  methodology  was proposed. It was demonstrated that all considered datasets can be described by the $\alpha$-stable distribution. However, for  Dataset 1 and Dataset 4  the corresponding (acceptable)  stability indices $\alpha$ should lie in $ [1.84,1.93]$ and  $[1.88,1.96]$, respectively. For Datasets 2 the Gaussian (or very close to the Gaussian) distribution was confirmed. Finally, for Dataset 3, the $\alpha=1.98$ was accepted. This may also suggest the \stab distribution with the stability index very close to $2$. 

To continue this research, in this paper, we have applied the estimation methodology proposed to the datasets described above.  In Table \ref{Tab:Bootstrap}, we present the estimated values of $\alpha$ together with empirical confidence intervals (CI) obtained using the bootstrap method. Namely, for each dataset, we have created $10\,000$ re-samples and calculated empirical $95\%$ confidence intervals by estimating $\alpha$ for each of re-sampled data set.

In Table \ref{Tab:Bootstrap}, we also demonstrate the estimated values of $\alpha$ using the McCulloch, regression, and maximum likelihood methods (denotes as MCH, REG and MLE, respectively).  It is worth recalling that all estimation techniques discussed are dedicated for the symmetric $\alpha$-stable distributions, i.e. when the skewness parameter satisfies $\beta=0$. However, as it is demonstrated in Figure~\ref{Fig1_real} (bottom panel), the considered datasets exhibit non-symmetric behavior, especially see histograms of Dataset 2, Dataset 3, and Dataset 4. Thus, for comparison, in Table \ref{Tab:Bootstrap} we demonstrate also the results of modified MCH, REG, and MLE estimators,  that simultaneously estimate $\alpha$ and $\beta$ parameters; in this case, the methods are fitted for generic \stab distribution. For transparency, we denote them as MCH2, REG2, and MLE2, respectively. Similarly as for the symmetric case, we construct the $95\%$ bootstrap confidence intervals. The only exceptions are the MLE and MLE2 methods, as the construction of confidence intervals was too time-consuming.  As demonstrated in Section \ref{S:resistance}, even though the $N_1$ and $N_2$-based methods presented in this paper are dedicated for the symmetric $\alpha$-stable distributed samples, the techniques are robust to the skewness parameter $\beta$, especially for the stability index close to $2$, which is the case considered in this section.

\begin{table}[ht]
\begin{center}
\scalebox{0.75}{
\begin{tabular}{|c|cccccccc|}
\multicolumn{9}{c}{Estimated $\alpha$ with $95\%$ confidence intervals (CI) }\\
  \hline
 & $N_1$ & $N_2$ & MCH & MCH2 & REG & REG2 & MLE & MLE2 \\ 
  \hline
Dataset 1 & 1.876 & 1.842 & 2.000 & 1.998 & 2.000 & 1.955 & 1.998 &1.930 \\ 
Dataset 1 CI & (1.800, 2.000) & (1.798, 1.964) & (1.786, 1.997) & (1.811, 2.000) &  (2.000, 2.000) & (1.876, 1.976) & &\\ 
   \hline
Dataset 2 & 2.000 & 1.982 & 2.000 & 2.000 & 2.000 & 2.0000 & 2.000 & 1.996 \\ Dataset 2 CI & (1.917, 2.000) & (1.907, 2.000) & (1.901, 2.000) & (1.905, 2.000) & (2.000, 2.000) & (1.966, 2.000) & &\\ 
    \hline
Dataset 3 & 1.973 & 1.997 & 1.924 &  1.922 & 2.000 & 1.985 & 1.922 & 1.990\\ 
Dataset 3 CI & (1.873, 2.000) & (1.917, 2.000) & (1.781, 2.000) & (1.783, 2.000)& (2.000, 2.000) & (1.941, 2.000) & &\\ 
       \hline
    Dataset 4 & 1.952 & 1.983 & 1.970 & 1.971 & 2.000 & 1.955 & 1.971 & 1.968 \\ 
    Dataset 4 CI & (1.848, 2.000) & (1.893, 2.000) & (1.795, 2.000) & (1.798, 2.000)& (2.000, 2.000) & (1.907, 2.000) & &\\ 
   \hline
\end{tabular} }
\end{center}
\caption{Estimation of $\alpha$ parameter with $95\%$ confidence intervals (CI) based on $10\,000$ bootstrap samples. For comparison we demonstrate the results for benchmark methods dedicated for symmetric (MCH, REG, MLE) and general (MCH2, REG2, MLE2) $\alpha$-stable distribution. The CI for MLE and MLE2 methods are not presented (the techniques are time consuming).}\label{Tab:Bootstrap}
\end{table}

The results obtained based on QCV approach confirm the previous conclusions. The results based on $N_1$ and $N_2$ statistics for Dataset 1 clearly indicate the stability index between $1.8$ and $1.96$. Let us recall that if the parameter $\alpha$ is close to $2$, then $N_2-$based should be more adequate; thus, the upper limit of the CI for the $N_1$-based method is not reliable here. The benchmark techniques dedicated for the symmetric \stab distribution (i.e. MCH, REG and MLE) for Dataset 1 fail, as they indicate the Gaussian distribution.  The methods that simultaneously estimate the $\alpha$ and $\beta$ parameters (i.e. MCH2, REG2 and MLE2) seem to overestimate the stability index. The maximum likelihood-based technique can be seen as the only exception. 

For Dataset 2, identified in the previous research as the Gaussian distributed sample, all analyzed methods confirm this hypothesis and give comparable results. However, the regression-based methods brings the most precise results. 

The estimated values of $\alpha$ for Dataset 3 coincide with the previous results. The $N_1$ and $N_2$-based techniques as well as regression-based methods confirm the Gaussianity of the data or the \stab distribution with $\alpha$ very close to $2$. The McCulloch-based methods (i.e. MCH and MCH2) seems to underestimate the results. The situation seems to be the same for the maximum likelihood-based approach (i.e. REG and REG2 methods). 

For Dataset 4, the $N_1$-based methodology as well as the REG2 algorithm confirm the previous hypothesis of \stab distribution with $\alpha\in [1.88, 1.96]$.  The other considered techniques seem to be less efficient as they overestimate the results.

%%%%%%%%%%%%%%%%%%%%%%%%%%%%%%%%%%%%%%%%%%%%%%%%%%%%%%%%%
\section{Summary and conclusions}
In this paper we show that the quantile conditional variances can be used to construct efficient fitting statistics for the symmetric $\alpha$-stable distributions. The proposed estimators show good performance both on the simulated and real data, and often outperform other benchmark frameworks. The methods proposed in this paper build on the results obtained recently in~\cite{PitCheWyl2021} in the context of goodness-of-fit testing. In fact, in this paper we show that the statistical QCV based procedures could be efficiently used also for a parameter estimation.  It is worth noting that while the standard moment-fit based methods cannot be directly applied in the \stab case due to the infinite variance problem, our method is based on quantile conditional variances which always exists and consequently can be directly used for fitting. In fact, the approach proposed can be seen as an extension of the method of moments to the infinite-moment framework via second moment conditioning scheme. Also, we want to highlight some form of similarity between our method and the classical McCulloch method, see \eqref{eq:McCulloch_nu}. In a nutshell, in the McCulloch framework the fitting statistic is based on the appropriate function of tail quantiles while the procedures proposed in this paper use the quantile tail variances, see~\eqref{eq:S1S2}. As expected, the quantile conditional variances are more informative than quantiles which results in better performance of the proposed method, see Section~\ref{S:comparison} for details. Next, it is worth mentioning that while we focus on the symmetric case, the QCV ratio based statistic is in fact insensitive to changes in the skewness parameter and our method could be applied to a more generic framework. Finally, we have shown that the approach proposed extract sample characteristics different to the ones extracted by the other benchmark frameworks, so that our approach could be used to refine existing methodologies e.g. via ensembling.

%%%%%%%%%%%%%%%%%%%%%%%%%%%%%%%%%%%%%%%%%%%%%%%%%%%%%%%%%%%%%%%%%%%%%%%%%

\section*{Acknowledgements}
Marcin Pitera and Agnieszka Wy\l{}oma\'{n}ska acknowledge support from the National Science Centre, Poland, via project 2020/37/B/HS4/00120. Part of the work of Damian Jelito was funded by the Priority Research Area Digiworld under the program Excellence Initiative – Research University at the Jagiellonian University in Krak\'{o}w.

\begin{footnotesize}
\bibliographystyle{agsm}
\bibliography{mybibliography}

@article{DEHAAN199939,
title = {Estimating the index of a stable distribution},
journal = {Statistics \& Probability Letters},
volume = {41},
number = {1},
pages = {39-55},
year = {1999},
author = {L. {de Haan} and T. {Themido Pereira}},
}

@article{iskander2020,
	year = {2020},
	author = {Wy{\l}oma{\'{n}}ska, Agnieszka and Burnecki, Krzysztof and Iskander, D. Robert},
	title = {Omnibus test for normality based on the Edgeworth expansion },
	journal = {PLoS ONE},
	number={15(6): e0233901} 
}

@article{Paulson,
    author = {Paulson, A. S. and Holcomb, E. W. and Leitch, R. A.},
    title = "{The estimation of the parameters of the stable laws}",
    journal = {Biometrika},
    volume = {62},
    number = {1},
    pages = {163-170},
    year = {1975},
    month = {04},
   }

@article{Brorsen,
author = {   B.   Wade Brorsen  and  Seung Ryong   Yang },
title = {Maximum Likelihood Estimates of Symmetric Stable Distribution Parameters},
journal = {Communications in Statistics - Simulation and Computation},
volume = {19},
number = {4},
pages = {1459-1464},
year  = {1990},

}

@article{leitch,
author = { R. A.   Leitch  and  A. S.   Paulson },
title = {Estimation of Stable Law Parameters: Stock Price Behavior Application},
journal = {Journal of the American Statistical Association},
volume = {70},
number = {351a},
pages = {690-697},
year  = {1975},
}

@article{Akgiray,
author = { Vedat   Akgiray  and  Christopher G.   Lamoureux },
title = {Estimation of Stable-Law Parameters: A Comparative Study},
journal = {Journal of Business \& Economic Statistics},
volume = {7},
number = {1},
pages = {85-93},
year  = {1989},
}

@ARTICLE{nikias1,
  author={Tsihrintzis, G.A. and Nikias, C.L.},
  journal={IEEE Transactions on Signal Processing}, 
  title={Fast estimation of the parameters of alpha-stable impulsive interference}, 
  year={1996},
  volume={44},
  number={6},
  pages={1492-1503},
}

@article{Sathe,
author = {Aastha M. Sathe and N. S. Upadhye},
title = {Estimation of the parameters of multivariate stable distributions},
journal = {Communications in Statistics - Simulation and Computation},
volume = {0},
number = {0},
pages = {1-18},
year  = {2020},
}

@article{teimouri,
author = {Teimouri, Mahdi and Rezakhah, Saeid and Mohammadpour, Adel},
year = {2018},
pages = {439},
title = {Parameter Estimation Using the EM Algorithm for Symmetric Stable Random Variables and Sub-{G}aussian Random Vectors},
volume = {17},
journal = {Journal of Statistical Theory and Applications},
}

@article{garcia2011estimation,
  title={Estimation of stable distributions by indirect inference},
  author={Garcia, Ren{\'e} and Renault, Eric and Veredas, David},
  journal={Journal of Econometrics},
  volume={161},
  number={2},
  pages={325--337},
  year={2011},
  }

@article{arad,
  author = {Ruth W. Arad},
 journal = {International Economic Review},
 number = {1},
 pages = {209--220},
  title = {Parameter Estimation for Symmetric Stable Distribution},
 volume = {21},
 year = {1980}
}

@ARTICLE{lombardi,
  author={Lombardi, M.J. and Godsill, S.J.},
  journal={IEEE Transactions on Signal Processing}, 
  title={On-line Bayesian estimation of signals in symmetric $\alpha$-stable noise}, 
  year={2006},
  volume={54},
  number={2},
  pages={775-779},
}

@Inbook{Nolan2001,
author="Nolan, John P.",
editor="Barndorff-Nielsen, Ole E.
and Resnick, Sidney I.
and Mikosch, Thomas",
title="Maximum Likelihood Estimation and Diagnostics for Stable Distributions",
bookTitle="L{\'e}vy Processes: Theory and Applications",
year="2001",
publisher="Birkh{\"a}user Boston",
address="Boston, MA",
pages="379--400",
}

@article{MITTNIK,
title = {Computing the probability density function of the stable Paretian distribution},
journal = {Mathematical and Computer Modelling},
volume = {29},
number = {10},
pages = {235-240},
year = {1999},
author = {Mittnik, S. and Doganoglu, T. and Chenyao, D.},
}

@article{Matsui,
author = { Muneya, M.  and  Akimichi, T. },
title = {Some Improvements in Numerical Evaluation of Symmetric Stable Density and Its Derivatives},
journal = {Communications in Statistics - Theory and Methods},
volume = {35},
number = {1},
pages = {149-172},
year  = {2006},
}

@article{dumouchel,
author = {DuMouchel, W.},
year = {1973},
month = {09},
pages = {},
title = {On the Asymptotic Normality of the Maximum-Likelihood Estimate When Sampling From a Stable Distribution},
volume = {1},
journal = {The Annals of Statistics},
}

@article{kogon,
  title={Characteristic function based estimation of stable distribution parameters},
  author={Kogon, S. M. and Williams, D. B.},
  journal={A practical guide to heavy tails: statistical techniques and applications},
  pages={311--338},
  year={1998},
  publisher={Birkh{\"a}user Boston, MA}
}

@article{HASSANNEJAD,
title = {Characteristic function based parameter estimation of skewed alpha-stable distribution: An analytical approach},
journal = {Signal Processing},
volume = {130},
pages = {323-336},
year = {2017},
author = {Mohammadreza, H. B. and Amindavar, H. and Amirmazlaghani, M.},
}

@article{Huixia,
author = { Huixia, J. W. and Deyuan, L. and Xuming, H.},
title = {Estimation of High Conditional Quantiles for Heavy-Tailed Distributions},
journal = {Journal of the American Statistical Association},
volume = {107},
number = {500},
pages = {1453-1464},
year  = {2012},
}

@INPROCEEDINGS{Maynon,

  author={Maymon, S. and Friedmann, J. and Messer, H.},
  booktitle={2000 IEEE International Conference on Acoustics, Speech, and Signal Processing. Proceedings (Cat. No.00CH37100)}, 
  title={A new method for estimating parameters of a skewed alpha-stable distribution}, 
  year={2000},
  volume={6},
  number={},
  pages={3822-3825 vol.6},

}

@article{press,
 author = {S. James Press},
 journal = {Journal of the American Statistical Association},
 number = {340},
 pages = {842--846},
 publisher = {[American Statistical Association, Taylor & Francis, Ltd.]},
 title = {Estimation in Univariate and Multivariate Stable Distributions},
 volume = {67},
 year = {1972}
}

@article{dominicy,
title = {The method of simulated quantiles},
journal = {Journal of Econometrics},
volume = {172},
number = {2},
pages = {235-247},
year = {2013},
author = {Yves Dominicy and David Veredas},
}

@article{fama,
 author = {Eugene F. Fama and Richard Roll},
 journal = {Journal of the American Statistical Association},
 number = {334},
 pages = {331--338},
 publisher = {[American Statistical Association, Taylor & Francis, Ltd.]},
 title = {Parameter Estimates for Symmetric Stable Distributions},
 volume = {66},
 year = {1971}
}

@article{phys33,
	Author = {P. D. Ditlevsen},
	Journal = {Geophysical Research Letters},
	Pages = {1441--1444},
	Title = {Observation of alpha-stable noise induced millennial climate changes from an ice-core record},
	Volume = {26},
	Year = {1999}}

@article{phys35,
	Author = {C. K. Peng and J. Mietus and J. M. Hausdorff and S. Havlin and H. E. Stanley and A. L. Goldberger},
	Journal = {Physical Review Letters},
	Pages = {1343},
	Title = {Long-range anticorrelations and non-{G}aussian behavior of the heartbeat},
	Volume = {70},
	Year = {1993}}

@article{phys21,
	Author = {P. Barthelemy and J. Bertolotti and D. S. Wiersma},
	Journal = {Nature},
	Pages = {495--498},
	Title = {A {L}\'evy flight for light},
	Volume = {453},
	Year = {2008}}

@article{phys22,
	Author = {I. M. Sokolov and J. Mai and A. Blumen},
	Journal = {Physical Review Letters},
	Pages = {857},
	Title = {Paradoxal Diffusion in Chemical Space for Nearest-Neighbor Walks over Polymer Chains},
	Volume = {79},
	Year = {1997}}

@article{phys23,
	Author = {M. A. Lomholt and T. Ambjornsson and R. Metzler},
	Journal = {Physical Review Letter},
	Pages = {260603},
	Title = {Optimal Target Search on a Fast-Folding Polymer Chain with Volume Exchange},
	Volume = {95},
	Year = {2005}}

@article{limit1,
	Author = {Jakubowski, A. and Kobus, M.},
	Journal = {Journal of Multivariate Analysis},
	Pages = {219-251},
	Title = {Alpha-Stable limit theorems for sums of dependent random vectors},
	Volume = {29(2)},
	Year = 1989}

@article{est3_new,
	Author = {McCulloch, J. H.},
	Journal = {Communications in Statistics-- Simulation and Computation},
	Pages = {1109--1136},
	Title = {Simple consistent estimators of stable distribution parameters},
	Volume = {15},
	Year = {1986}}

@article{est4_new,
	Author = {Koutrouvelis, I. A.},
	Journal = {Journal of the Americal Statistical Association},
	Pages = {918--928},
	Title = {Regression type estimation of the parameters ofstable laws},
	Volume = {75},
	Year = {1980}}

@article{mon_new1,
	Author = {Yu, G. and Li, C. and Zhang, J.},
	Journal = {Mechanical Systems and Signal Processing},
	Pages = {155--175},
	Title = {A new statistical modeling and detection method for rolling element bearing faults based on alpha-stable distribution},
	Volume = {41},
	Year = {2013}}

@article{mon_new2,
	Author = {G. {\.Z}ak and A. Wy{\l}oma{\'n}ska and R. Zimroz},
	Journal = {Shock and Vibration},
	Pages = {Article ID 3698370},
	Title = {Data driven iterative vibration signal enhancement strategy using alpha stable distribution},
	Volume = {2017},
	Year = {2017}}

@article{mon_new3,
	Author = {G. {\.Z}ak and A. Wy{\l}oma{\'n}ska and R. Zimroz},
	Journal = {Journal of Vibroengineering},
	Pages = {826--837},
	Title = {Data-driven vibration signal filtering procedure based on the alpha-stable distribution},
	Volume = {18(2)},
	Year = {2016}}

@book{fin_new1,
	Address = {New York},
	Author = {Rachev, S. T. and Mittnik, S.},
	Title = {Stable Paretian Models in Finance},
	Year = {2000}}

@article{fin_new2,
	Author = {Bidarkota, P. V. and Dupoyet, B. V. and McCulloch, J. H.},
	Journal = {Journal of Economic Dynamics and Control},
	Pages = {1314--1331},
	Title = {Asset pricing with incomplete information and fat tails},
	Volume = {33(6)},
	Year = {2009}}

@article{biol_new1,
	Author = {Durrett, R. and Foo, J. and Leder, K. and Mayberry, J. and Michor, F.},
	Journal = {Genetics},
	Pages = {1--17},
	Title = {Intratumor heterogeneity in evolutionary models of tumor progression},
	Volume = {188},
	Year = {2011}}

@article{biol_new2,
	Author = {Lan, B. L. and M. Toda},
	Journal = {Europhysics Letters},
	Pages = {18002--p1--p6},
	Title = {Fluctuations of healthy and unhealthy heartbeat intervals},
	Volume = {102(1)},
	Year = {2013}}

@article{phys_new1,
	Author = {M. Majka and P.F. G{\'o}ra},
	Journal = {Physical Review E},
	Pages = {052602},
	Title = {Non-{G}aussian polymers described by alpha-stable chain statistics: Model, effective interactions in binary mixtures, and application to on-surface separation},
	Volume = {91},
	Year = {2015}}

@book{alek_book,
	Address = {New York},
	Author = {A. Janicki and A. Weron},
	Publisher = {Marcel Dekker, Inc.},
	Title = {Simulation and Chaotic Behavior of Alpha-stable Stochastic Processes},
	Year = {1994}}

@article{phys_new2,
	Author = {Bart Kosko and Sanya Mitaim},
	Journal = {Physical Review E},
	Pages = {031911},
	Title = {Robust stochastic resonance for simple threshold neurons},
	Volume = {70},
	Year = {2004}}

@article{levy1924theorie,
	Author = {L{\'e}vy, Paul},
	Journal = {Bulletin de la Soci{\'e}t{\'e} math{\'e}matique de France},
	Pages = {49--85},
	Title = {Th{\'e}orie des erreurs. La loi de {G}auss et les lois exceptionnelles},
	Volume = {52},
	Year = {1924}}

@article{khinchine1936lois,
	Author = {Khinchine, Alexander Ya and L{\'e}vy, Paul},
	Journal = {CR Acad. Sci. Paris},
	Pages = {374--376},
	Title = {Sur les lois stables},
	Volume = {202},
	Year = {1936}}

@book{nolan_book,
	Author = {Nolan, J. P.},
	Publisher = {Springer},
	Title = {Univariate Stable Distributions. Models for Heavy Tailed Data},
	Year = {2020}}

@book{weronr,
	Address = {Berlin},
	Author = {Cizek, P. and Haerdle, W. and Weron, R.},
	Publisher = {Springer},
	Title = {Statistical Tools for Finance and Insurance},
	Year = {2005}}

@book{feller1966,
	Author = {William Feller},
	Publisher = {New York:John Wiley and Sons},
	Title = {An Introduction to Probability Theory and Applications},
	Year = {1966}}

@book{non_gauss,
	Editor = {Edward J. Wegman and Stuart C. Schwartz and John B. Thomas},
	Publisher = {New York: Springer},
	Title = {Topics in Non-{G}aussian Signal Processing},
	Year = {1989}}

@article{beletskii,
	Author = {A. Beletskii and L.I. Grigor'eva and E. L. Sorokovoy and V. S. Romanov},
	Journal = {Plasma Physics Reports},
	Pages = {818-823},
	Title = {Spectral and statistical analysis of fluctuations in the SOL and diverted plasmas of the Uragan-3M torsatron},
	Volume = {35},
	Year = {2009}}

@book{shao22,
	Address = {New York},
	Author = {Nikias, C.L. and Shao, M.},
	Publisher = {John Wiley and Sons},
	Title = {Signal Processing with Alpha-Stable Distributions and Applications},
	Year = {1995}}

@article{krzysiek1,
	Author = {Burnecki, K. and Wy{\l}oma{\'{n}}ska, A. and Chechkin, Aleksei},
	Issue = {12},
	Journal = {PLOS ONE},
	Pages = {e0145604},
	Title = {Discriminating between light- and heavy-tailed distributions with limit theorem},
	Volume = {10},
	Year = {2015}}

@article{krzysiek0,
	Author = {Burnecki, K. and Wy{\l}oma{\'{n}}ska, A. and Beletskii, A. and Gonchar, V. and Chechkin, A.},
	Journal = {Phys. Rev. E},
	Pages = {056711},
	Title = {Recognition of stable distribution with {L}{\'e}vy index $\alpha$ close to 2},
	Volume = {85},
	Year = {2012}}

@article{stuck,
	Author = {J. M. Chambers and C. L. Mallows and B. W. Stuck},
	Journal = {Journal of the American Statistical Association},
	Number = {354},
	Pages = {340--344},
	Title = {A Method for Simulating Stable Random Variables},
	Volume = {71},
	Year = {1976}}

@article{shao,
	Author = {Shao, M. and Nikias, C.L.},
	Journal = {Proceedings of the IEEE},
	Pages = {986--1010},
	Title = {Signal processing with fractional lower order moments: Stable processes and their application},
	Volume = {81},
	Year = {1993}}

@article{mand,
	Author = {Benoit Mandelbrot},
	Journal = {International Economic Review},
	Number = {2},
	Pages = {79--106},
	Title = {The {P}areto-{L}{\'e}vy {L}aw and the Distribution of Income},
	Volume = {1},
	Year = {1960}}

@book{stable,
	Address = {New York},
	Author = {Samorodnitsky, G and Taqqu, MS},
	Publisher = {Chapman and Hall},
	Title = {Stable Non-{G}aussian Random Processes: Stochastic Models with Infinite Variance},
	Year = {1994}}

@article{JawPit2020,
	Author = {Jaworski, P. and Pitera, M.},
	Journal = {Statistics \& Probability Letters},
	Volume ={163},
	Pages = {108800},
	Title = {A note on conditional variance and characterization of probability distributions},
	Year = {2020}}

@article{JelPit2018,
	Author = {Jelito, D. and Pitera, M.},
	Journal = {Statistical Papers},
	Volume = {62},
	Pages = {2083--2108},
	Title = {New fat-tail normality test based on conditional second moments with applications to finance},
	Year = {2021}}

@article{JawPit2015,
	Author = {Jaworski, P. and Pitera, M.},
	Journal = {Discrete \& Continuous Dynamical Systems-Series B},
	Number = {4},
	Title = {The 20-60-20 Rule},
	Volume = {21},
	Year = {2016}}

@article{HebZimWyl2020,
	Author = {Hebda-Sobkowicz, J. and Zimroz, R. and Wy\l{}oma\'nska, A.},
	Journal = {Applied Science},
	Number = {8},
	Pages = {2657},
	Title = {{Selection of the Informative Frequency Band in a Bearing Fault Diagnosis in the Presence of Non-{G}aussian Noise -- Comparison of Recently Developed Methods}},
	Volume = {10},
	Year = {2020}}

@article{HebZimPitWyl2019,
	Author = {Hebda-Sobkowicz, J. and Zimroz, R. and Pitera, M. and Wy\l{}oma\'nska, A.},
	journal = {Mechanical Systems and Signal Processing},
	volume = {145},
	pages = {106971},
	Title = {Informative frequency band selection in the presence of non-{G}aussian noise -- a novel approach based on the conditional variance statistic with application to bearing fault diagnosis},
	Year = {2020}}

@Book{Mit1970,
  author    = {D. S. Mitrinovi\'{c}},
  publisher = {Springer},
  title     = {Analytic Inequalities},
  year      = {1970},
}

@article{BedLocMar2021,
  title={On tails of symmetric and totally asymmetric $alpha$-stable distributions},
  author={Bednorz, Witold and {\L}ochowski, Rafa{\l} and Martynek, Rafa{\l}},
  journal={Probability and Mathematical Statistics},
  volume={41},
  number={2},
  pages={321--345},
  year={2021}
}

@article{Zie2001,
  title={A Reparameterization of the Symmetric $\alpha$-Stable Distributions and Their Dispersive Ordering},
  author={Zieli\'{n}ski, R},
  journal={Theory of Probability \& Its Applications},
  volume={45},
  number={2},
  pages={357--358},
  year={2001},
  publisher={SIAM}
}

@article{PitCheWyl2021,
  title={Goodness-of-fit test for a-stable distribution based on the quantile conditional variance statistics},
  author={Pitera, Marcin and Chechkin, Aleksei and Wy{\l}omanska, Agnieszka},
  journal={Statistical Methods \& Applications},
  pages={387--424},
  year={2022},
  volume={31}
}

@article{OrtLanPet2016,
title = {Asymptotic stochastic dominance rules for sums of i.i.d. random variables},
journal = {Journal of Computational and Applied Mathematics},
volume = {300},
pages = {432--448},
year = {2016},
issn = {0377-0427},
author = {Sergio Ortobelli and Tommaso Lando and Filomena Petronio and Tomas Tich\'{y}}
}

@Book{Picket1998,
title = {Hill, bootstrap and jacknife estimators for heavy tails. A Practical Guide to Heavy Tails},
pages = {283–310},
year = {1998},
author = {Pictet, O.V. and Michel, M.D. and Muller, U.A.}
}

@article{Pickands1975,
title = {Statistical inference using extreme order statistics},
journal = {The Annals of Statistics},
volume = {3},
pages = {119-131},
year = {1975},
author = {Pickands, J.}
}

@article{DeHaan1980,
title = { A simple asymptotic estimate for the index $\alpha$ of a stable distribution},
journal = {Journal of Royal Statistics Society},
volume = {B},
pages = {83-87},
year = {1980},
author = {de Haan, L. and Resnick, S.I.}
}

@article{Dekkers1990,
title = {A moment estimator for the index of an extreme value distribution},
journal = {The Annals of Statistics},
volume = {17},
pages = {1795-1832},
year = {1990},
author = {Dekkers, A.L.M. and Einmahl, J.H.J. and de Haan, L.}
}

@article{Resnick1997,
title = {Heavy Tail modeling and teletraffic data},
journal = {The Annals of Statistics},
year = {1997},
volume = {25},
number= {5},
pages = {1805-1869},
author = {Resnick, S.I.}
}

@article{Pol2006,
  title={Ensemble based systems in decision making},
  author={Polikar, R.},
  journal={IEEE Circuits and systems magazine},
  volume={6},
  number={3},
  pages={21-45},
  year={2006},
  publisher={IEEE}
}

@article{Ghoudi2018,
  title={Serial independence tests for innovations of conditional mean and variance models},
  author={Ghoudi, K.},
  journal={TEST},
  volume={27},
  pages={3–26},
  year={2018},
  publisher={Springer}
}

@article{Escobar2016,
  title={Bias-corrected and robust estimation of the bivariate stable tail dependence function},
  author={Escobar-Bach, M. and Goegebeur, Y. and Guillou, A. and You, A.},
  journal={TEST},
  volume={26},
  pages={284–307},
  year={2017},
  publisher={Springer}
}

@article{Matusi2018,
  title={Goodness-of-fit tests for symmetric stable distributions—Empirical characteristic function approach},
  author={Matsui, M. and Takemura, A.},
  journal={TEST},
  volume={17},
  pages={546–566},
  year={2008},
  publisher={Springer}
}
\end{footnotesize}

%%%%%%%%%%%%%%%%%%%%%%%%%%%%%%%%%%%%%%%%%%%%%%%%%%%%%%%%%%%%%%%%%%%%
\appendix
\section{Additional tables}\label{S:app_figures}

In this section we present additional tables with numerical results. This complements the analysis in Section~\ref{S:comparison} and Section~\ref{S:resistance}.

In Table~\ref{tab:RMSE_MLE} we present the RMSE metrics for $N_1$, $N_2$, the McCulloch method (MCH), the regression method (REG), and the  maximum likelihood method (MLE). The results complements the outcomes presented in Table~\ref{table:MC} by adding the metrics for the maximum likelihood method. As seen in  Table~\ref{tab:RMSE_MLE}, the MLE method exhibits relatively poor performance, particularly for $\alpha$ in moderate to high range. Taking into account this observation and the fact that the results for MLE requires high number of time-consuming calculations, we decided not to include the outcomes for MLE in Section~\ref{S:comparison}.

\begin{table}[!ht]
\centering
\scalebox{0.7}{

\begin{tabular}{|r|rrrrr|}
\multicolumn{5}{c}{$n=250$}\\
  \hline
 $\alpha$ & N1 & N2 & MCH & REG & MLE \\ 
  \hline
1.0 & {\bf 0.089} & 0.123 & 0.091 & 0.091 & 0.091 \\ 
  1.1 & 0.096 & 0.124 & 0.097 & {\bf 0.094} & 0.099 \\ 
  1.2 & {\bf 0.100} & 0.126 & 0.106 & {\bf 0.100} & 0.108 \\ 
  1.3 & {\bf 0.101} & 0.125 & 0.114 & 0.103 & 0.114 \\ 
  1.4 & {\bf 0.102} & 0.121 & 0.118 & 0.105 & 0.118 \\ 
  1.5 & 0.104 & 0.115 & 0.129 & {\bf 0.102} & 0.135 \\ 
  1.6 & {\bf 0.111} & 0.112 & 0.145 & 0.113 & 0.146 \\ 
  1.7 & 0.112 & 0.105 & 0.154 & {\bf 0.104} & 0.156 \\ 
  1.8 & 0.115 & 0.099 & 0.150 & {\bf 0.098} & 0.150 \\ 
  1.9 & 0.102 & 0.083 & 0.134 & {\bf 0.078} & 0.131 \\ 
  2.0 & 0.103 & 0.070 & 0.148 & {\bf 0.042} & 0.145 \\ 
  \hline
\end{tabular}

\begin{tabular}{|r|rrrrr|}
\multicolumn{5}{c}{$n=500$}\\
  \hline
$\alpha$  & $N_1$ & $N_2$ & MCH & REG & MLE \\ 
  \hline
1.0 & {\bf 0.061} & 0.085 & 0.063 & 0.064 & 0.063 \\ 
  1.1 & 0.068 & 0.088 & 0.069 & {\bf 0.067} & 0.071 \\ 
  1.2 & {\bf 0.067} & 0.084 & 0.070 & 0.069 & 0.070 \\ 
  1.3 & {\bf 0.070} & 0.084 & 0.078 & 0.071 & 0.077 \\ 
  1.4 & {\bf 0.074} & 0.087 & 0.083 & 0.075 & 0.083 \\ 
  1.5 & {\bf 0.073} & 0.080 & 0.090 & 0.076 & 0.091 \\ 
  1.6 & {\bf 0.074} & 0.077 & 0.103 & 0.080 & 0.103 \\ 
  1.7 & 0.080 & {\bf 0.070} & 0.116 & 0.076 & 0.116 \\ 
  1.8 & 0.084 & {\bf 0.069} & 0.121 & 0.070 & 0.120 \\ 
  1.9 & 0.083 & 0.067 & 0.110 & {\bf 0.061} & 0.108 \\ 
  2.0 & 0.075 & 0.053 & 0.108 & {\bf 0.031} & 0.105 \\ 
   \hline
\end{tabular}

\begin{tabular}{|r|rrrrr|}
\multicolumn{5}{c}{$n=1000$}\\
  \hline
$\alpha$  & N1 & N2 & MCH & REG & MLE \\ 
  \hline
1.0 & {\bf 0.043} & 0.060 & 0.045 & 0.045 & 0.044 \\ 
  1.1 & {\bf 0.046} & 0.060 & 0.049 & 0.049 & 0.050 \\ 
  1.2 & {\bf 0.048} & 0.059 & 0.050 & 0.049 & 0.050 \\ 
  1.3 & {\bf 0.050} & 0.061 & 0.055 & 0.052 & 0.054 \\ 
  1.4 & {\bf 0.051} & 0.060 & 0.059 & 0.054 & 0.059 \\ 
  1.5 & {\bf 0.052} & 0.056 & 0.062 & {\bf 0.052} & 0.063 \\ 
  1.6 & {\bf 0.052} & 0.053 & 0.072 & 0.056 & 0.072 \\ 
  1.7 & 0.054 & {\bf 0.049} & 0.086 & 0.053 & 0.085 \\ 
  1.8 & 0.059 & {\bf 0.046} & 0.092 & 0.050 & 0.092 \\ 
  1.9 & 0.067 & 0.052 & 0.090 & {\bf 0.045} & 0.089 \\ 
  2.0 & 0.060 & 0.050 & 0.080 & {\bf 0.023} & 0.077 \\ 
   \hline
\end{tabular}
}
\caption{RMSE for 1000 Monte Carlo samples with sizes  $n \in \{250, 500, 1000\}$ and the stability parameter $\alpha \in \{1, 1.1, \ldots, 2\}$. The symbols $N_1$ and $N_2$ refer to the estimators defined by \eqref{eq:hat_alpha}. The symbols MCH, REG, and MLE represent the McCulloch, the regression, and the maximum likelihood methods, respectively. The best performance is marked in bold.}\label{tab:RMSE_MLE}
\end{table}

In Table~\ref{tab:MC_beta_RMSE} we present the RMSE metrics of the estimated stability index for various combinations of  $\alpha$ and  $\beta$. The results are used to produce Figure~\ref{fig_res_2}.

\begin{table}[ht!]
\begin{center}
\scalebox{0.68}{
\begin{tabular}{|r|rrrrrrrrrrr|}
\multicolumn{12}{c}{$N_1$}\\
 \hline
$\alpha$ \textbackslash \, $\beta$
 & 0.0 & 0.1 & 0.2 & 0.3 & 0.4 & 0.5 & 0.6 & 0.7 & 0.8 & 0.9 & 1.0 \\ 
  \hline
1.1 & 0.046 & 0.047 & 0.048 & 0.051 & 0.055 & 0.057 & 0.060 & 0.060 & 0.060 & 0.061 & 0.061 \\ 
  1.2 & 0.048 & 0.049 & 0.051 & 0.052 & 0.055 & 0.058 & 0.059 & 0.061 & 0.061 & 0.062 & 0.062 \\ 
  1.3 & 0.049 & 0.050 & 0.052 & 0.053 & 0.056 & 0.058 & 0.061 & 0.062 & 0.064 & 0.066 & 0.066 \\ 
  1.4 & 0.050 & 0.050 & 0.052 & 0.054 & 0.056 & 0.059 & 0.061 & 0.064 & 0.066 & 0.069 & 0.072 \\ 
  1.5 & 0.051 & 0.051 & 0.052 & 0.053 & 0.056 & 0.059 & 0.062 & 0.065 & 0.069 & 0.074 & 0.078 \\ 
  1.6 & 0.051 & 0.052 & 0.053 & 0.055 & 0.057 & 0.059 & 0.064 & 0.067 & 0.071 & 0.076 & 0.082 \\ 
  1.7 & 0.054 & 0.055 & 0.055 & 0.056 & 0.058 & 0.061 & 0.064 & 0.068 & 0.071 & 0.076 & 0.081 \\ 
  1.8 & 0.062 & 0.062 & 0.062 & 0.062 & 0.064 & 0.064 & 0.066 & 0.069 & 0.072 & 0.075 & 0.078 \\ 
  1.9 & 0.066 & 0.066 & 0.067 & 0.066 & 0.068 & 0.068 & 0.069 & 0.069 & 0.071 & 0.071 & 0.072 \\ 
  2.0 & 0.064 & 0.064 & 0.066 & 0.065 & 0.065 & 0.064 & 0.065 & 0.065 & 0.064 & 0.065 & 0.064 \\ 
   \hline
\end{tabular}
\begin{tabular}{|r|rrrrrrrrrrr|}
\multicolumn{12}{c}{$N_1$}\\
 \hline
$\alpha$ \textbackslash \, $\beta$
 & 0.0 & 0.1 & 0.2 & 0.3 & 0.4 & 0.5 & 0.6 & 0.7 & 0.8 & 0.9 & 1.0 \\ 
\hline
1.1 & 0.060 & 0.060 & 0.063 & 0.066 & 0.070 & 0.074 & 0.078 & 0.080 & 0.081 & 0.082 & 0.081 \\ 
  1.2 & 0.060 & 0.060 & 0.062 & 0.064 & 0.069 & 0.072 & 0.075 & 0.078 & 0.079 & 0.078 & 0.078 \\ 
  1.3 & 0.059 & 0.059 & 0.061 & 0.063 & 0.066 & 0.069 & 0.072 & 0.075 & 0.076 & 0.077 & 0.076 \\ 
  1.4 & 0.058 & 0.057 & 0.059 & 0.061 & 0.063 & 0.067 & 0.069 & 0.071 & 0.072 & 0.073 & 0.075 \\ 
  1.5 & 0.054 & 0.055 & 0.056 & 0.057 & 0.060 & 0.062 & 0.065 & 0.068 & 0.070 & 0.073 & 0.075 \\ 
  1.6 & 0.052 & 0.052 & 0.053 & 0.054 & 0.056 & 0.059 & 0.062 & 0.064 & 0.067 & 0.071 & 0.075 \\ 
  1.7 & 0.049 & 0.049 & 0.049 & 0.050 & 0.052 & 0.055 & 0.058 & 0.062 & 0.064 & 0.069 & 0.073 \\ 
  1.8 & 0.048 & 0.048 & 0.048 & 0.049 & 0.050 & 0.052 & 0.055 & 0.057 & 0.062 & 0.064 & 0.069 \\ 
  1.9 & 0.052 & 0.052 & 0.052 & 0.052 & 0.053 & 0.053 & 0.054 & 0.055 & 0.058 & 0.058 & 0.059 \\ 
  2.0 & 0.054 & 0.054 & 0.054 & 0.054 & 0.054 & 0.054 & 0.055 & 0.054 & 0.054 & 0.054 & 0.053 \\ 
   \hline
\end{tabular}}
\end{center}
\caption{RMSE of the estimated stability index for various combinations of  $\alpha \in \{1.1,1.2,\ldots,2\}$ and  $\beta \in \{0, 0.1,\ldots,1\}$; the number of strong Monte Carlo simulations is $10\, 000$ and the sample size is $n=1000$. }
\label{tab:MC_beta_RMSE}
\end{table}

\end{document}